\documentclass[twocolumn]{aa} 

\usepackage{graphicx}
\usepackage{booktabs}
\usepackage{amsmath}
\usepackage{color}
\usepackage{txfonts}

\newcommand{\msol}{\ensuremath{\rm M_{\odot}}\xspace}

\newcommand{\lsol}{\ensuremath{\rm L_{\odot}}\xspace}
\providecommand{\lum}{\ensuremath{{L}}}

\newcommand{\mstar}{\ensuremath{M_{\star}}\xspace}

\newcommand{\ngc}{NGC\,2477}

\newcommand{\lightkurve}{{\tt LightKurve }}
\newcommand{\teff}{\ensuremath{T_{\rm eff}}}
\newcommand{\mh}{[M/H]}

\newcommand{\feh}{[Fe/H]}
\newcommand{\logg}{\ensuremath{\log g}}
\newcommand{\ebr}{\ensuremath{E({\rm BP} - {\rm RP})}}
\newcommand{\ag}{\ensuremath{A_{G}}}
\newcommand{\ebv}{\ensuremath{E(B - V)}}
\newcommand{\ebvbkg}{\ensuremath{E(B - V)_{bkg}}}
\newcommand{\gbprp}{\ensuremath{(G_{BP}-G_{RP})}}
\newcommand{\ebprp}{\ensuremath{E(BP-RP)}}
\newcommand{\deltanu}{\ensuremath{\Delta \nu}}

\newcommand{\gaia}{{Gaia}}
\newcommand{\vsini}{$v\sin i$}
\usepackage[linkcolor=blue]{hyperref}%
\usepackage{multirow}

\begin{document}

\title{
Asteroseismic age constraints on the open cluster NGC 2477 using oscillating stars identified with TESS FFI}
 \titlerunning{Age constraints for open cluster NGC 2477} 
   \author{D. B. Palakkatharappil
          \inst{1}\thanks{   \email{dinilbose@oca.eu}}
          \and
          O. L. Creevey\inst{1}\fnmsep
          }

   \institute{Université Côte d'Azur, Observatoire de la Côte d'Azur, CNRS, Laboratoire Lagrange, Bd de l'Observatoire, CS 34229, 06304 Nice cedex 4, France
             }

   \date{Received September 15, 1996; accepted March 16, 1997}

 
  \abstract
   {Asteroseismology is one of the few methods to derive ages of individual stars due to the high precision of their observations.   Isochrone fitting is a powerful alternative method for deriving ages by studying clusters of stars. Pulsating stars in clusters should therefore allow for detailed studies of the stellar models.  }
   {Our objectives are to exploit the NASA TESS data along with ESA Gaia data to search for and detect oscillations in cluster member stars.   We analyse the intermediate-age open cluster \ngc, known to suffer from differential extinction, to explore if asteroseismology and cluster characteristics can help understand the metallicity and extinction, as well as  result in better age determinations than isochrone-fitting alone.}
   {We combined a multitude of recent observations from \gaia, high-resolution spectroscopy, and extinction maps to analyse the cluster and then search for and detect variability in the member stars using TESS full frame images (FFIs) data. To interpret all of these data, we used stellar structure, evolution and oscillation codes. }
   {We conducted an in-depth analysis of the extinction and metallicity of \ngc, using the most recent spectroscopic, photometric, and extinction observations for the cluster. Analysis of dust and extinction maps confirmed that the differential extinction in the direction of the cluster is not due to the background. The cluster's metallicity from high-resolution spectroscopy varies from 0.06 to 0.16~dex. We performed an isochrone fitting to the cluster using publically available isochrones (BASTI, MIST, and PARSEC), which provides a cluster age of between 0.6 to 1.1~Ga. 
   Then using TESS FFI, we analysed the time dimension of the members of this cluster. We created optimised pixel light curves using the {\tt tessipack} package which allows us to consider possible contamination by nearby stars. Using these light curves, we identified many interesting levels of variability of stars in this cluster, including binaries and oscillating stars. For the asteroseismic analysis, we selected a few uncontaminated  {A--F type} oscillating stars 
   and used the MESA and GYRE codes to interpret the frequency signals.
   By comparing the theoretical and the observed spectra, we identified frequency separations, \deltanu, for {four} stars. Then using the identified \deltanu\ and imposing that the best matched theoretical models have the same age, metallicity, and background extinction, we constrained the cluster's age to 1.0 $\pm$ 0.1~Ga. } 
   {{We conclude that using the TESS FFI data}, we can identify oscillating stars in clusters and constrain the age of the cluster using asteroseismology. }

   \keywords{Asteroseismology
 -- Stars: fundamental parameters --  open clusters and associations: individual: NGC 2477 -- Stars: evolution -- Stars: oscillations -- dust, extinction}
   \maketitle
%

\section{Introduction}

Knowledge of the ages of stars is a crucial element to understand evolutionary processes in astrophysical systems, such as the evolution of our Milky Way populations \citep{gallart_2019} or stellar evolutionary processes such as rotation and activity \citep{garcia_2014,epstein_2014}. The age of a star can only be measured in some very specific cases. In general, we rely on stellar evolution models to inform us of this quantity; for more details, readers can refer to  \cite{soderblom_2010} and \cite{lebreton2014a}. However, the stellar evolution models also need to be tested and corrected for missing ingredients, such as diffusion in stars \citep{hidalgo_2018} which impacts the stellar lifetime and the star's observable properties.

The best candidates for exploring these stellar interior processes are stellar Galactic clusters where members are assumed to be born at the same time (in stellar evolution timescales) and from the same molecular cloud and thus share the same initial chemical conditions. {The shape of a cluster in the colour-magnitude diagram gives information on potential missing ingredients in the models.  Studying several members at once also removes some of the {free parameters} involved when aiming to determine the age of the member stars.}

Asteroseismology is the technique of interpreting stellar oscillation frequencies which probe
the interior of stars (\citealt{aerts_2021} and references therein). In particular, as a star evolves, it builds up a steep chemical gradient in its core as H is converted to He, and stellar oscillation frequencies are sensitive to these structural changes. This allows us to determine a precise age.
Applying this technique to the age of Sun using hundreds of available high mode degrees 
\citep{deubner_1983,isaak_1989,demarque_1999,chaplin_2013} has allowed for the determination of the Sun's age of 4.57~Ga; for more details, readers can refer to \cite{bonanno2002}, for example, who derived 4.57 $\pm$ 0.11~Ga.
Even reducing the observational constraints to those similar for other stars, the Sun can still be derived with a precision of a few percent; for more details, readers can refer to \citet{creevey17}, for example, who derived an age of 4.38 $\pm$ 0.22~Ga, by studying frequency ratios considering star-like uncertainties. 

The detection of oscillations on stars in clusters provides the possibility to perform studies on the ages of stars from stellar models. For example, a direct comparison of two independently determined ages of a cluster may bring insights into missing physics or the treatment of physics in the models.  Or, using the cluster age and metallicity constraints {may} allow one to test the computation of oscillation models of stars. 
The possibilities are open. {For example, \cite{miglio_2012} studied two open clusters (NGC 6791 and 6819) using $Kepler$ \citep{gilliland_2010} data and identified oscillating red giant stars in them.} They used seismic parameters characterising solar-like oscillations ($\Delta \nu$ and $\nu_{max}$) to estimate the masses of stars in the red clump and on the low-luminosity red giant branch.

In this paper, we use data from NASA TESS satellite launched in 2018 \citep{ricker_2014}. It collects time series for a minimum of 27 days to a maximum of one year across 80\% of the sky, just above and below the ecliptic plane. The availability of these time series provides an excellent opportunity to search for and characterise oscillating stars in clusters. At the same time, the Gaia satellite \citep{gdr2,gdr3} provides complementary information on cluster member properties, such as proper motions (giving membership probabilities), parallax combined with photometry providing the observable quantities for interpreting clusters with models.  

In order to exploit the TESS data, we have developed an analysis tool {\tt tessipack} to explore the TESS full frame images (FFIs) to detect oscillations \citep{palakkatharappil_2021}. In this paper, we apply this methodology to the intermediate-aged open cluster \ngc\ with the aim to better characterise the age of the cluster with asteroseismology. 

In Section~\ref{sec:data} we discuss our current knowledge of \ngc.  Then, in Section~\ref{sec:extinction} we perform a new thorough analysis of the extinction towards the cluster to properly correct for the differential reddening. This allows us to perform new isochrone fits (Sect.~\ref{sec:isochrone}). We then describe the analysis of the TESS data for this cluster (Sect.~\ref{sec:tess}), along with the detection of variability and oscillations of stars (Sect.~\ref{sec:tessvariability}). Asteroseismic modelling to estimate the age of the cluster using standard  and rotating models is discussed in Sections \ref{sec:models} and \ref{sec:rotation}.  We summarise our findings and conclude in  Sect.~\ref{sec:conclusions}.

\section{NGC 2477 \label{sec:data}}

\subsection{General properties}\label{ssec:data_general}

\begin{figure}
\centering
\includegraphics[width=\hsize]{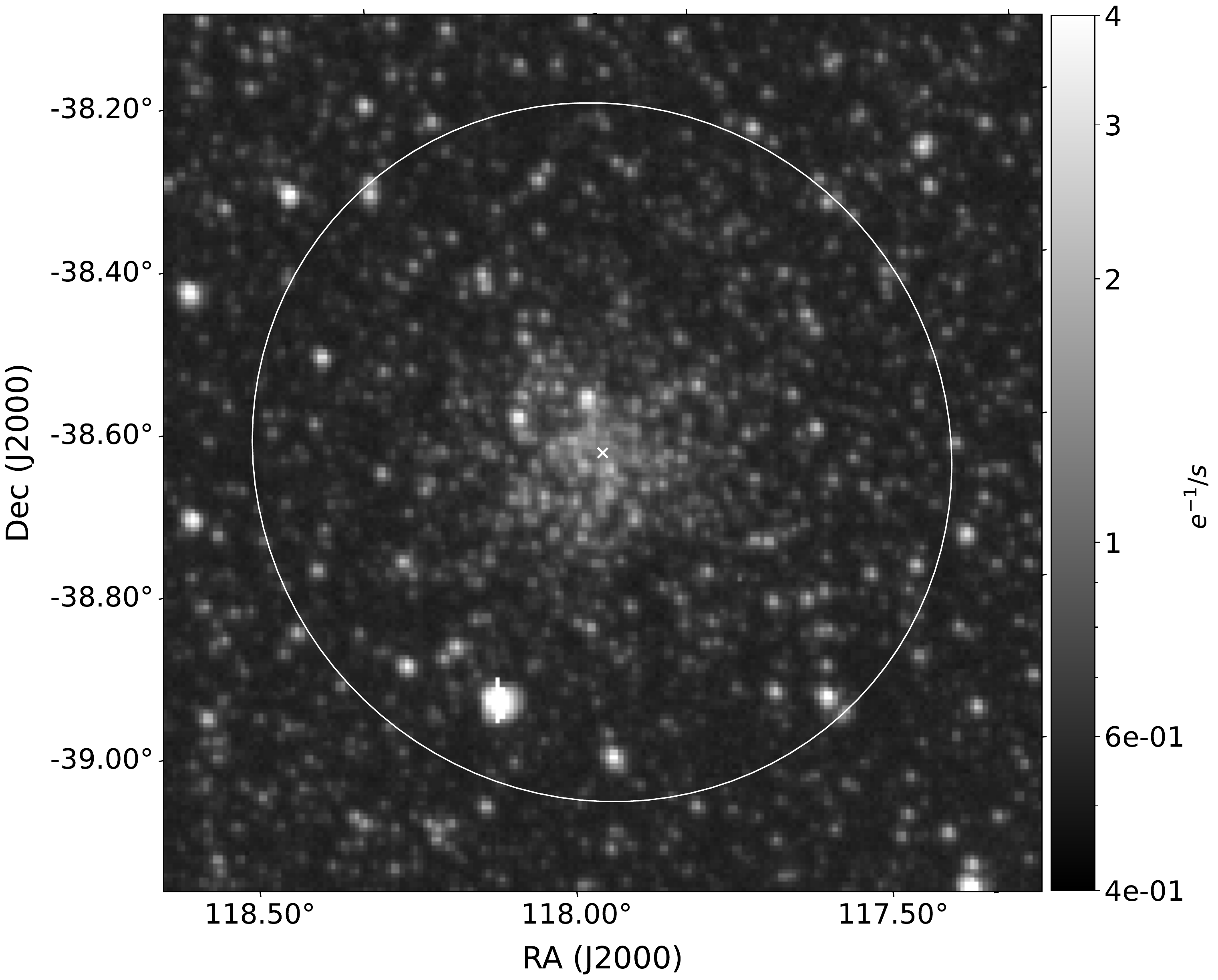}

    \caption{TESS FFI image of \ngc\ observed at sector 7 (camera 3, CCD 1) at 2019-01-15 12:03:39.542 UTC for 30 minutes. The radius of cluster (marked by the white circle) is 26\arcmin and the centre is marked by 'x'. }
        \label{fig:FFI_ngc2477}
\end{figure}

\begin{figure}
\centering
    \includegraphics[width=0.49\textwidth]{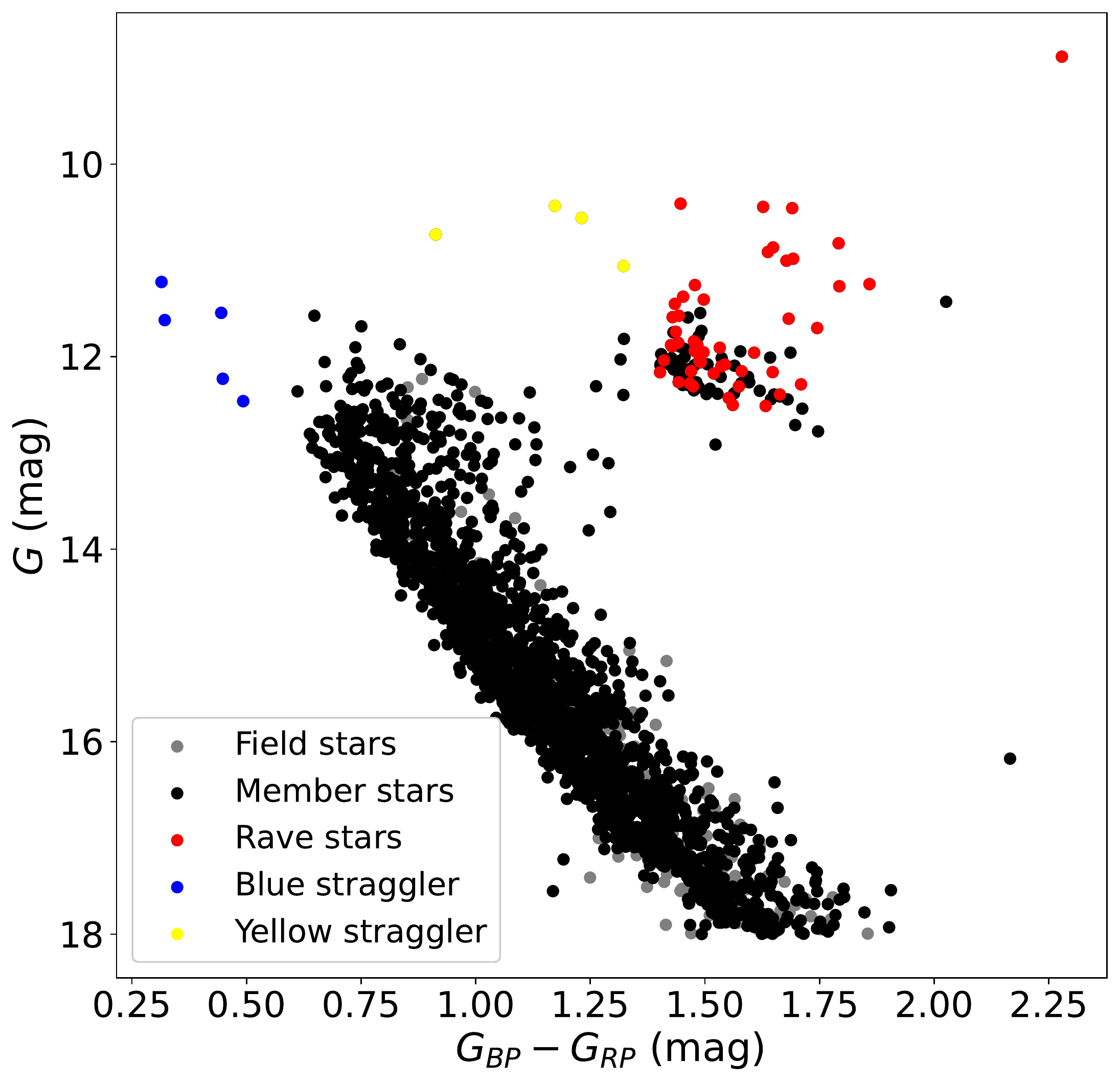}

         \caption{Observed colour-magnitude diagram of  NGC 2477. Stars observed by RAVE are given in red. Blue and yellow stragglers reported by \cite{rain_2020} shown by the blue and yellow points, respectively.}
         \label{fig:cmd_cluster_obs}
\end{figure}

The cluster \ngc\ is one of the richly populated open cluster located in Puppis' constellation ($l$=253.6$^{\circ}$; $b$=-5.8$^{\circ}$). The distribution of stars in the cluster is shown in Fig \ref{fig:FFI_ngc2477}. The first comprehensive photometric study was done by \cite{hartwick_1972}. They obtained DDO\footnote{David Dunlap Observatory} photometry for 26 giants stars and UBV photometry for 2000 stars. They found that the metallicity of the cluster should be 1.5 times the metallicity of the Hyades cluster and the  age to be 1.5 $\pm$ 0.2~Ga. The authors also pointed out strong differential reddening, later confirmed by \cite{hartwick_1974}.  They estimated a differential reddening $E(B-V)$ from about 0.2 to 0.4~mag by intermediate-band photometry. The broadening of the main sequence as seen in the colour-magnitude diagram (CMD, Fig.~\ref{fig:cmd_cluster_obs}) is  due to this differential reddening. 

\cite{eigenbrod_2004} obtained a mean radial velocity of 7.32$\pm$0.13 km s$^{-1}$ by studying 49 constant velocity member red giant stars. They also found 13 spectroscopic red giant binaries in the cluster. The Gaia DR2 catalogue \citep{katz_2019} has obtained radial velocities for 119 member stars and the mean radial velocity for the cluster is found to be 8.42$\pm$0.39 km~s$^{-1}$.

\cite{jeffery_2011} used the WFPC2 instrument on the Hubble Space Telescope to identify seven-candidate white dwarfs in the cluster. Recently, it has been shown that the cluster also harbours five blue stragglers and four yellow straggler candidates \citep{rain_2020} marked in Fig.~\ref{fig:cmd_cluster_obs}.

\subsection{Cluster membership}\label{ssec:data_membership}

Using the Gaia DR2 catalogue \citep{gdr2}, \citet{cantat-gaudin_2018} have characterised a large number of Galactic open clusters. They used proper motion ($\mu_{\alpha}, \mu_{\beta}$) and parallax ($\varpi$) to derive the membership probabilities using the Unsupervised Photometric Membership Assignment in Stellar Clusters code (UPMASK) described by \citet{krone-martins_2014}. From the catalogue of \citeauthor{cantat-gaudin_2018} we have identified  2039 stars with membership probabilities greater than or equal to 0.5 as cluster members. We have adopted our own naming convention for each star in the cluster. The adopted name along with position, probability membership, \gaia\ $G$ magnitude, colour and DR3 ID is given in Table~\ref{tab:sourceinfo}. 

\citeauthor{cantat-gaudin_2018} also derived a distance to the cluster of 1441 $\pm$ 1 pc through a maximum likelihood procedure using parallax and its uncertainties. This is in agreement with that of \cite{bailer-jones_2018} who derived distances to the individual stars.  {Considering only the member stars, the distance we obtained from that catalogue is 1444$\pm$ 81 pc. We use the latter reference for the distance in this paper. A summary of the global properties of the cluster are given in Table~\ref{tab:literatureages}}.

\begin{table*}[]
\caption{Gaia source identification, position, magnitude, and observed colours of the member stars \cite{gdr2,gdr3}. 
}
\begin{tabular}{lccccccc}
\toprule
ID &                  \gaia\ DR2/3 ID &     RA &    Dec &       $G$ &     $BP - RP$ &  PMemb & Code\tablefootmark{*} \\
&                   &     (deg) &    (deg) &       (mag) &     (mag) &   &  \\
\midrule
       N77-0 &  5538869956136670976 &  118.079 &  -38.473 &   8.88 &   2.28 &    0.9 &      \\
       N77-1 &  5538866176565512960 &  118.074 &  -38.557 &  10.41 &   1.45 &    0.9 &      \\
       N77-2 &  5538869956136673792 &  118.082 &  -38.482 &  10.43 &   1.17 &    1.0 &      \\
       N77-3 &  5538869058489205504 &  118.030 &  -38.556 &  10.44 &   1.63 &    1.0 &  ORD \\
       N77-4 &  5538866623242071936 &  118.094 &  -38.529 &  10.46 &   1.69 &    0.9 &      \\
       N77-5 &  5538493579557251072 &  117.983 &  -38.614 &  10.56 &   1.23 &    0.9 &  ORD \\
       N77-6 &  5538866344059037952 &  118.126 &  -38.542 &  10.73 &   0.91 &    0.8 &      \\
       N77-7 &  5538868100710944640 &  118.178 &  -38.485 &  10.82 &   1.79 &    0.9 &  ORD \\
       N77-8 &  5538490349748201856 &  118.062 &  -38.647 &  10.87 &   1.65 &    0.5 &  ORD \\
       N77-9 &  5538865798608422400 &  118.078 &  -38.604 &  10.91 &   1.64 &    0.9 &      \\
      N77-210 &  5538864797870824448 &  118.139 &  -38.626 &  12.77 &   0.72 &    0.9 &    O \\   
      N77-255 &  5538506911142214272 &  117.918 &  -38.490 &  12.95 &   0.99 &    0.9 &    O \\
      N77-325 &  5538510415835055616 &  117.680 &  -38.436 &  13.25 &   0.67 &    0.7 &    O \\
      N77-342 &  5538870639026317312 &  118.037 &  -38.480 &  13.30 &   0.75 &    1.0 &    O \\
\bottomrule
\end{tabular}
\tablefoot{Membership Probability (PMemb) is from \cite{cantat-gaudin_2018} while Column "Code" indicates the type of variability we detect, see Section \ref{sec:tessvariability}. 
The entire table~is also available in machine-readable form in the electronic edition of A $\&$ A.
\tablefoottext{*}{O: Oscillating stars. B: Candidate binaries. D: Stars contaminated by nearby source. R: Oscillating red-giants. U: Unknown variable Stars}}
\label{tab:sourceinfo}
\end{table*}

\begin{table*}[]
\caption{ Metallicity, reddening, age, and distance of \ngc\ according to the literature, and the values adopted in this work.   }
\begin{tabular}{lccccccc} \toprule
 &  \multicolumn{4}{c}{Determination of metallicity} & \\
Paper & Value &Technique &Spectral &No. & \ebv\ & Age & Distance \\
&   (dex)& & Resolution & Stars& (mag) & (Ga) & (pc) \\ \midrule
 \cite{hartwick_1972} &    &&&& 0.2-0.4     & 1.5  $\pm$  0.2 &       \\ 
 \cite{hartwick_1974} &    &&&& 0.2-0.4     &0.9&       \\ 
\cite{geisler_1992}   & -0.13 $\pm$ 0.18 & PHOT & -    &&&&\\
 \cite{friel_1993} & -0.05 $\pm$ 0.11 & SPEC &1000&7&      &1.3&       \\ 
 \cite{vonhippel_1995} & Solar   &&&&0.3& 1  $\pm$  0.2 &       \\ 
 \cite{kassis_1997} & -0.05 $\pm$ 0.11 &&&& 0.2-0.4     & 1$^{+0.3}_{0.2}$ &   1300    \\ 
 \cite{friel_2002} & -0.13 $\pm$ 0.1 & SPEC &1250&28&0.29&1& 1260      \\ 
 \cite{bonatto_2005} & Solar   &&&& 0.06  $\pm$  0.03 & 1.1  $\pm$  0.1 & 1200      \\ 
 \cite{carraro_2005} & Solar   &&&&0.24&0.6& 1300      \\ 
\cite{bragaglia_2008} & 0.07 $\pm$ 0.03  & SPEC &45000&4&&&\\
 \cite{monteiro_2010} &-0.07&&&& 0.29  $\pm$  0.03 & 0.79  $\pm$  0.2 & 1385  $\pm$  64  \\ 
 \cite{jeffery_2011} &  &&&&      & 1.035  $\pm$  0.054 &       \\ 
 \cite{dias_2012} \tablefootmark{a} &0.118&&&& 0.29  $\pm$  0.02 & 0.67  $\pm$  0.1 & 1565  $\pm$  103  \\ 
 \cite{dias_2012} \tablefootmark{b}&0.118&&&& 0.27  $\pm$  0.09 & 0.89  $\pm$  0.1 & 1300  $\pm$  117  \\ 
 \cite{oliveira_2013} & -0.10 $\pm$ 0.12 &&&& 0.31  $\pm$  0.03 & 0.7  $\pm$  0.15 & 1341  $\pm$  106  \\ 
\cite{caffau_2014}    & 0.09 $^{*}$ & SPEC &47000&4&&&\\
\cite{mishenina_2015} & 0.18 $\pm$ 0.11  & SPEC &47000&5&&&\\
 \cite{jeffery_2016} & -0.24 $\pm$ 0.04 &&&& 0.27  $\pm$  0.009 & 1.02  $\pm$  0.2 &       \\ 
\cite{kunder_2017}    & -0.07 $\pm$ 0.1 $^{*}$  & SPEC &7500&55&&&\\

This work & $0.10 \pm 0.05 $  & & & & {$0.29 \pm 0.05$}  & $1.0 \pm 0.1 $ & $1444 \pm 81$ \\
\bottomrule
\end{tabular}
\tablefoot{Some authors fit isochrones to the full cluster, others use only parts of the cluster.  
The solar iron abundance values adopted by different authors vary between 7.50 to 7.67. {$^{*}$Metallicity values reported as [M/H] instead of [Fe/H]}.   PHOT and SPEC imply photometric and spectroscopic measurements, respectively; if blank the value was either fixed from the literature or fitted to isochrones.
\tablefoottext{a} UBV photometric data used for fitting isochrones.  
\tablefoottext{b} 2MASS JHK photometric data used for fitting isochrones.}
\label{tab:literatureages}
\end{table*}

\subsection{Metallicity}
\label{ssec:metallicity}
Initial photometric abundance measurements towards the cluster were carried out by \cite{geisler_1992} and they found a mean {[Fe/H]} of --0.13 $\pm$ 0.18~dex, although such measurements have been shown to have errors of the order of 0.15 -- 0.20~dex \citep{gratton_2000}. Low-resolution spectroscopic metallicity measurements (R$\sim$1000) were carried out by \cite{friel_1993} and \cite{friel_2002} where they found {[Fe/H] to be} --0.05 $\pm$ 0.11 and --0.13 $\pm$ 0.1~dex, respectively.  

\begin{figure}
 \centering
    \includegraphics[width=0.49\textwidth]{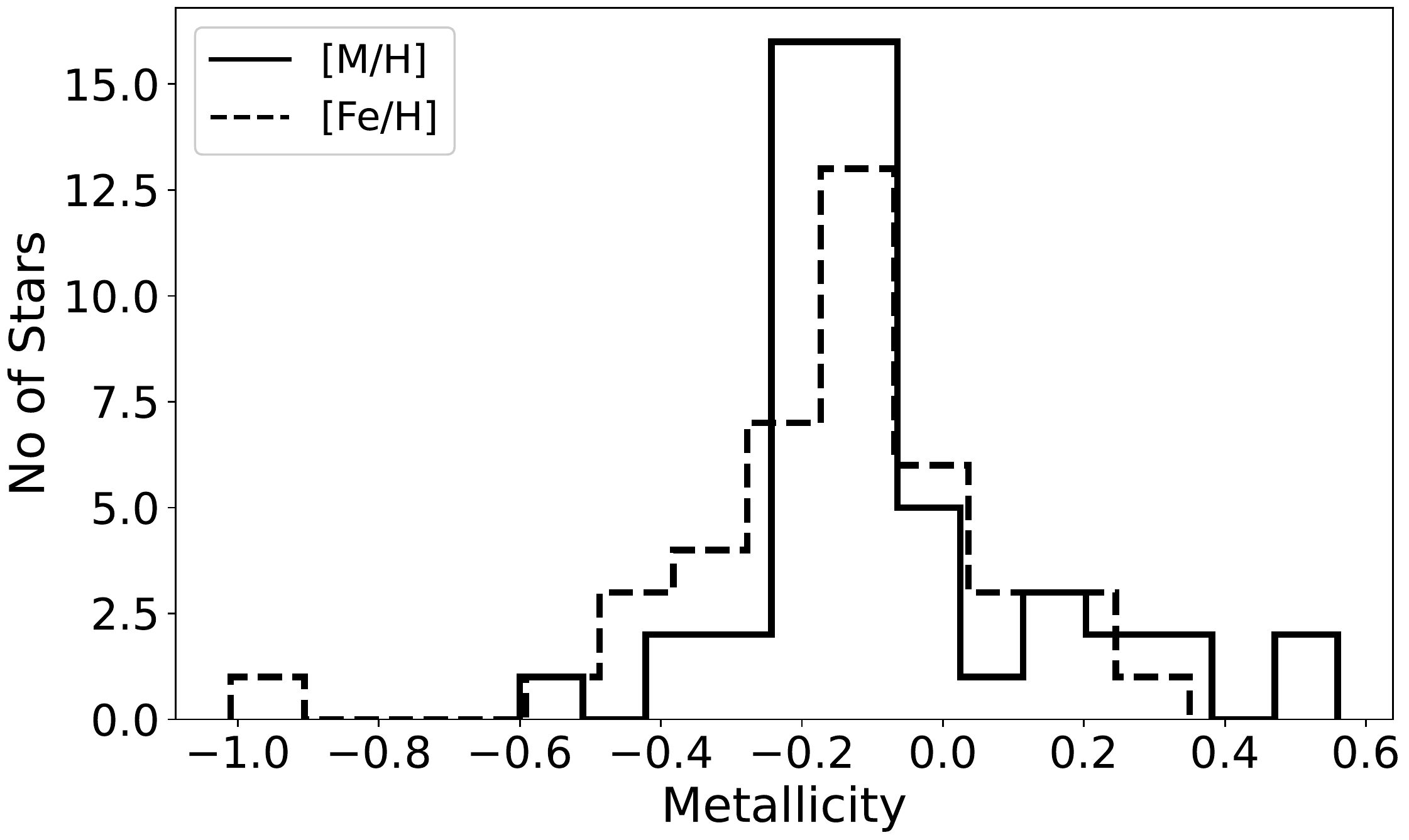}    
\caption{Distribution of \mh\ and \feh\ derived from the RAVE survey for the red giant members of \ngc.}
\label{fig:distribution_of_mh}
\end{figure}

More recently, the Radial Velocity Experiment (RAVE) survey obtained medium-resolution spectra (R$\sim$7\,500) covering the Ca-triplet region ($841.0 - 879.4~$nm) using the Six Degree Field multi-object spectrograph on the 1.2 m UK Schmidt Telescope of the Anglo-Australian Observatory. RAVE has observed 53 member stars of the cluster \ngc\ which all 
lie on the red giant branch shown as red points in Fig.~\ref{fig:cmd_cluster_obs}. {Their distribution of [M/H] and [Fe/H]  from the fifth data release of RAVE \citep{kunder_2017} are shown in   Fig.~\ref{fig:distribution_of_mh} }.  The mean \mh\ and [$\alpha$/Fe] are -0.07 and +0.1~dex, respectively, with a scatter of 0.21~dex. 
{\cite{bragaglia_2008} also reported the [$\alpha$/Fe] to be $0.02 \pm 0.03$~dex.}
{ Since the $\alpha$-enhancement is negligible, we approximate \mh\ $\sim$ \feh\ in this section.} 

High-resolution spectroscopy (R $\geq$ 25\,000) has been shown to be  more reliable compared to other methods to derive the metallicity; for more details, readers can refer to \cite{heiter_2014,magrini_2009}. High-resolution ($R \sim 45\,000$) spectroscopic observations  for some giant cluster members have been carried out by \cite{bragaglia_2008,caffau_2014,mishenina_2015} using the multi-object fibre-fed FLAMES facility mounted at the European Southern Observatory (ESO)-VLT/UT2 telescope. The derived parameters of these members are given in Table~\ref{tab:spectrsocopicmetallicity}. 

\begin{figure}
 \centering
    \includegraphics[width=0.49\textwidth]{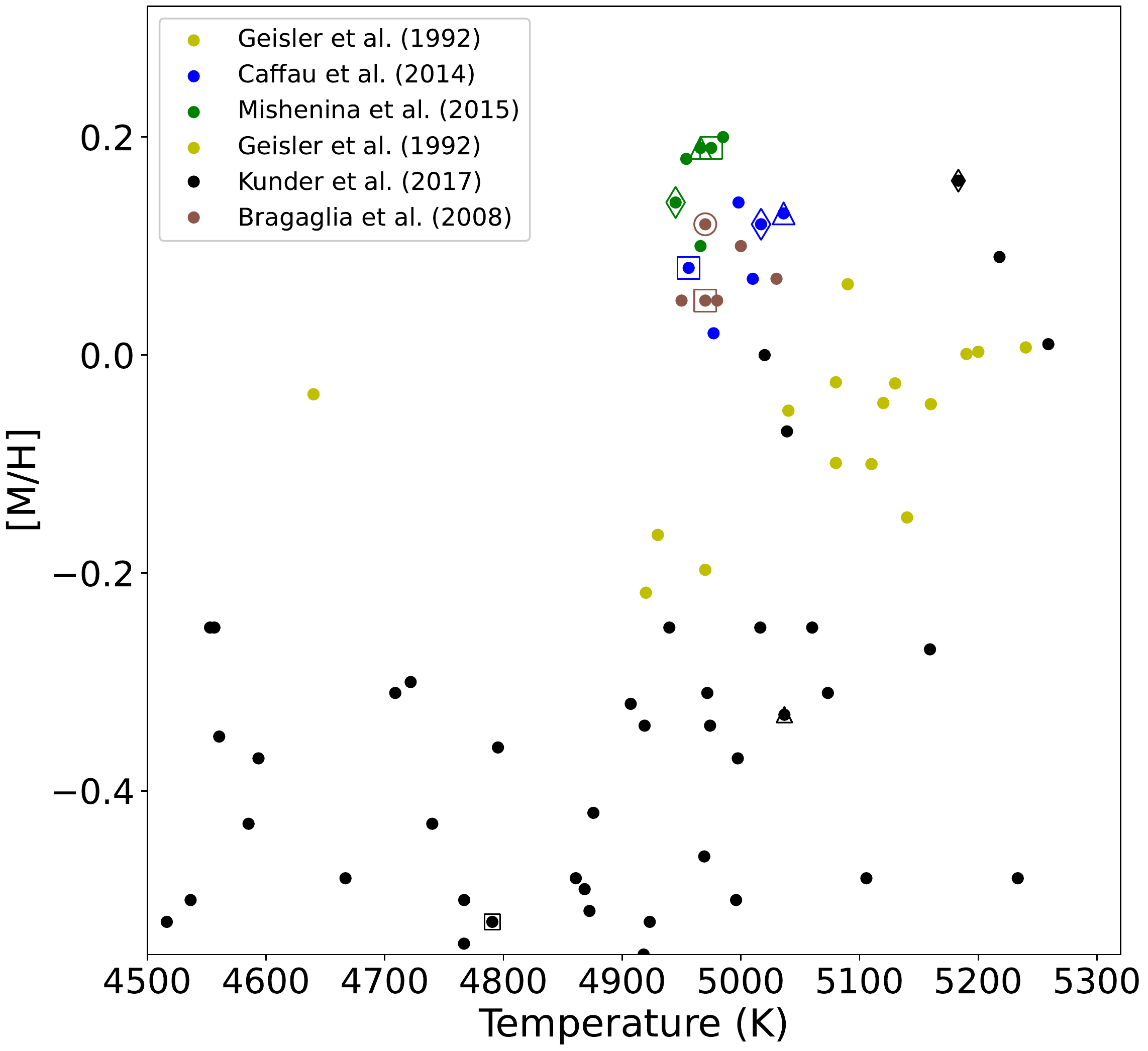}
\caption{Distribution of high- and low-resolution spectroscopic determinations of \teff\ and \mh.
The common stars are denoted by open symbols: N77-39 (square), N77-42 (triangle), N77-47  (circle), and N77-49 (diamond).  Only the latter is observed by all surveys.
}
\label{fig:temp_feh_comparison}
\end{figure}

\begin{table*}[]
\caption{Basic parameters and spectroscopic determinations of the metallicity and the \teff\ of several members of \ngc.}
\label{tab:spectrsocopicmetallicity}
\resizebox{18cm}{!}{%
\begin{tabular}{@{}lccccccccccccc}
\toprule
ID           & WEBDA ID & {RA}    & {Dec}   & $G$  & \gbprp & {$V$}     & {$B$}     & \teff & {$\log g$} & \feh & $\rm \sigma([Fe/H])$ & {$E(B-V)$}  \\ 
                        &           & {(deg)} & {(deg)} & {(mag)} & {(mag)} & {(mag)} & {(mag)} & (K)  & (dex)      &(dex)                    & (dex)                & (mag)         \\
                        \hline 
\multicolumn{6}{l}{\cite{bragaglia_2008}} \\
N77-119                & 13385     & 118.1055                  & -38.6441                  & 12.334                    & 1.511                     & 12.77                     & 14.01                     & 4980 & 2.80                      & 0.05               & 0.11            & 0.31                             \\
N77-39                 & 4221      & 118.1522                  & -38.6318                  & 11.88                     & 1.426                     & 12.23                     & 13.43                     & 4970 & 2.68                      & 0.05               & 0.11            & 0.27                            \\
N77-46                 & 5035      & 118.0448                  & -38.5878                  & 11.919                    & 1.455                     & 12.31                     & 13.52                     & 5000 & 2.70                      & 0.10               & 0.11            & 0.30                         \\
N77-50                 & 3206      & 118.2293                  & -38.5535                  & 11.955                    &                        & 12.32                     & 13.54                     & 4950 & 2.66                      & 0.05               & 0.11            & 0.28                            \\
N77-113                & 2061      & 118.1743                  & -38.5369                  & 12.309                    & 1.474                     & 12.71                     & 13.91                     & 5030 & 2.67                      & 0.07               & 0.11            & 0.30                         \\
N77-47                 & 8039      & 118.0033                  & -38.4803                  & 11.94                     & 1.478                     & 12.32                     & 13.55                     & 4970 & 2.65                      & 0.12               & 0.11            & 0.30                            \\
\multicolumn{6}{l}{\cite{mishenina_2015}}       \\
N77-33                 & 4027      & 118.0878                  & -38.5772                  & 11.748                    & 1.431                     & 12.15                     & 13.35                     & 4966 & 2.70                      & 0.10               &                 &                                 \\
N77-39                 & 4221      & 118.1522                  & -38.6318                  & 11.88                     & 1.426                     & 12.27                     & 13.44                     & 4975 & 2.80                      & 0.19               &                 &                                 \\
N77-54                 & 5076      & 118.0615                  & -38.6292                  & 12.004                    & 1.456                     & 12.41                     & 13.63                     & 4954 & 2.70                      & 0.18               &                 &                                 \\
N77-42                 & 7266      & 117.955                   & -38.5357                  & 11.894                    & 1.428                     & 12.25                     & 13.45                     & 4966 & 2.80                      & 0.19               &                 &                                \\
N77-60                 & 7273      & 117.9478                  & -38.5434                  & 12.032                    & 1.418                     & 12.39                     & 13.56                     & 4985 & 2.80                      & 0.20               &                 &                                 \\
N77-49                 & 8216      & 118.0647                  & -38.4573                  & 11.955                    & 1.497                     & 12.33                     & 13.61                     & 4945 & 2.70                      & 0.14               &                 &                                \\
\multicolumn{6}{l}{\cite{caffau_2014}}      \\
N77-33                 & 4027      & 118.0878                  & -38.5772                  & 11.748                    & 1.431                     & 12.15                     & 13.35                     & 4998 & 2.78                      & 0.14               &                 &                                \\
N77-39                 & 4221      & 118.1522                  & -38.6318                  & 11.88                     & 1.426                     & 12.27                     & 13.44                     & 4956 & 2.70                      & 0.08               &                 &                                 \\
N77-54                 & 5076      & 118.0615                  & -38.6292                  & 12.004                    & 1.456                     & 12.41                     & 13.63                     & 5010 & 2.80                      & 0.07               &                 &                                 \\
N77-42                 & 7266      & 117.955                   & -38.5357                  & 11.894                    & 1.428                     & 12.25                     & 13.45                     & 5036 & 2.92                      & 0.13               &                 &                                \\
N77-60                 & 7273      & 117.9478                  & -38.5434                  & 12.032                    & 1.418                     & 12.39                     & 13.56                     & 4977 & 2.67                      & 0.02               &                 &                                 \\
N77-49                 & 8216      & 118.0647                  & -38.4573                  & 11.955                    & 1.497                     & 12.33                     & 13.61                     & 5017 & 2.84                      & 0.12               &                 &                               \\
\multicolumn{6}{l}{\cite{kunder_2017}}\\
N77-39$^{*\dagger}$    & 4221      & 118.1522                  & -38.6318                  & 11.88                     & 1.426                     & 12.23                     & 13.43                     & 4791 & 2.52                      & -0.2               & 0.09            & 0.32                            \\
N77-47$^{*}$           & 8039      & 118.0033                  & -38.4803                  & 11.94                     & 1.478                     & 12.32                     & 13.55                     & 5079 & 4.04                      & 0.5                & 0.14            & 0.08                           \\
N77-50$^{*}$           & 3206      & 118.2293                  & -38.5535                  & 11.955                    &                        & 12.32                     & 13.54                     & 4578 & 2.27                      & -0.5               & 0.12            & 0.10                           \\
N77-113$^{*}$          & 11389     & 118.1743                  & -38.5369                  & 12.309                    & 1.474                     & 12.71                     & 13.91                     & 5159 & 3.91                      & -0.1               & 0.29            & 0.14                        \\
N77-39$^{*\dagger}$    & 4221      & 118.1522                  & -38.6318                  & 11.88                     & 1.426                     & 12.27                     & 13.44                     & 4791 & 2.52                      & -0.2               & 0.09            & 0.13                            \\
N77-42$^{\dagger}$     & 7266      & 117.955                   & -38.5357                  & 11.894                    & 1.428                     & 12.25                     & 13.45                     & 5037 & 3.35                      & -0.1               & 0.09            & 0.19                           \\
N77-49$^{\dagger}$     & 8216      & 118.0647                  & -38.4573                  & 11.955                    & 1.497                     & 12.33                     & 13.61                     & 5183 & 3.49                      & 0.27               & 0.09            & 0.19                          \\ \bottomrule
\end{tabular}
}
\tablefoot{RA, DEC, $G$, and \gbprp\ from \cite{gdr2}.  The $V$ and $B$ are taken from the respective papers, originally from WEBDA but typically with no uncertainties. Typical errors in \teff\ and \logg\ are $\pm$100~K and $\pm$0.2~dex, respectively.
Common stars observed by RAVE: $^{*}$\cite{bragaglia_2008} and $^{\dagger}$\cite{mishenina_2015,caffau_2014}.{\cite{caffau_2014} reported metallicity as [M/H].}}
\end{table*}

A global comparison of the temperature and metallicity derived from photometric measurements \citep{geisler_1992}, medium-resolution spectroscopy \citep[RAVE;][]{kunder_2017} and high-resolution spectroscopy \citep{bragaglia_2008,caffau_2014,mishenina_2015} is shown in Fig.~\ref{fig:temp_feh_comparison}. The common stars  N77-39, N77-42, N77-47, and N77-49 studied by these authors are marked as a square, triangle, circle, and diamond, respectively, in Fig.~\ref{fig:temp_feh_comparison}. As one can appreciate we see a large dispersion in the observed metallicity of \ngc, and the mean metallicity, considering the most recent publications, varies between --0.07 and 0.18~dex, see Table~\ref{tab:spectrsocopicmetallicity}. This can be attributed to various factors; quality of spectra, the number of stars observed, solar metallicity value, atomic data, and method of analysis. Additionally, interpretation of these data is based on the selection of a model for theoretical stellar atmospheres and spectral lines. Different authors adopt different physical assumptions and methods of analysis resulting in different chemical abundance scales. \cite{caffau_2014} and \cite{mishenina_2015} have used the same high-resolution spectroscopic data from FLAMES/VLT, but there are small disparities of the order of 0.07~dex in individual metallicity, and this is because of differences in the method of analysis. Additionally, we can not ignore the possibility that these metallicity variations are indeed intrinsic.

According to the literature the metallicity value is between --0.13 to 0.18~dex (Table~\ref{tab:literatureages}). Considering only the high resolution spectroscopy the average metallicity is 0.10~$\pm$~0.05~dex. {The high-resolution spectroscopy metallicity will be further used for asteroseismic modelling  (Sect.~\ref{ssec:obs_osc_pul_model}).}

\subsection{Age of the cluster}

\begin{figure}

        \includegraphics[width=0.49\textwidth]{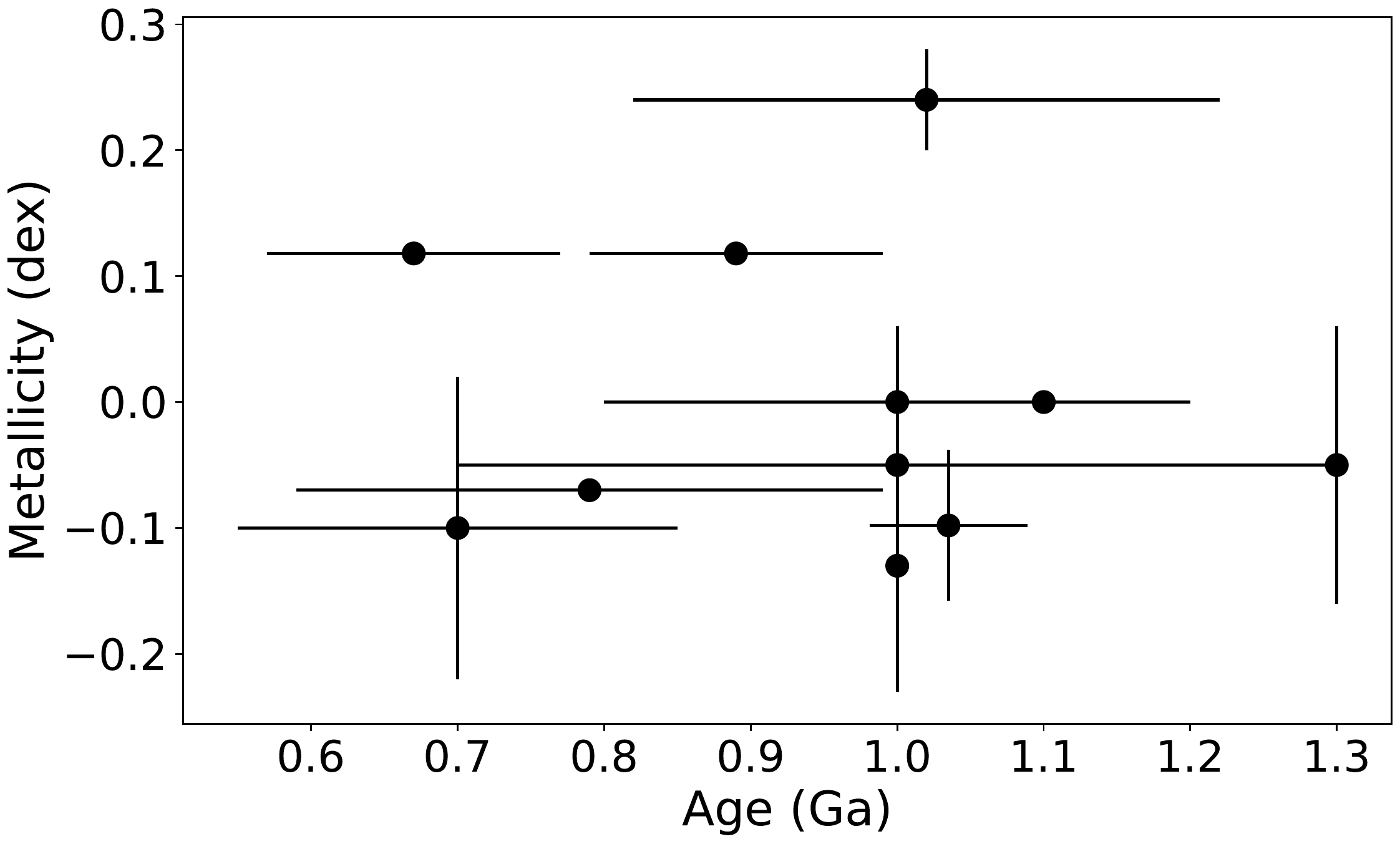}
         \caption{Distribution of age and metallicity found in the literature (Table~\ref{tab:literatureages}). Solar metallicities are plotted as zero. }
         \label{fig:litr_age_feh_dist}
\end{figure}

The age of the cluster reported in the literature has been derived through isochrone fitting, which depends upon the choice of metallicity, distance, and extinction. 
Table~\ref{tab:literatureages} summarises the derived ages, along with the corresponding distances, metallicities, and the adopted extinction from literature. 
Most of these works fit specific sets of stars in the cluster to derive their ages, without giving further details about the masses, \teff, and luminosities of the individual members. 
The distribution of age and metallicity from the literature is shown in Fig.~\ref{fig:litr_age_feh_dist}, showing the reported ages varying between 0.6 and 1.5~Ga.
The most precise age of the cluster is determined by \cite{jeffery_2011} 
using the white dwarf sequence with Hubble Space Telescope data. They infer an age of the cluster of 1.035~$\pm$~0.054~$\pm$~0.087~Ga.

\section{Extinction\label{sec:extinction}}

 \begin{figure}
\centering
  \includegraphics[width=8.5cm]{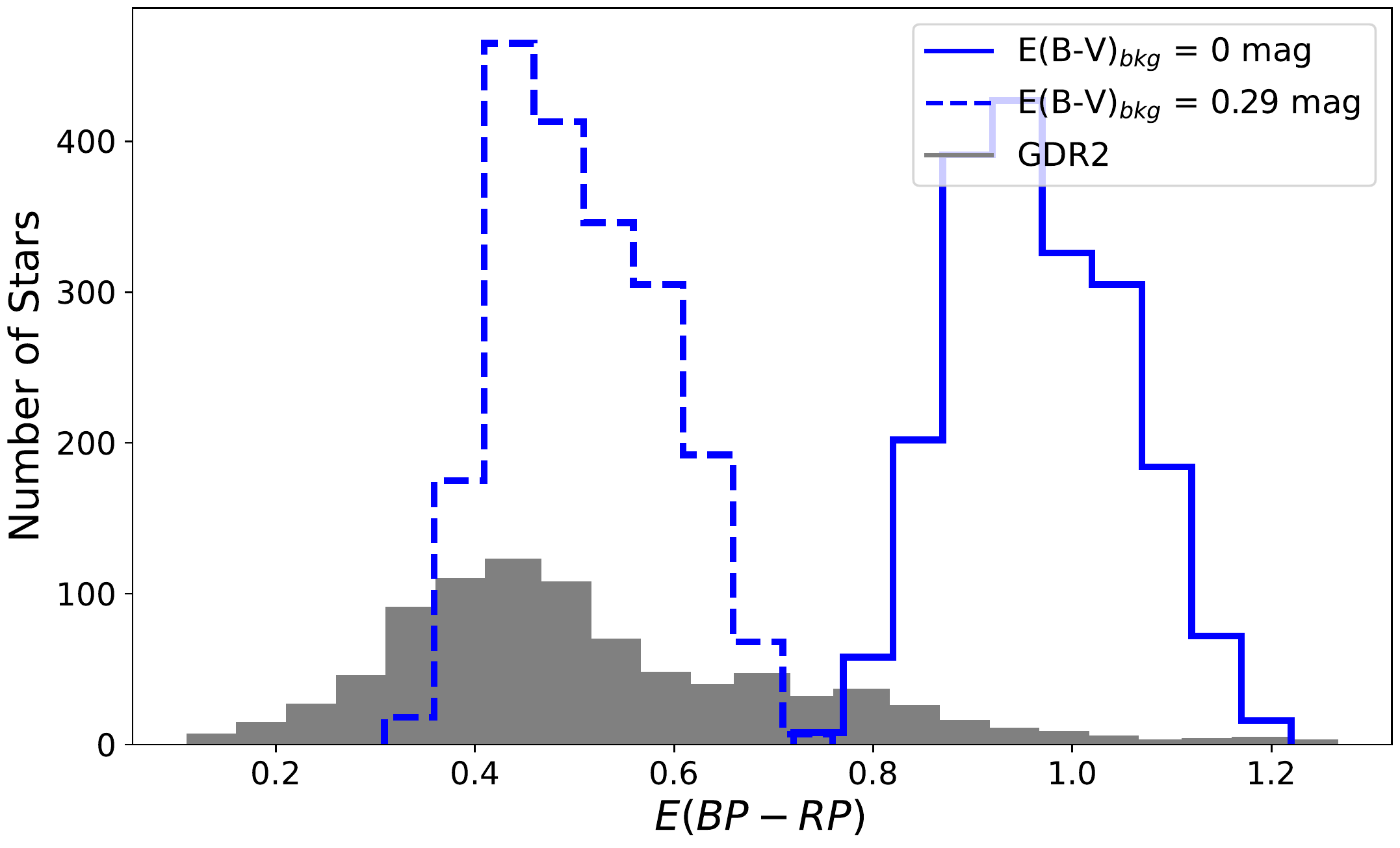}
     \caption{Distribution of the calculated and measured \ebr. The filled grey histogram shows the measured \ebr\ values from GDR2. The distribution of calculated \ebr\ values using Eq.~\ref{eqn:extinctioncoefficients} with $X = (G_{BP}-G_{RP})_0$ and adopting \ebv\ from the SFD map are shown as blue histograms, with (dashed) and without (solid) a correction of the background reddening, see Sect.\ref{ssec:background_extinction} for details.
     } 
     \label{fig:ag_ebprp_distribution}
\end{figure}

Extinction in star clusters arises due to the general interstellar medium (ISM) in the foreground of the cluster and the localised cloud associated with the cluster \citep{pandey_2003}. The ISM in the line of sight absorbs and scatters around 30$\%$ of the light in the UV, optical, and NIR wavelengths, thus dimming the source \citep{Bernstein_2002}. If the distribution of dust is not homogeneous across the field of view, for example if the cluster is quite extended, or it is near the Galactic plane, then this causes differential reddening (DR; \citealt{rain_2020}). The DR causes the broadness or dispersion in the CMD, as can be seen in Fig.~\ref{fig:cmd_cluster_obs}. For our work, it is necessary to characterise this extinction as accurately as possible and on a star-by-star basis, so that we can derive individual \teff\ and \lum\ for each of the stars we intend to inspect with detailed models. In this section, we explore several methods of deriving the extinction.  

 \begin{figure*}
\centering
   \includegraphics[width=0.49\textwidth]{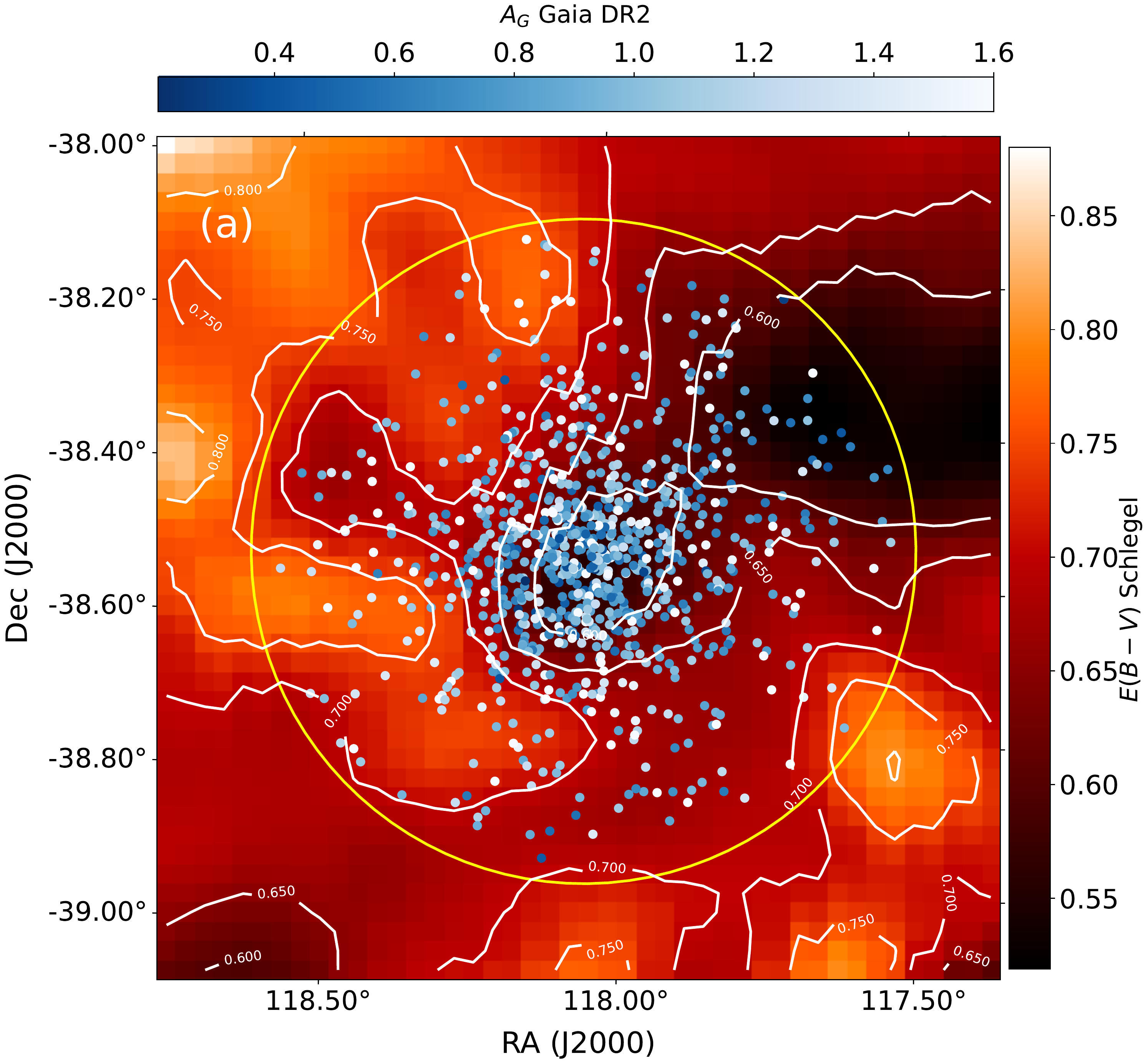}
     \includegraphics[width=0.49\textwidth]{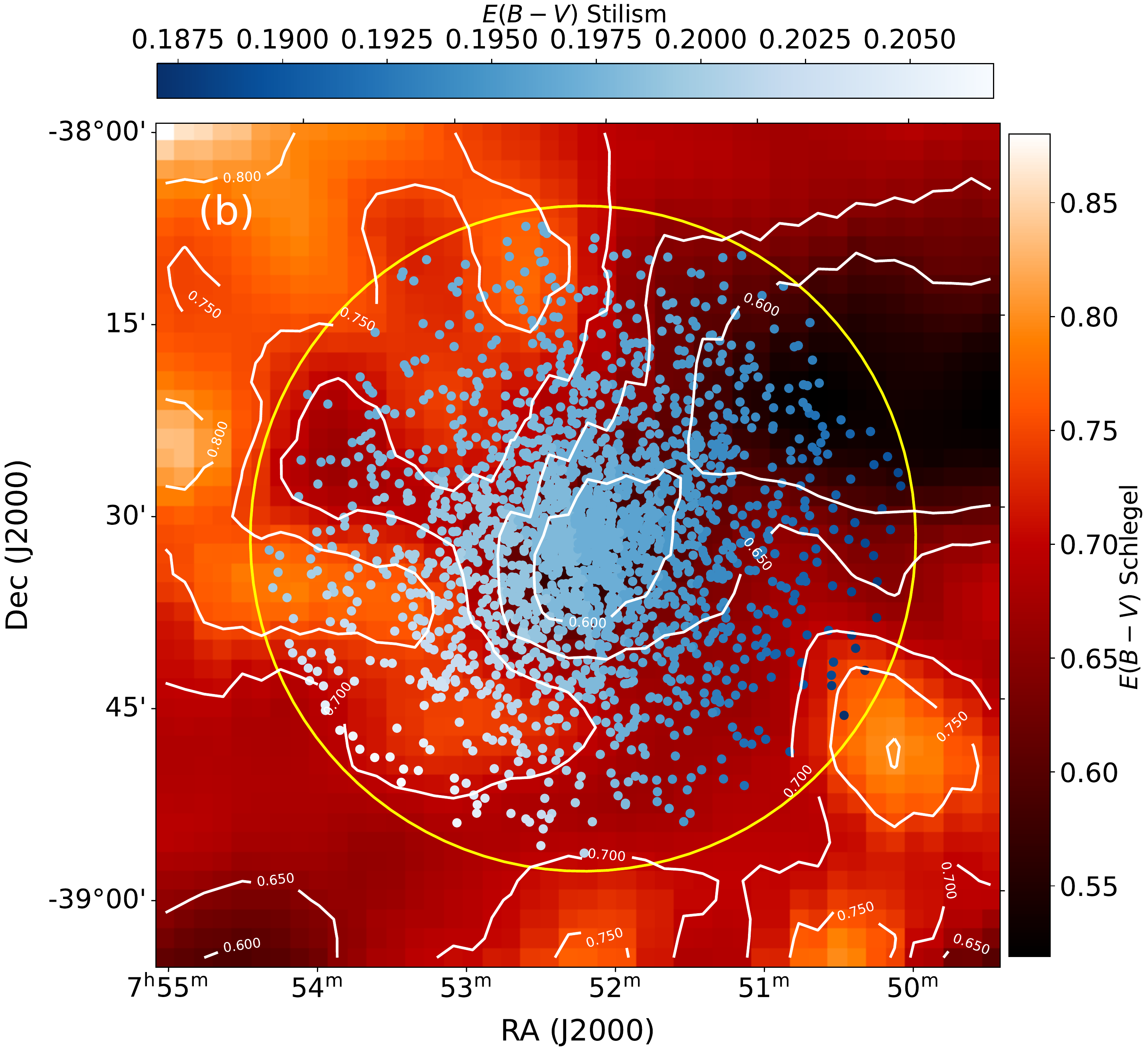}
     
     \caption{\cite{schlegel_1998} 2D reddening map for the cluster. The contour levels are 0.65, 0.75, 0.8, 0.85, and 0.9~mag in $E(B-V)$. The yellow circle represents the extent of the cluster. (a) Overplotted are the stars with measured \ag\ values (838 stars) from \gaia\ DR2.  (b) Distribution of \ebv\ across the cluster from \cite{lallement_2018}.}
     \label{fig:SFDmap_stilism_map}
\end{figure*}

\subsection{Individual Measurements and 3D maps}

\cite{hartwick_1972} and \cite{hartwick_1974} reported the presence of DR in the cluster, which increases from the north-west towards the south-east. They measured \ebv\ for 68 stars across the cluster and reported values varying from 0.22 to 0.47~mag. \cite{friel_2002} using H$\beta$ lines deduced reddening for eight stars in the cluster and found similar results that differed only by $-0.02$~mag compared to \citeauthor{hartwick_1972}.

Line-of-sight extinction $A_{G}$ in the \gaia\ $G$-band and reddening values $E(BP-RP)$ are available for 838 member stars from the \gaia\ DR2 catalogue. These values also show DR, with less extinction towards the centre of the cluster. Fig.~\ref{fig:ag_ebprp_distribution} shows the distribution of \ebr\ as a grey filled histogram, which peaks at 0.8~mag and 0.4~mag, respectively, while 
Fig.~\ref{fig:SFDmap_stilism_map}a shows a 2D map of the total galactic extinction (background, red shades, see next section) and the GDR2 measured values as blue shaded filled circles which correlate well with the measurements.

We also investigated the 3D dust map provided by \cite{lallement_2018}, by exploiting their \href{https://stilism.obspm.fr/}{\tt Stilism}\footnote{\url{https://stilism.obspm.fr/}} package, to obtain another estimate of the individual reddening values for all the member stars. We calculated their \ebv\ at the adopted distance (1441 $\pm$ 81 pc). The median reddening value is 0.197~mag with a large uncertainty of up to 0.48~mag. The spatial distribution of the \ebv\ values for the individual stars is shown in  Fig.~\ref{fig:SFDmap_stilism_map}b, and a smooth gradient can be seen, while varying by a mere 0.02~mag across the field.   These values did not correlate well with the total galactic extinction nor the Gaia ones, and were also inconsistent with earlier authors.  We therefore decided not to adopt these values. 

 \begin{figure*}
\centering
   \includegraphics[width=\textwidth]{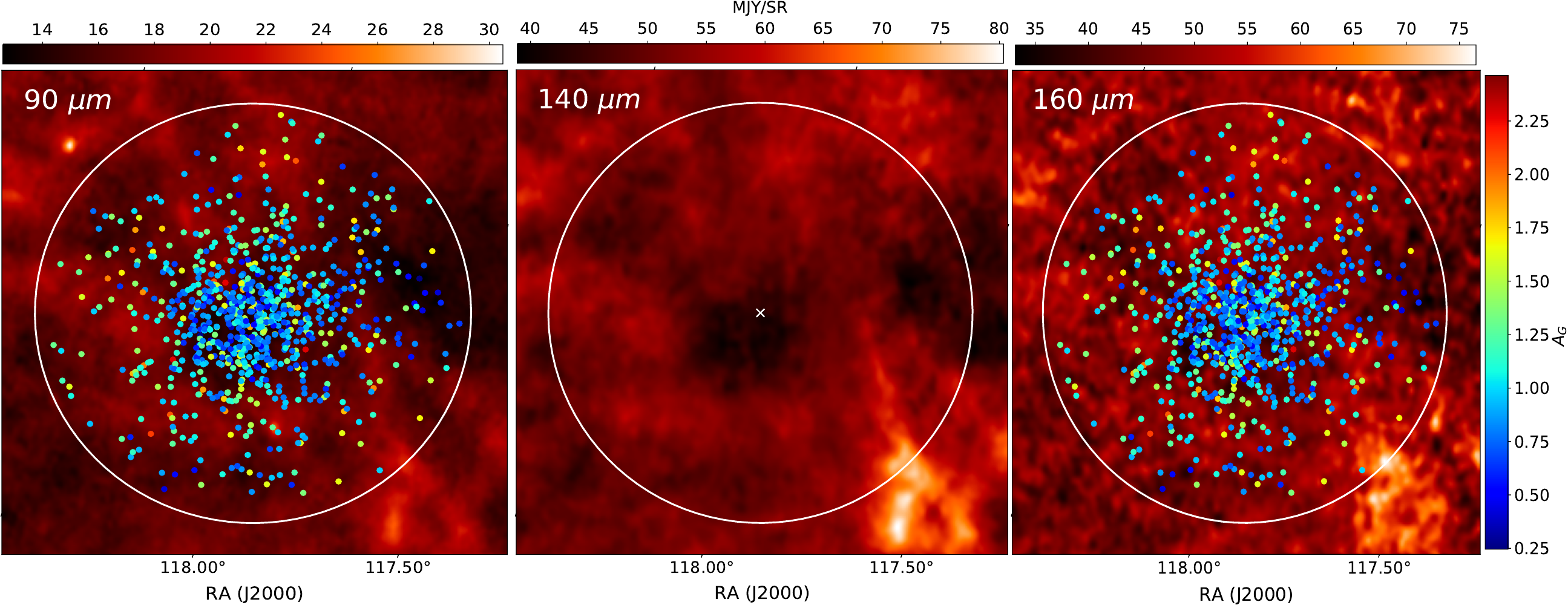}
     \caption{AKARI 90, 140, and 160~$\mu$m maps showing the presence of dust emission from the vicinity of the cluster. For the 90 and 160~$\mu$m maps, $A_{G}$ from the \gaia\ catalogue is overlayed. The circle represents the extent of the cluster and the cross represents the centre of the cluster.} 
     \label{fig:akari_maps}
\end{figure*}

\subsection{2D maps}\label{ssec:2dmaps}

To investigate the DR more, we used the full-sky 
2D reddening map by \citet{schlegel_1998}, known as the  SFD. The reddening values are derived using temperature effects from the 100~$\mu$m DIRBE \citep{silverberg_1993} and 200~$\mu$m IRAS maps  \citep{neugebauer_1984}. Later, \cite{schlafly_2011} re-calibrated the SFD using photometry from the Sloan Digital Sky Survey \cite[SDSS;][]{york_2000}, and reported a multiplicative correction factor of 0.86. We used the re-calibrated SFD map for \ngc\footnote{https://irsa.ipac.caltech.edu/applications/DUST/}. The SFD map for the cluster's coordinates shows DR, with the centre and north-west part having less extinction compared to the south-east part, see Fig.~\ref{fig:SFDmap_stilism_map}a~or~b. Even though this map is a total extinction map along the line-of-site, this is in agreement with the results reported by \cite{hartwick_1972,hartwick_1974} and from GDR2. Unlike the 3D Stilism map, the 2D SFD map may suffer from background extinction because we do not know the location (in distance) of the interstellar cloud. This is discussed further in Sect.~\ref{ssec:background_extinction}. 

\begin{figure}
    \centering

        \includegraphics[width=0.49\textwidth]{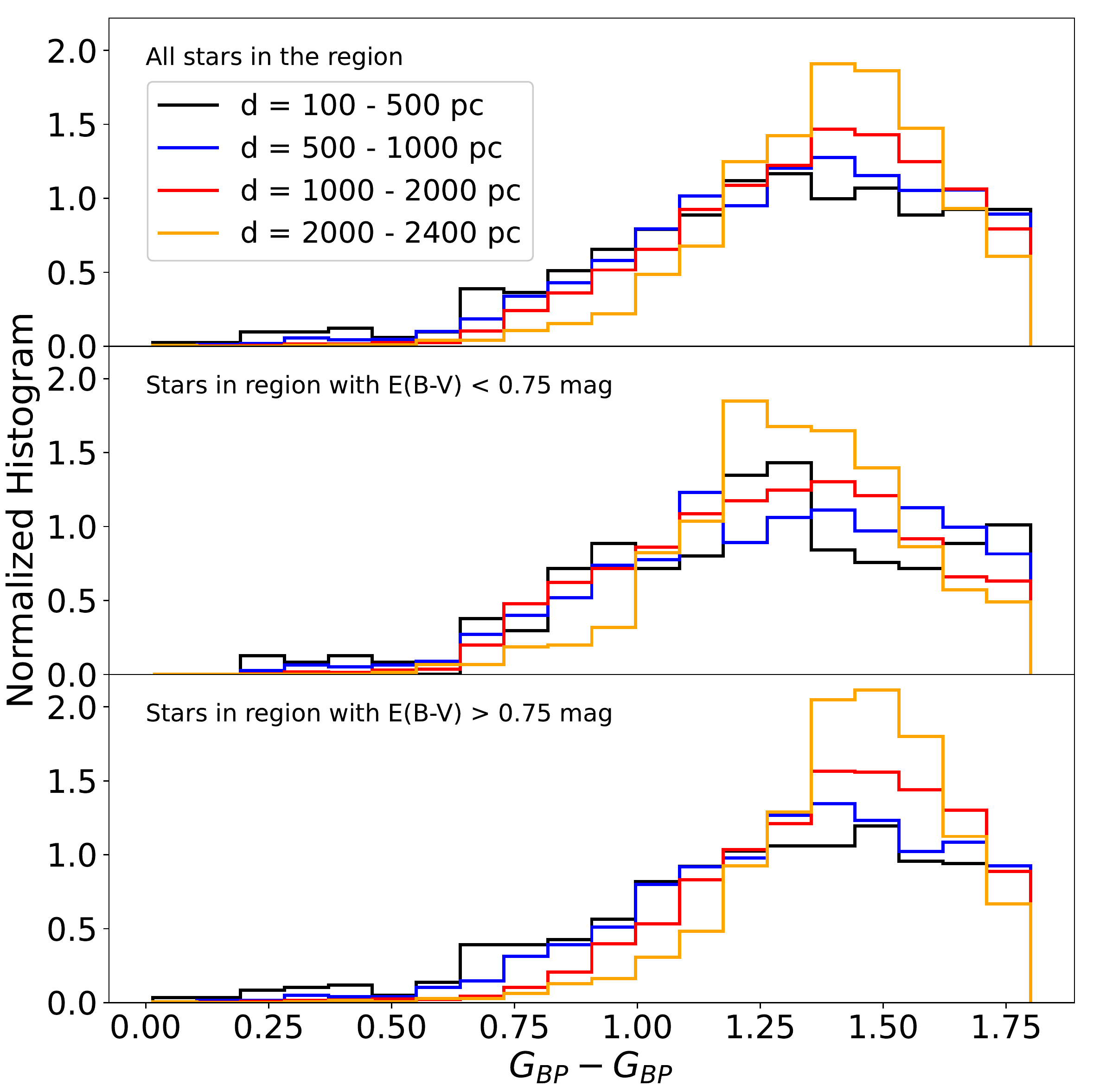}

    \caption{Normalised number histogram of stars as a function of \gbprp\ for each distance sub-sample. (a) All the \gaia\ stars in the region of the cluster, (b) \gaia\ stars which fall in the regions where there is less extinction ($\ebv\ < 0.75$~mag) (c) \gaia\ stars which fall in more extincted region ($\ebv\ > 0.75$~mag).}
    \label{fig:dust_distance_hist}
\end{figure}

The presence of dust in the line-of-sight of the region can be clearly seen in the AKARI 90, 140 and 160~$\mu$m maps \citep{doi_2015}, see Fig.~\ref{fig:akari_maps}. The dust seen in the AKARI maps can, however, be in the foreground, background or associated with the cluster. \cite{dutra_2000} has reported that the cluster \ngc\ is inside the dust cloud, and the edge of the dust layer extends to 2.1 kpc along the cluster line-of-sight from the true Galactic plane. To substantiate this, we first identified all the \gaia\ catalogued stars in the vicinity of the cluster. We removed the stars with \gbprp\ $>1.8$~mag to have a uniform distribution in magnitude -- these stars 
are too faint to be observed by \gaia\ beyond 2~kpc. 
Fig.~\ref{fig:dust_distance_hist} shows the normalised distribution of \ebv\ with the colour-code indicating different distances (where we take 1/parallax as the distance). 
The full selected sample is shown in the top panel, and middle and lower panel separate this sample into the less extinct region (\ebv\ $< 0.75$~mag, middle) and more extinct region (\ebv\ $> 0.75$~mag, lower). 
The stars within a distance of 1000~pc (black and blue) in less extinct regions and more extinct regions have similar morphology and have peak from 1.00 to 1.75~mag. But the stars in the distance 2000 to 2400~pc (yellow) peak at 1.25~mag in less extinct regions and 1.50~mag in more extinct regions. This shift in peak would indicate the presence of dust between 1000 to 2000 pc, thus associating the dust directly with the cluster.

\subsection{Conversion of $E(B-V)$ to \ebprp\ and $A_G$}
\label{ssec:ebv_to_ag}

As we are dealing with extinction in different passbands, we require a correct conversion between \ebv\ and \ebprp, which we address in this section.   
\gaia\ has a wide pass-band ranging from 330 - 1050~nm \citep{Evans_2018} and the extinction in the \gaia\ band  will be different from the blue end to the red end 
of the spectrum. 
The extinction coefficient and reddening $A_G$ not only depend upon the amount of extinction, but also the intrinsic colour of the star.
For other narrow photometric bands like UBV and 2MASS, the extinction coefficient is nearly a constant for stars with different intrinsic characteristics. 
To compute the temperature-dependent extinction coefficient $k_{m} = A_m / A_0$ where $m$ is any photometric band and $A_{0}=3.1E(B-V)$ \citep{babusiaux_2018}, we used the expression 

\begin{equation}
k_m  = c_1+c_2X+c_3X^2+c_4X^3+c_5A_0+c_6A^2_0+c_7XA_0
\label{eqn:extinctioncoefficients}
    \end{equation}
where the coefficients are found in \cite{babusiaux_2018} and \cite{danielski_2018} respectively, $X = (G_{BP}-G_{RP})_0$ and $(G-K_{s})_0$, and $K_s$ is the 2MASS magnitude \citep{skrutskie_2006}. 
We then calculate \ag, $A_{BP}$ and $A_{RP}$, and from these latter two, 
the colour excess: $E(BP-RP)$ ($ = A_{BP} - A_{RP}$) which gives a 
relationship of $\ag \sim 0.54 \ebprp$ with a large scatter due to the different \teff\ of the stars.    
The relation between \ag\ and \ebprp\ can also be checked by plotting the Gaia DR2 values of \ag\ and \ebprp, and we obtain 
$\ag \sim 0.58 \ebprp$ which is in global agreement.

\subsection{Background Extinction  $E(B-V)_{\rm bkg}$} \label{ssec:background_extinction}
The conversion between \gaia\ bands and \ebv\ allows us to check the 
consistency of the actual values of extinction between the different approaches for estimating extinction.   
The results between GDR2 and the SFD maps (module a background correction) and previous literature values are in agreement, and therefore for the rest of this work we adopt the \ebv\ values from the SFD maps because these values are available for all of the cluster members.   However, as the SFD map is a 2D extinction, we need to make some correction for the background extinction,  $E(B-V)_{\rm bkg}$.  This is what we present here.

The analysis presented in section \ref{ssec:2dmaps} concluded that the dust is mostly in front of or within the cluster.
Since the majority of the DR comes from the dust (in the foreground), we assume the $E(B-V)_{\rm bkg}$ to be a constant. 

\begin{figure}
    \centering
    \includegraphics[width=8.5cm]{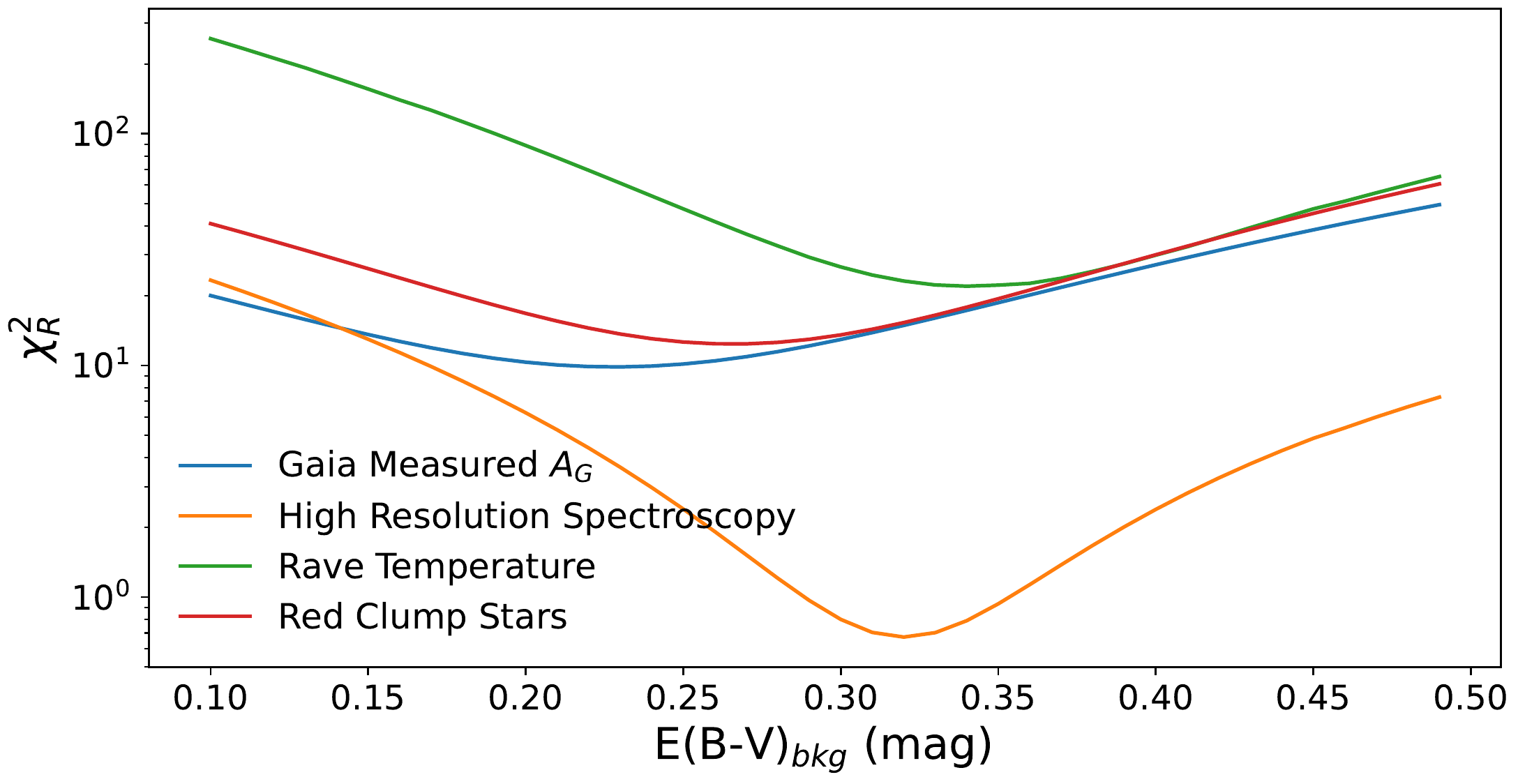}

    \caption{Reduced chi-squared values for different background reddening using the different techniques discussed in section \ref{ssec:background_extinction}.}
    \label{fig:bkg_estimation_chisquare}
\end{figure}

In order to estimate  $E(B-V)_{\rm bkg}$, we performed several analyses. 
The first of these consists of comparing the observed spectroscopic temperature of 53 RAVE stars, whose derived values  are not affected by dust, with photometric temperatures that rely on dereddening the colours, that is they are affected by dust.  By comparing the two determinations of \teff, we should be able to derive the  $E(B-V)_{\rm bkg}$ that minimises the difference between the two.  
We used the relations from \cite{casagrande_2020} to calculate the photometric temperatures.
For each $E(B-V)_{\rm bkg}$ we calculated the dereddened colours of the stars and converted these to photometric \teff.  We then calculated a mean $\chi^2$ value based on the differences between the two \teff.
We repeated this for a range of $E(B-V)_{\rm bkg}$.   
Fig.~\ref{fig:bkg_estimation_chisquare} illustrates $\chi^2$ versus  $E(B-V)_{\rm bkg}$ and we can see that the best fit value is 0.34~mag. 

The above analysis is also applied to the 11 stars with observed high-resolution spectroscopic temperatures (see Table~\ref{tab:spectrsocopicmetallicity}), and results in an optimal $E(B-V)_{\rm bkg}$ = 0.32~mag (Fig.~\ref{fig:bkg_estimation_chisquare}, {orange curve}).

Instead of comparing the \teff, 
we can also compare the observed \ag\ by \gaia\ and the computed one (Sect.~\ref{ssec:ebv_to_ag}) to estimate $E(B-V)_{\rm bkg}$.  The computed one also 
requires \ebv\ of individual stars. 
The $\chi^2$ for this comparison is shown in Fig \ref{fig:bkg_estimation_chisquare} by the blue curves and we obtain
a best $E(B-V)_{\rm bkg}$ = 0.27~mag.

A final method we use to estimate the background extinction is to look at the core helium burning red clump (RC) stars which should all have similar luminosities i.e. $M_G = 0.44$~mag, where %
an intrinsic dispersion of 0.20~mag
is estimated by \cite{hawkins_2017}.
However, as we can not yet say which stars are red giant or red clump, 
we selected all the stars in the {\it clumpy} region.   By varying $E(B-V)_{\rm bkg}$ we calculated different values of $M_G$ for each of these stars by using $G$, $A_G$ and the adopted distance.  The $\chi^2$ value now compares the expected $M_G$ to the calculated ones, and this is illustrated by the red curve in  Fig.~\ref{fig:bkg_estimation_chisquare}.  From these values we estimate
estimate $E(B-V)_{\rm bkg}$ = {0.24~mag}. 

In summary, the mean background extinction $E(B-V)_{\rm bkg}$ considering the different observables  varies between 0.24 and 0.34~mag, which is in agreement with \cite{dutra_2000} who found a background extinction of 0.36~mag. 
{In the rest of our analyses in this paper, we therefore, consider three background extinction values:  0.24, 0.29, and 0.34~mag. These lower, middle, and upper background values helps us to study the effects of extinction in our analysis.}

We compare the derived \ebprp\ from using the SFD maps while varying the \ebvbkg\ directly with the GDR2 values in Fig.~\ref{fig:ag_ebprp_distribution}.   The latter are shown by the grey histogram, while the resulting \ebprp\ assuming a \ebvbkg\ of 0.0 and 0.29 magnitudes are shown by the continuous and dashed lines, respectively.   
The middle 0.29~mag value gives a peak \ebprp\ in agreement with the Gaia measured one.

\begin{table*}[]
\caption{Best fit isochrones in Age--[M/H] space for three values of 
$E(B-V)_{\rm bkg}$ for each set of public isochrones. We note that $Y$ is fixed by a metallicity enrichment law given by the models and is not a free parameter. }
\begin{tabular}{lcccccc}
\toprule
Isochrone               & \ebvbkg & Z      & Y      & \mh & Age  & \begin{tabular}[c]{@{}r@{}}Turn off\\ mass\end{tabular} \\ 
                        & (mag)                                                            &        &        & (dex)      & (Ga) & (\mstar)                                 \\ \midrule
          & 0.24                                                                                                                                     & 0.0095 & 0.2654 & -0.2                & 0.95 & 2.064                                                                                                                           \\
PARSEC    & 0.29                                                                                                                                     & 0.0118 & 0.2695 & -0.1                & 0.95 & 2.105                                                                                                                           \\
          & 0.34                                                                                                                                     & 0.0199 & 0.2839 & 0.14                & 1.0  & 2.157                                                                                                                           \\
          & 0.24                                                                                                                                     & 0.0187 & 0.2715 & 0.1                 & 0.65 & 2.232                                                                                                                           \\
BASTI     & 0.29                                                                                                                                     & 0.0195 & 0.2725 & 0.12                & 0.7  & 2.143                                                                                                                           \\
          & 0.34                                                                                                                                     & 0.0269 & 0.2822 & 0.27                & 0.75 & 2.230                                                                                                                           \\
          & 0.24                                                                                                                                     & 0.0105 & 0.2646 & -0.14               & 0.85 & 2.105                                                                                                                           \\
MIST      & 0.29                                                                                                                                     & 0.0143 & 0.2703 & 0.0                 & 0.9  & 2.125                                                                                                                           \\
          & 0.34                                                                                                                                     & 0.019  & 0.2773 & 0.13                & 0.95 & 2.144                                                                                                                           \\ \hline
\end{tabular}
\label{tbl:isochrone_fitting}
\end{table*}

\section{Isochrone fitting}\label{sec:isochrone}

 \begin{figure*}[t!]
\centering
   \includegraphics[width=\textwidth]{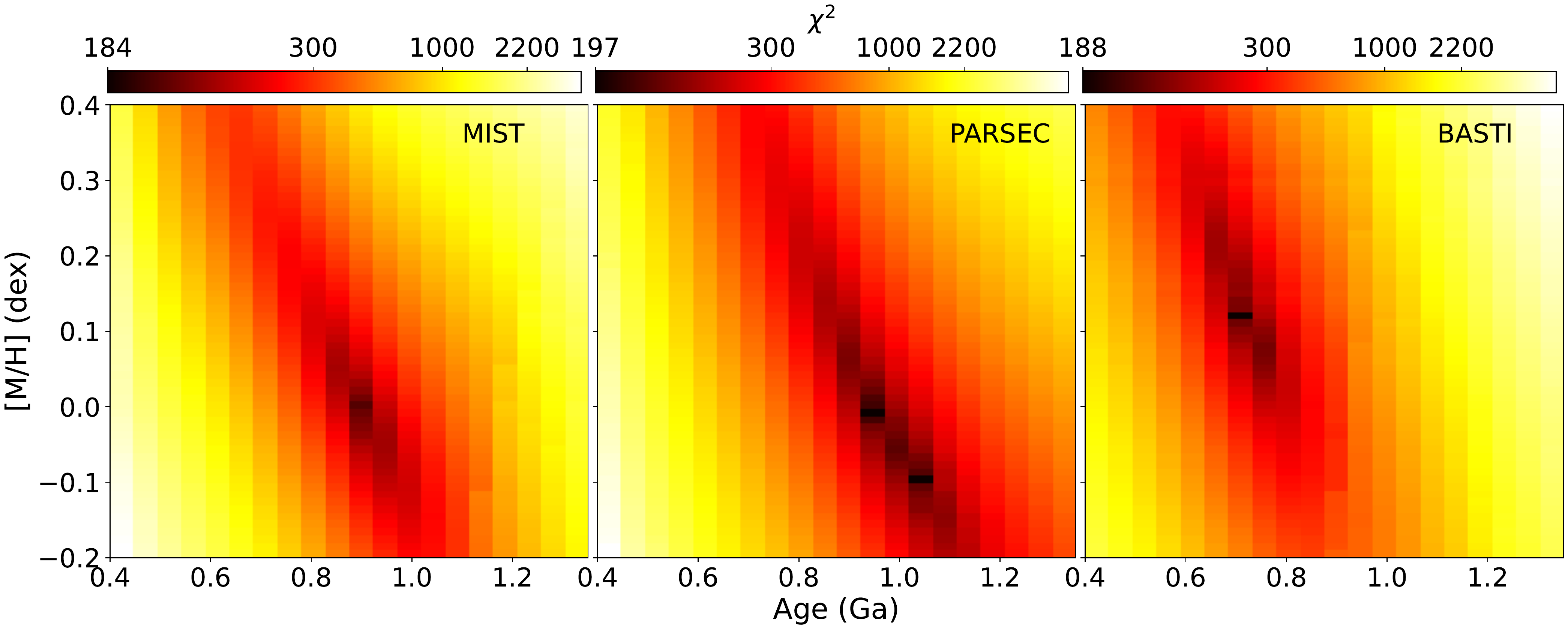}
     \caption{$\chi^2$ distribution as a function of age and [M/H] using MIST (left), PARSEC (centre), and BASTI (right) isochrones for \ebvbkg\ =  0.29~mag. }
     \label{fig:isochrone_chisquare}
\end{figure*}

In order to estimate the cluster's age, we have used three publicly available sets of isochrones from the literature, PARSEC \citep{bressan_2012}, BASTI \citep{hidalgo_2018}, and MIST  \citep{dotter_2016a,choi_2016a}. Using our assumed distance and correcting the individual extinction values (Sect.~\ref{sec:extinction}) for the different background  extinctions we derived, we converted the \gaia\ photometric magnitudes to absolute magnitudes $M_{\rm G}$ and dereddened colour \ebr$_0$ in order to fit the data in the CMD space.

To fit the data to the isochrones, we used all of the stars from the cluster, but applied higher weight to the turn-off stars and giant branch stars in the fitting function.  In particular, given the relatively large number of main sequence stars, it was critical to down-weight their contribution to the fit, because it would entirely dominate it. The weights that we applied to each of the main sequence, turn off and giant stars are {0.0001, 1, and 0.5} respectively. We performed a $\chi^2$ analysis to find the optimal combination of metallicity and age from the isochrones that best fit our corrected colour-magnitude diagram. {In the fitting procedure, we used ages from 0.4 to 1.2~Ga with a step size of 0.05~Ga and metallicity from -0.2 to 0.4~dex with a step size of 0.01~dex.} 

\begin{figure}
\centering
   \includegraphics[width=0.49\textwidth]{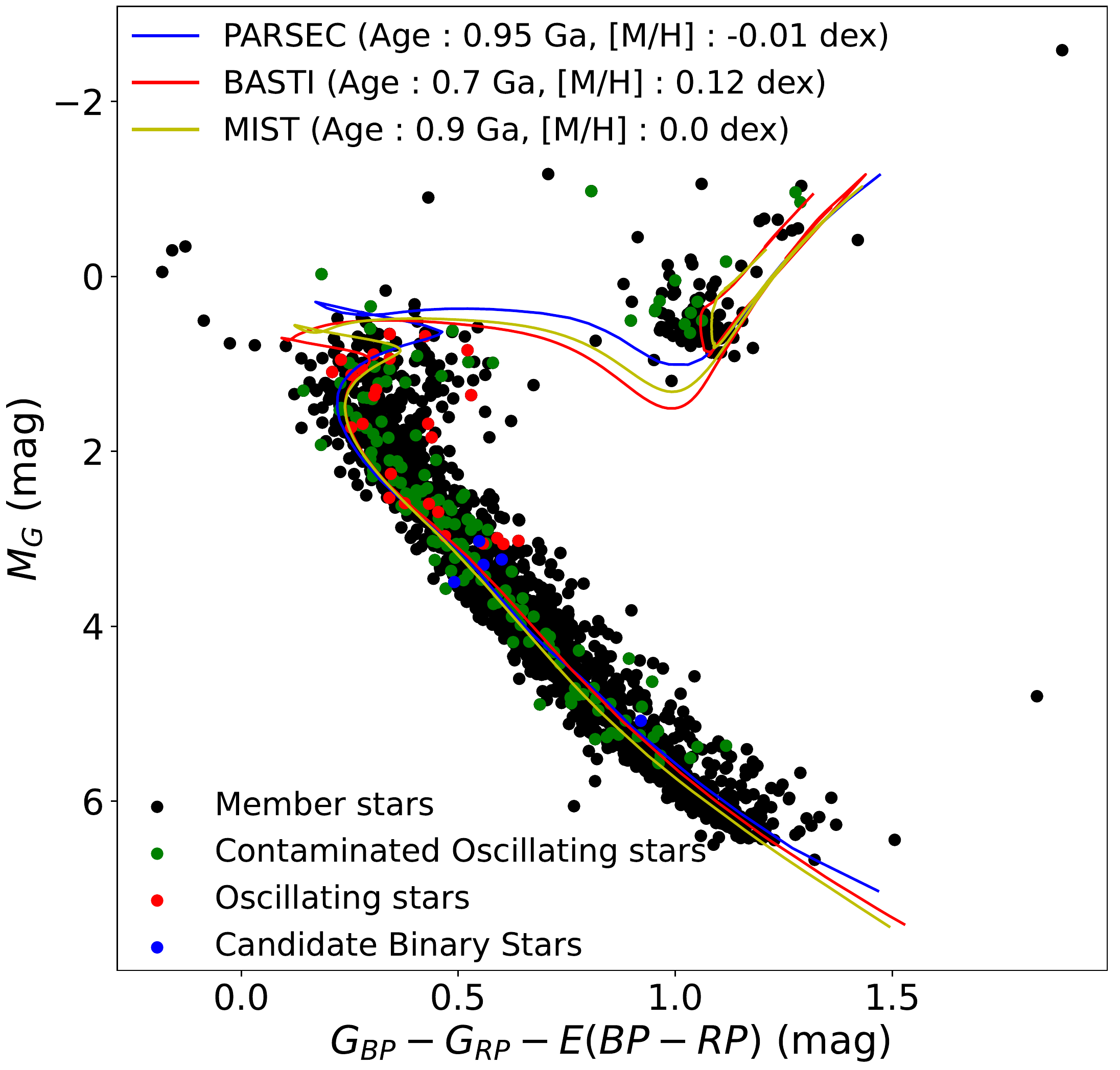}
     \caption{Dereddened colour-magnitude diagram of NGC 2477 showing  the best fitted isochrones for \ebvbkg\ = 0.29~mag (Sect.~\ref{sec:isochrone}) along with contaminated and non-contaminated oscillating stars and candidate binaries, see Sect,~\ref{sec:tessvariability} for details on the variable stars.}

    \label{fig:isochrone_fit}
\end{figure}

\begin{figure}
\centering
   \includegraphics[width=0.49\textwidth]{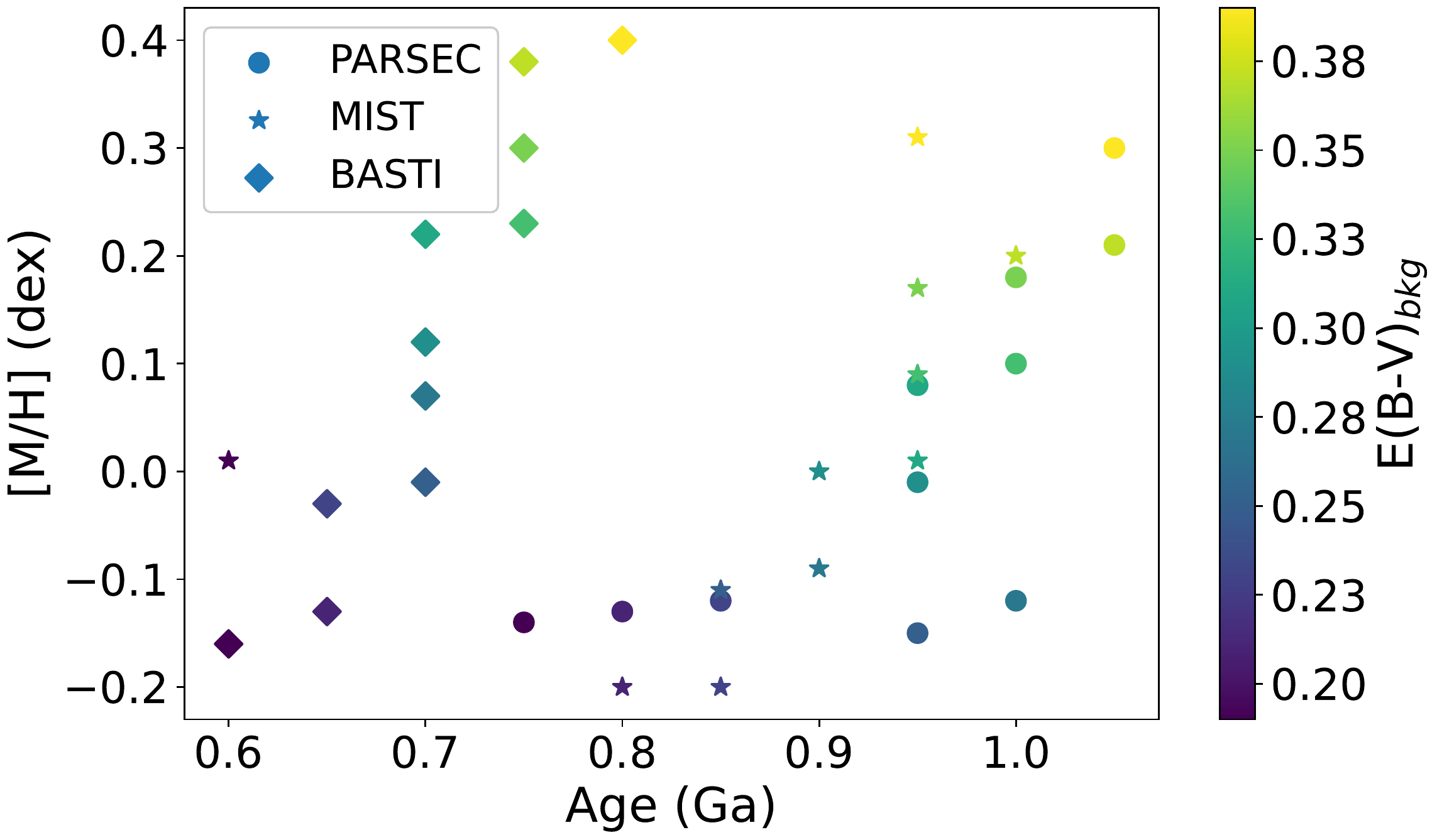}
     \caption{Distribution of age and metallicity as a function of \ebvbkg\ for different isochrone fitted models.}
     \label{fig:isochrone_fit_ext}
\end{figure}

We fitted the data to the three sets of isochrones assuming the three $E(B-V)_{\rm bkg}$ (0.24, 0.29, and 0.34~mag). 
In Fig.~\ref{fig:isochrone_chisquare} we show an example of the 2D $\chi^2$ map as a function of the two free parameters (age, [M/H]), for \ebvbkg~=~0.29~mag for each isochrone set. The best fitted PARSEC, BASTI and MIST isochrones for the cluster with \ebvbkg~=~0.29~mag is shown in CMD in Fig.~\ref{fig:isochrone_fit}.
As expected, a higher background extinction, which implies a lower foreground extinction, shifts the CMD down and to the right (fainter and cooler), resulting in higher ages and higher metallicities. In Fig.~\ref{fig:isochrone_fit_ext} we show the distribution of age and \mh\ as function of \ebvbkg\ for different best fitted isochrone models. The metallicity of the best fitted isochrones varies from -0.16 to 0.38~dex, and the age between 0.6 to 1.0~Ga.  Table~\ref{tbl:isochrone_fitting} lists the best fitting isochrone for each $E(B-V)_{\rm bkg}$ and isochrone. Their metallicities and ages are comparable with reported values in the literature (Table~\ref{tab:literatureages}) however the latter does not contain any information on the correlations in the parameters, which are important to consider.  

Furthermore, we used the fitted isochrones to determine the stellar parameters of each of the member stars.  The model values of mass, surface gravity, \teff, and \lum\ were  determined by applying a minimum distance method to the interpolated isochrone. We also give the turn-off mass of the fitted isochrones in Table~\ref{tbl:isochrone_fitting}. The masses determined here are  used as initial values for modelling in Sect.~\ref{sec:models}. From here on we use all of our previous analysis in this paper to help interpret the light curve variations that we discuss in the next section.

\section{TESS light curve extraction \label{sec:tess}}

Now we turn our attention to the analysis of the light curves of cluster members using the Transiting Exoplanet Survey Satellite (TESS). TESS collects all-sky high-precision photometry; In cycles 1 and 2 photometry with 2-minute cadence (SC) was collected for about  200\,000 selected stars and FFI were taken with a 30-minutes cadence.
The Extended Mission (Cycle 3 and 4), which started in  July 2020, collected 20-second cadence data for 1000 pre-selected targets, 2-minute cadence for the main sample and the FFI cadence was reduced from 30 to 10 minutes. 
TESS has excellent photometric precision, and combined with its fine time sampling and long intervals of uninterrupted observations asteroseismology is possible \citep{campante_2016, schofield_2019, hekker_2017}. Recently, \cite{sahoo_2020} identified pulsating subdwarf B stars from TESS FFI, and \cite{stello_2021} compared the seismic quantities $\Delta \nu$ and $\nu_{max}$ of red giant stars observed both by TESS and $Kepler$ and confirmed that the results were consistent.

\subsection{Light curves from FFIs}

\ngc\ was observed in TESS sectors 7 and 8 in cycle one and in sectors 34 and 35 in cycle three as part of the extended mission. None of the member stars are observed in SC mode and  
the processed light curves from FFI are not available from the Science Processing Operations Center (SPOC).   There are a number of packages available to extract light curves from the FFI, such as the Delta Function Difference Imaging Code \citep{oelkers_2018}, {\tt TESScut} \citep{brasseur_2019}, {\tt Eleanor} \citep{feinstein_2019}, \lightkurve \citep{lightkurvecollaboration_2018}, Cluster Difference Imaging \citep[CDPIS,][]{bouma_2019}, the MIT Quick Look Pipeline\citep[QLP,][]{huang_2020}, and A PSF-based Approach to TESS High quality data of Stellar clusters \citep[PATHOS][]{nardiello_2019}. 

The centre of \ngc\ is very dense. Since the TESS pixel size is 21$''$, there can be two or more stars in an aperture. To obtain a light curve for a target we have to ensure that it is not contaminated by a nearby source. 
We used the {\tt Eleanor} package to help extract the light curves.
However, because the default aperture from Eleanor would result in some contamination in dense regions, we developed an interactive package \cite[{\tt tessipack},][]{palakkatharappil_2021} which helps to define custom apertures around the source and to then obtain the light curves from {\tt Eleanor}. 

We acquired target pixel files (TPFs) with square target masks of 13 pixels on each side and a square background mask of 31 pixels on each side using {\tt Eleanor}.  We  used the \gaia\ DR2 catalogue to search for nearby stars in the vicinity. If a star with similar or brighter magnitude falls in the aperture of a target star, the target star is disregarded since the nearby star's light contaminates the target star. Different apertures are tested for finding the best light curve, and the PCA (Principal Component Analysis) corrected light curve is then extracted.

\subsection{Comparison of light curves}

To validate the TESS FFI light curve from the  {\tt tessipack} package, we first compared existing light curves from the K2 mission \citep{howell_2014}.  One example of these stars is TIC~187129574, an F0 spectral type star which was observed by both TESS and the K2 mission. K2 observed the star in long cadence mode (30 minutes) in  campaigns 111 and 112, while TESS observed the star in sector 12 in  FFI mode only.  We compare the K2 light curve using the pipeline aperture from the \lightkurve package, with our light curve using the {\tt tessipack} package and the same custom aperture.  
Fig.~\ref{fig:TIC_187129574_TESS} illustrates this comparison.  The left panels show the apertures from K2 (top) and the same aperture but on the TESS FFI (lower).  The top right panel shows the K2 light curve that is obtained from the K2 pipeline, while the lower right shows the normalised folded light curves at the dominant periodicity of both the K2 and tessipack-generated light curves.   The same variability is contained in both light curves, thus validating our pipeline.

\begin{figure*}
\centering
   \includegraphics[width=\textwidth]{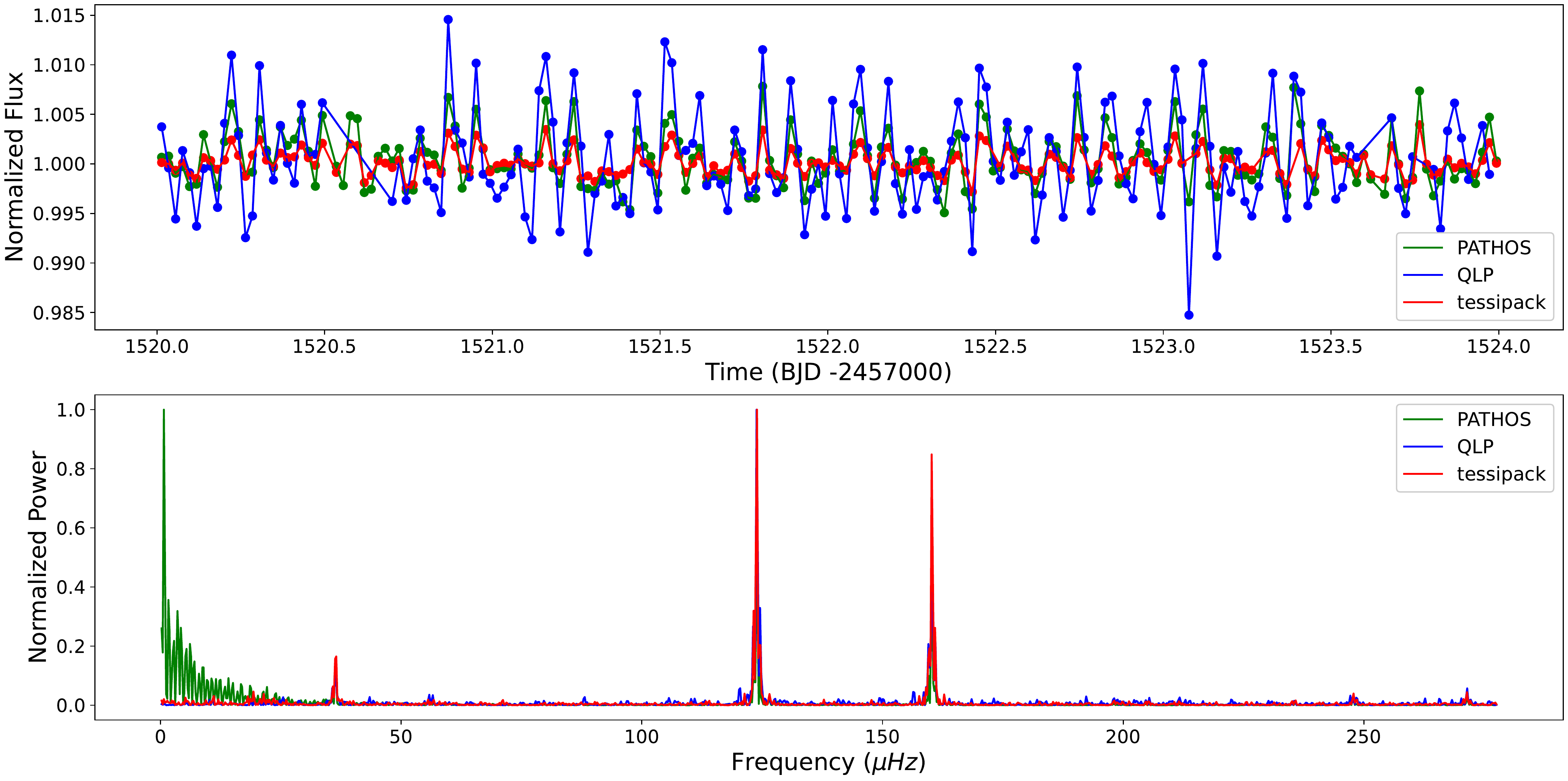}
     \caption{(a) Two day light curve obtained for variable star N77-135 from TESS FFI using {\tt tessipack}, PATHOS and QLP . (b) Periodogram obtained from the light curve for entire time series.}
     \label{fig:other_package_TESS_FFI}
\end{figure*}

We also compared the light curve from {\tt tessipack} with other existing pipelines. In other pipelines, we cannot determine a custom aperture so we used the default apertures. Fig.~\ref{fig:other_package_TESS_FFI} shows the comparison of the light curve of one of the oscillating stars (N77-135) in \ngc\ using PATHOS, QLP and our {\tt tessipack} packages. As can be seen, the light curves that are produced by {\tt tessipack} show similar variability. 

As we are confident in the light curves produced by our own package, for the rest of this work we used {\tt tessipack} to extract the light curves.  The package also allows us to take into account the environment of the star to ensure that we know where the stellar light originates from. We analysed the full content of the cluster \ngc, to search for variability and oscillating stars, which we describe in the next few sections.

\section{Detection of variability and oscillating stars \ngc } 
\label{sec:tessvariability}

\subsection{Identification of variability and contamination}

To search for variability, we analysed the light curves {from Sectors 7, 8, 34, and 35} that we extracted from the TESS data, as described in the previous section. To determine the type of variable stars we performed the following steps. {Firstly, we constructed a Lomb-Scargle periodogram \citep{lomb_1976,scargle_1982}, by searching up to the frequency limit (Nyquist frequency) of 277~$\mu$Hz and 833~$\mu$Hz for the 30- and 10-minutes data, respectively.} Then, we produced folded light curves (phase diagrams) using the dominant frequencies to determine the oscillation type.  Oscillating stars show sinusoidal phase diagrams, while other variability such as binaries show different form of phase diagrams.

Apart from extracting the optimal light curve, the environment of each member star was also inspected using the help of the {\tt tessipack} package to determine contaminating sources. Of the 2039 member stars, we found 185 oscillating stars and 13 binaries, with  only 26 oscillating stars not contaminated by neighbouring sources. This is due to the overdensity of cluster members towards the centre of the cluster. We then  only considered 16 oscillating stars since the nature of the variability of the other ten stars is unknown. An example of the light curve and periodogram of the uncontaminated oscillating star N77-320 is shown in Fig.~\ref{fig:n77_320_osc}, while the power spectra of all of the uncontaminated variable stars are shown in Fig.~\ref{fig:all_osc_star}.  
For the binaries, only 5 were not contaminated by the other sources and these are shown in Fig.~\ref{fig:all_binary_star}.

\begin{figure*}
\centering
   \includegraphics[width=\textwidth]{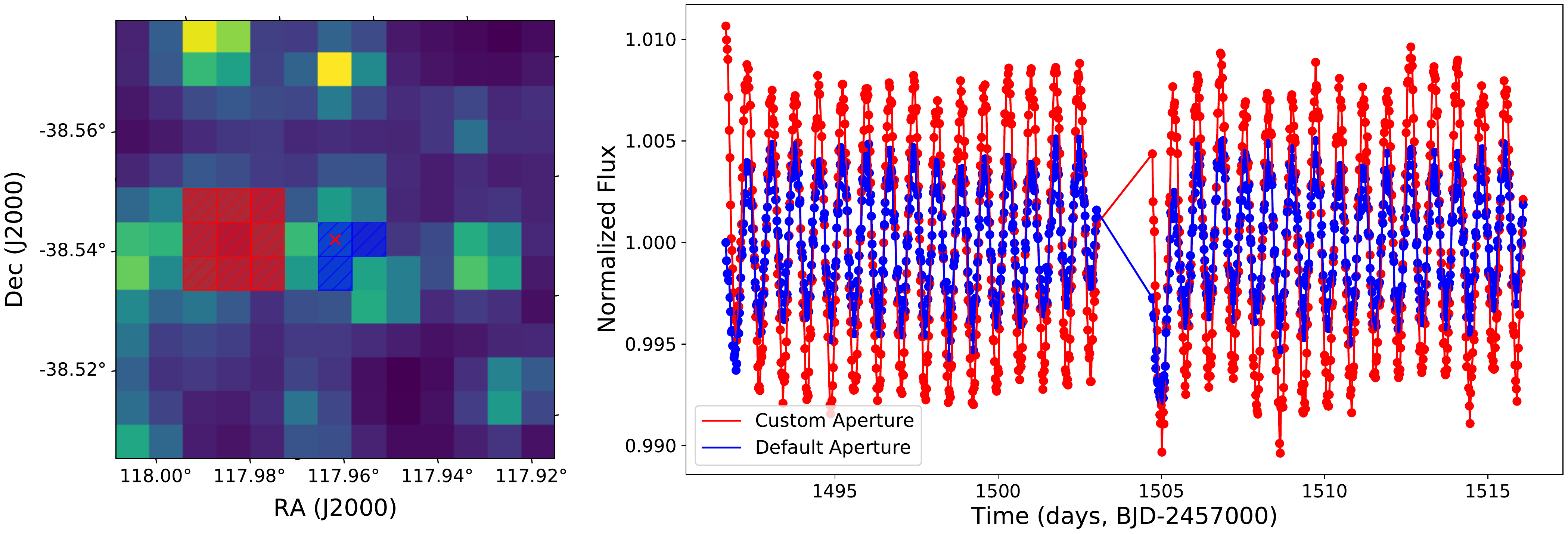}
     \caption{Target pixel file and light curve of star N77-42. The default aperture in blue shows variation in the light curve. The custom aperture from the nearby source in red also shows variation in the light curve.}
     \label{fig:duplicated_source}
\end{figure*}

As an example of the effect of contamination, the light curve of N77-42 is shown in blue in Fig.~\ref{fig:duplicated_source} using the default aperture.  Our custom aperture (red) of a nearby source also shows very similar oscillation behaviour, and thus  indicates that the target source is contaminated by a nearby star. 

In Fig.~{\ref{fig:isochrone_fit}}, we show the extinction-corrected HR diagram, with the identification of variable stars in \ngc, and the optimal isochrones.  We highlight newly identified binaries by blue dots, and the oscillating stars by red and green. The uncontaminated stars (in red) are more concentrated near the turn off region in the CMD. Four of the five binaries are located above the  main sequence as expected. The variable type that we identify here is given in Table~\ref{tab:sourceinfo} along with the star's ID, $G$ magnitude, and membership probability.

\subsection{Extraction of frequencies from the (uncontaminated) light curves}
\label{ssec:extraction_of_freq}
To further analyse the oscillating stars, we need to identify a detection threshold for which we are confident that the oscillation is not due to noise, S/N$_{\rm cons}$. In order to identify this value, we do the following: We first construct a  periodogram for the target source from the light curve. From the periodogram we identify the frequency with the maximum amplitude. We generate a simulated signal using the identified frequency and amplitude along with the real time series data points from the source. Gaussian noise with a standard deviation equal to the flux error of the signal is added to the simulated light curve. Then, a new periodogram is generated from this simulated light curve. By dividing by the background median value, we obtain a periodogram in units of signal-to-noise ratio (S/N). We adopt the 99.9 percentile of the distribution of the S/N for the full frequency range as an initial S/N threshold. This threshold ensures that any signal detected below this threshold is considered as noise, while still allowing to detect the input frequency. We repeat the simulation of the light curve $N = 2\,000$ times, and then adopt the median value of the S/N initial thresholds and denote this as S/N$_{\rm init}$.

In the second step of our analysis, we identify all of the frequency signals in the real light curve above S/N$_{\rm init}$, where again the noise is defined as the median background. We then repeat the simulation of the periodogram, as explained above, and obtain an updated more conservative S/N threshold S/N$_{\rm cons}$, defined as the 99.99 percentile. This value is more realistic because the signal now contains also low amplitude frequencies.

For each of the detected frequencies above S/N$_{\rm init}$ we have, in fact, a distribution  of S/N from the simulations corresponding to that detected frequency. From the distributions, we calculate the fraction of times that the signal is above S/N$_{\rm cons}$, and this is listed in Table~\ref{tab:oscillation_source} under the column heading $f_{\rm det}$. 

{
We repeated this signal simulation for each oscillating star for ten-minute and thirty-minute data and identified frequencies with $f_{\rm det}> 0.99$, while ignoring frequencies higher than  416~$\mu$Hz, half of the Nyquist frequency; this ensures the removal of reflected frequencies. We only used common frequencies identified with both 30- and 10-minute cadence data up to the Nyquist frequency of the 30-minute data (277~$\mu$Hz). 
 The list of detected frequencies, amplitudes and S/N, detection probabilities for each of the non-contaminated stars is given in the  appendix Table~\ref{tab:oscillation_source}. 
 In the next section, we only use the frequencies that have a $f_{\rm det}>0.99$.
}

\section{Structure and evolutionary models \label{sec:models}}
In this section we perform a modelling analysis of the observed oscillating stars with the aim to investigate if we can constrain any further the age, metallicity or extinction of \ngc. In order to compare the observed oscillations, we need to construct stellar structure models of our target stars and calculate oscillation frequencies for these models. This requires the use of a stellar structure, evolution and pulsation code (Sect.~\ref{ssec:obs_osc_pul_model}). However, the first step in finding adequate models for the oscillating stars, is to  constrain the global parameters of the star that is temperature and luminosity, see Sect.~\ref{ssec:templum}. We then also need an estimate of the stellar mass and chemical composition, and here we rely on our results based on publicly available isochrones from Sect.~\ref{sec:isochrone}.

\subsection{Temperature and Luminosity determination} \label{ssec:templum}

Effective temperature (\teff) and luminosity ($L$) are essential parameters to constrain the stellar models for asteroseismology. We therefore, need to convert our observed dereddened absolute magnitude $M_G$ and intrinsic colour \gbprp$_0$, to these quantities. We used the method described in \cite{casagrande_2020} to estimate \teff\footnote{{We use this approximation for \gaia\ DR2 data. We have compared the calculated \teff\ using \gaia\ DR2 data with \gaia\ DR3 and they are in agreement for our stars within the uncertainties.}} from our intrinsic colours. $L$ is calculated using the standard equation:
\begin{equation}
-2.5 \log_{10}  L/\lsol = {M_{\rm G}} + {BC_{\rm G}}(T_{\rm eff}) - {\rm 
 M_{{\rm bol{\odot}}}} 
\label{eqn:lum}
\end{equation}
where $BC_{\rm G}$ is the bolometric correction and is taken from \citep{andrae_2018} and M$_{{\rm bol{\odot}}}$ is 4.74~mag.

\subsection{MESA and GYRE}

We compute the evolutionary sequences and oscillations using the combination of the MESA stellar evolution code \citep{paxton_2011} and the GYRE pulsation code \citep{townsend_2013}. {We adopt a standard input physics\footnote{{We use the mesa version 12778. The default physics inputs can be found from 
\url{http://mesa.sourceforge.net}.}}, the summary of adopted physics is listed in Table~\ref{tab:meas_parameters}}. 
To run MESA, the following model parameters are needed: mass, initial chemical composition, and age. For the initial chemical composition, we need two of $X, Y$, and $Z$ where these are the mass fractions of hydrogen, helium, and other elements, respectively, with the constraint of $X+Y+Z=1$. The output of MESA is a stellar structure defined by the above parameters, and then this is used as an input to GYRE to calculate the oscillation frequencies.

To determine the mass, we use the initial values obtained from our analysis in Sect.~\ref{sec:isochrone}. The adopted initial $Z$ and $Y$ for different \mh\ is chosen based on the metallicity enrichment law adopted by the BASTI isochrones \citep[Sect.~4][]{hidalgo_2018}, but we note that we did not further explore the impact of $Y$. A mixing-length parameter is also needed as input, but as this has little or no effect on the stars we are interested in, that is stars in the range of $M>1.5$~\msol, in we set it to the default value (1.9179) obtained from calibrating a solar model. The age range of the models corresponds to the full range of plausible ages from the cluster analysis presented in Sect.~\ref{sec:isochrone}, that is 0.6 -- 1.1~Ga.

\subsection{Methodology to search for optimal models for each star}
\label{ssec:meth_of_search}
\begin{figure}
    \centering
    \includegraphics[width=8.5cm]{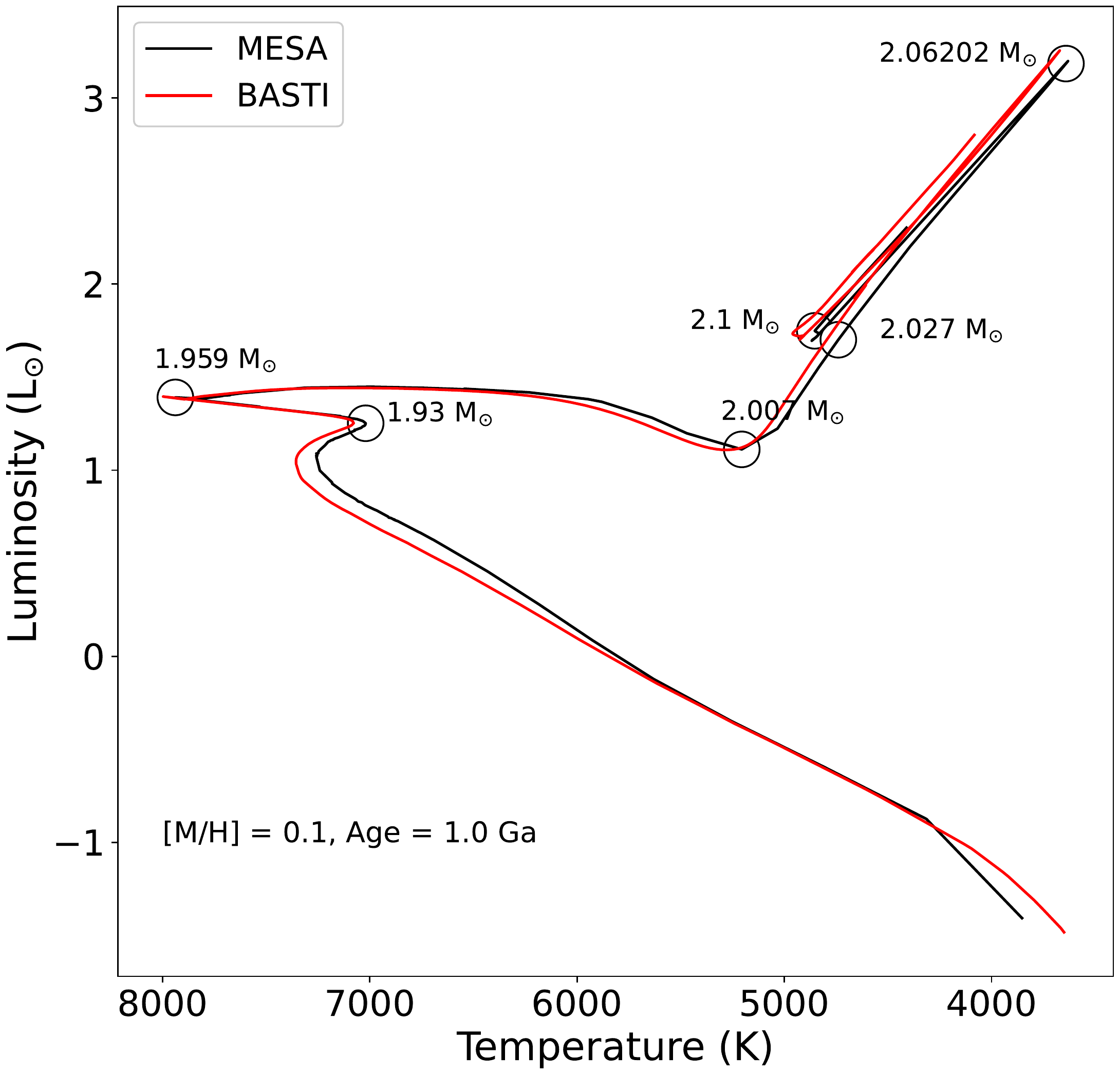} 
    \caption{Comparison of an isochrone created from stellar evolution tracks generated using the MESA code with an isochrone from BASTI.  We indicate the masses at the critical turning points of the 1.0~Ga isochrone with \mh\ = 0.1~dex.}
    \label{fig:mesa_basti_iso.pdf}
\end{figure}

{We created a grid of isochrones with ages from 0.6 to 1.2~Ga with a step size of 0.1~Ga and \mh\ with values 0.06, 0.1, and 0.16~dex, corresponding to the values we obtained for high resolution spectroscopy in Sect.~\ref{ssec:metallicity}.} The isochrones that we constructed were obtained by running many evolution models for various masses with different mass resolution in order to well cover the fast steps in an evolution track (that is around the turn-off and on the way to the giant branch). A comparison of one of our isochrones with BASTI is shown in Fig.~\ref{fig:mesa_basti_iso.pdf} for \mh\ = 0.1~dex and an age of 1.0~Ga. The MESA isochrone during main-sequence is very similar to the BASTI isochrone. But during the turn-off phase and at the red giant phase the isochrones differ slightly due to the difference in physics between the two models.

\begin{figure}
    \centering
    \includegraphics[width=8.5cm]{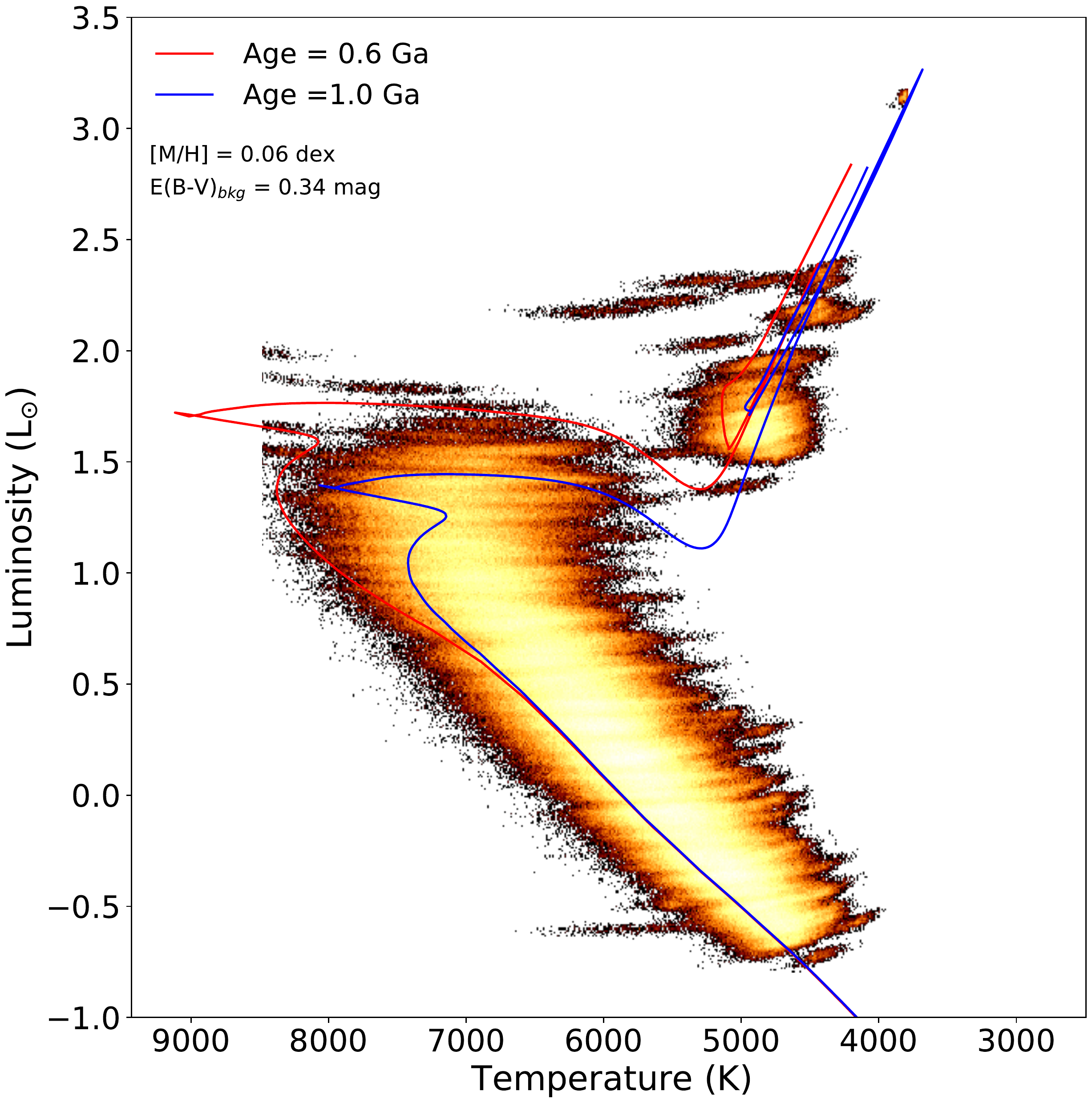}
    \caption{HR diagram for NGC 2477 with simulated LT error for background reddening of 0.34~mag for all stars in the cluster. BASTI isochrone of age 0.6 and 1.0~Ga with \mh\ = 0.06~dex is shown in the figure. For each star, Monte-Carlo simulation of 1000 samples is done.}
    \label{fig:sim_LT_comp_iso}
\end{figure}

In order to find the optimal models for each of the 16 stars, we performed our search on these isochrones in a 3D space: \ebv$_{bkg}$ (which impacts \teff\ and \lum), \mh, and age. The \ebv$_{bkg}$\ ranges from 0.24 -- 0.34~mag (see Sect.~\ref{ssec:background_extinction}), and this free parameter has a direct impact on the \ebv\ for each star.  We consider the impact of \ebv\ on the stellar properties by calculating, for each star, the \lum\ and \teff\ (Sect.~\ref{ssec:templum}) using a Monte-Carlo simulation with 1000 samples.  { This allows us to conserve the correlations among these properties for a given extinction and uncertainty.} By searching among the 3D parameter space (for each star), we could further exclude some combinations of the three parameters since the global \lum\ and \teff\ did not fit the cluster. 
As an example, Fig.~\ref{fig:sim_LT_comp_iso} illustrates an isochrone with an age of 0.6~Ga and \mh\ = 0.06~dex which does not fit the HR diagram; hence we exclude it from further analysis. The parameter combinations of isochrones that were further used subsequently are listed in Table~\ref{tab:iso_modelling}.

\begin{table}[]
\centering
\caption{Isochrones used for asteroseismic modelling}
\label{tab:iso_modelling}
\begin{tabular}{@{}ccccc@{}}
\toprule
\mh & Z      & Y      & \begin{tabular}[c]{@{}c@{}}\ebv$_{bkg}$\end{tabular} & Age           \\ 
(dex)               &        &        & (mag)                                                            & (Ga)          \\ \midrule
                    &        &        & 0.24                                                             & 0.6           \\
0.06                & 0.0171 & 0.2694 & 0.29                                                             & 0.8, 0.9, 1, 1.1 \\
                    &        &        & 0.34                                                             & 1, 1.1         \\
                    &        &        & 0.24                                                             & 0.6           \\
0.1                 & 0.0187 & 0.2715 & 0.29                                                             & 0.8, 0.9, 1, 1.1 \\
                    &        &        & 0.34                                                             & 0.8, 0.9, 1, 1.1 \\
                    &        &        & 0.24                                                             & 0.6           \\
0.16                & 0.0213 & 0.2748 & 0.29                                                             & 0.8, 0.9, 1, 1.1 \\
                    &        &        & 0.34                                                             & 0.8, 0.9, 1, 1.1 \\ \bottomrule
\end{tabular}
\end{table}

The mass of a star for a  \{\mh, age, \ebvbkg\}  combination is estimated by finding models within the luminosity-temperature error correlation, hereon the LT error.
To illustrate this, we show the position of the star N77-342 in the HR diagram for $\ebvbkg=0.29$~mag in Fig.~\ref{fig:N77_342_lower_age} upper left panel.  Its central \lum\ and \teff\ are 15.7~\lsol\ and 7531~K, respectively, and the LT error is shown by the green dots.  By inspecting the masses of the models that intersect the LT error, we derive the lower and upper limit of the mass of 1.90 and 1.95~\msol.

GYRE is then used to compute the theoretical pulsations for all of the masses along the track for the given chemical composition and age, where  restrict the theoretical frequency up to angular degree $l=2$ and radial order $n_{pg}$ > --10, where $n_{p}$ and $n_{g}$  represent the order of $p$ and $g$ modes, respectively. $n_{pg}$ is the radial order within the Eckart-Scuflaire-Osaki-Takata scheme \citep[see ][]{takata_2006}. This criterion avoids most higher order g-modes that will not have high amplitudes and the detection of modes with $l>2$ or 3 are more difficult with photometric data \citep{stamford_1981}. 

\begin{figure}
    \centering
    \includegraphics[width=0.5\textwidth]{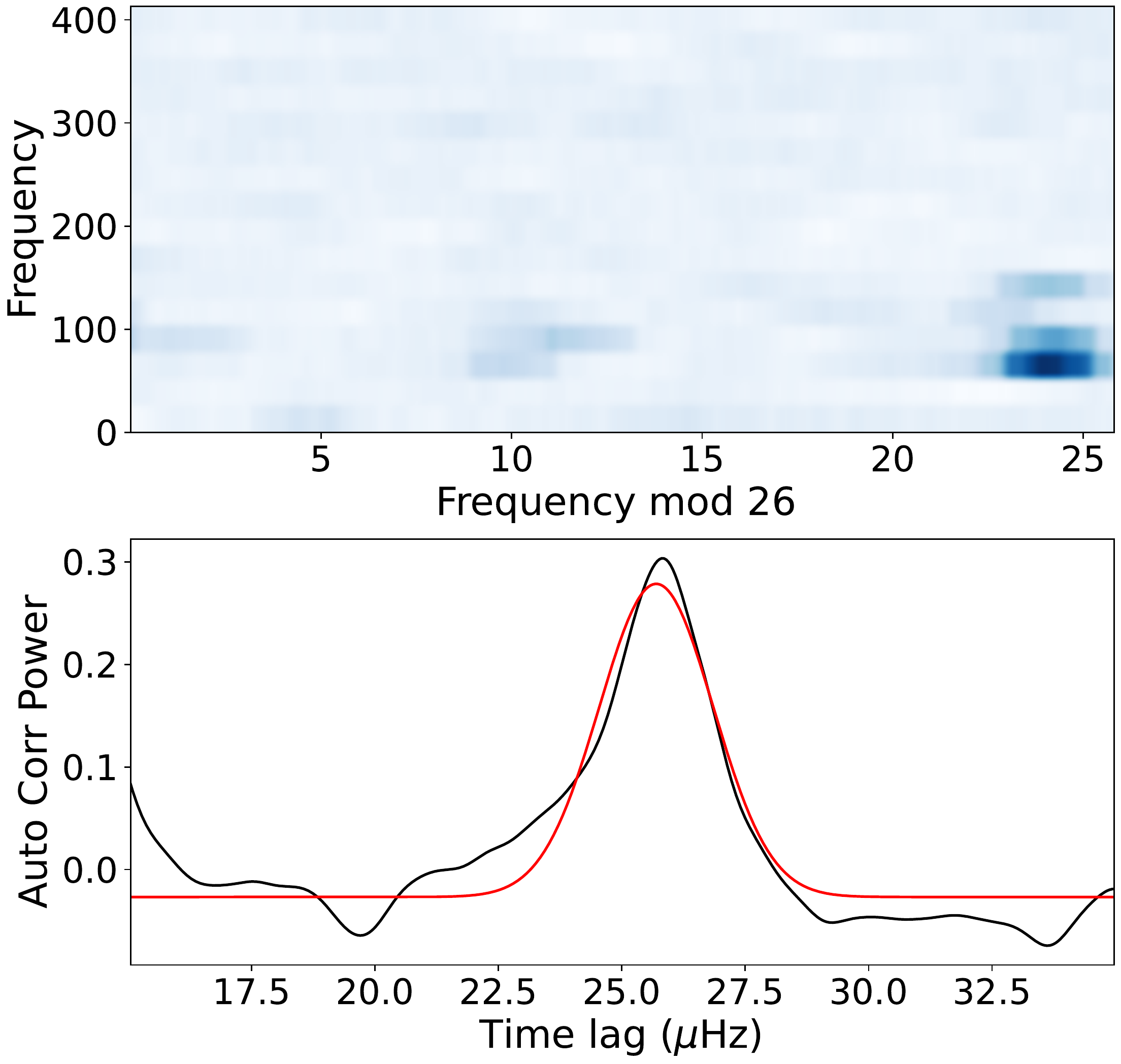} \\
    \caption{Echelle diagram (top) and auto correlation function (bottom) for N77-255.
    }
    \label{fig:echelle_255}
\end{figure}

{For each oscillating star, \ebvbkg, age, and metallicity, we find an optimal mass  within the upper and lower limits by comparing the theoretical and observed pulsations.   In Fig.~\ref{fig:N77_342_lower_age} upper left panel this is shown as the yellow filled circle.   The theoretical frequency spectrum of this model is illustrated on the right panel, where the colour-code represents different angular degrees ($l$) and the observed spectrum is shown in grey.  The confirmed oscillation peaks (Sect.~\ref{ssec:extraction_of_freq}) are indicated by black dots.  On this panel, which is a 0.6~Ga star, there is no agreement with the observed and theoretical frequencies.  
In the middle and lower panels, we show optimal models for different ages (0.8~Ga and 0.9~Ga) with the same metallicity.  There is more agreement found for the 0.9~Ga model.
We repeat this exercise for each star and obtain a list of optimal models {\it for each} combination of \{\mh, age, \ebvbkg\}.

\subsection{Optimised models with cluster constraints} 
\label{ssec:obs_osc_pul_model}
From here on, we approach the problem from a combined \{\mh, age, \ebvbkg\} approach.  
After finding the optimum model for all the stars for different combinations of parameters, we compared the theoretical and observed frequencies of all 16 stars. We found that the models with lower ages are less likely to match the theoretical frequencies.  To illustrate this, we take as an example the star N77-342 which was shown in Fig.~\ref{fig:N77_342_lower_age}. 
The observed and theoretical frequencies for the younger age models (0.6 and 0.8~Ga) are not in as good agreement compared to the higher age models. This is further confirmed from the analysis of \deltanu\ in the next section.

\begin{figure*}
    \centering
    \includegraphics[width=\textwidth]{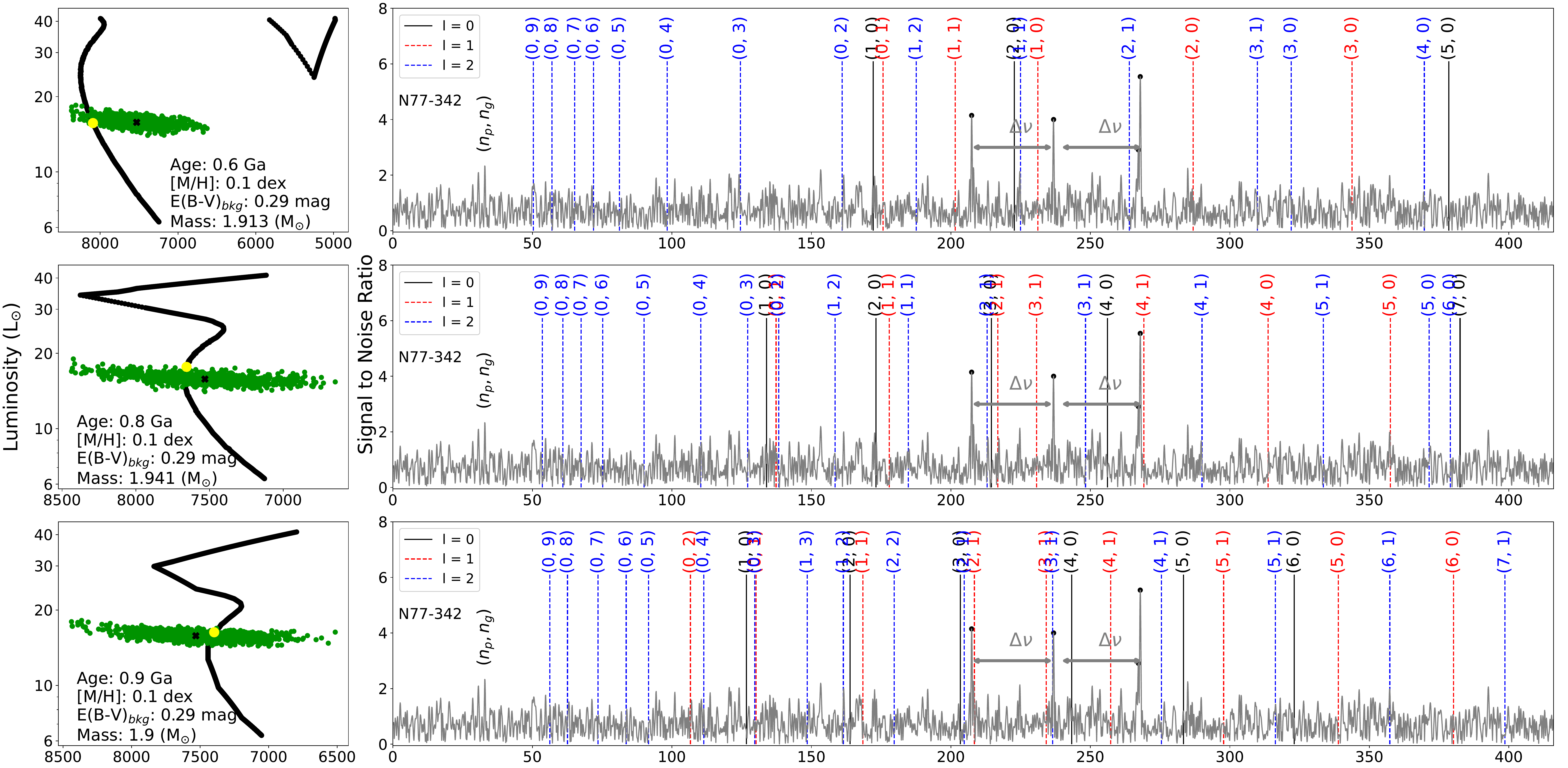}\\

    \caption{{\sl Left:} Isochrone tracks (black) with ages of 0.6, 0.8, and 0.9~Ga with \mh\ = 0.1~dex and \ebvbkg\ = 0.29~mag. The LT error for the star N77-342 is shown as the {green} points which crosses the isochrones, with the central position indicated by the black cross. The best-fitted model is indicated by the yellow filled circle. 
    {\sl Right:} 
    Observed oscillation spectrum (grey) for the star N77-342.
The black dots over the observed peaks indicate the observed frequencies with $f_{det}>0.99$.  
Theoretical frequencies corresponding to the best-fitted model (red dot on the left panel) are overplotted in each of the panels. 
The angular degree and the radial order of each theoretical frequency is also indicated above each frequency, colour-coded by the angular degree.}
    \label{fig:N77_342_lower_age}
\end{figure*}

\subsubsection{Identification of \deltanu}
\label{ssec:iden_deltnu_age}

Identifying \deltanu\ for solar-type stars can be straight forward if the S/N is good enough since they are known to follow a pattern \citep{garcia_2019}. Since we look into non-solar type oscillations that do not follow a pattern, it is much more challenging to identify \deltanu, but one is expected to exist \citep{garciahernandez_2015}. 
{ Moreover, TESS FFI 10 minutes data covers only low-frequencies (0 - 833~$\mu$Hz) which is not the asymptotic regime for these stars and so we can expect that the \deltanu\ have some non-equal-spaced frequencies, and this makes it harder to identify \deltanu.}
To search for and identify \deltanu, we investigated all of the pulsation models for each star for a given combination of \{\mh, age, \ebvbkg\}. The pulsation models allow one to calculate a frequency separation, and we used this theoretical value to search for an observable signature of \deltanu, while allowing the optimal mass to vary slightly. {The large frequency separation from the pulsation models is computed by finding the average difference of $l=0$ modes up to 416~$\mu$Hz}. To find the observed large frequency separation, we made use of the computed models for each star. The observed frequencies were compared with different models of combination of \{\mh, age, \ebvbkg\}.  

{ As an example, in Fig.~\ref{fig:N77_342_lower_age}, we show models of three different ages for a star N77-342. If we compare the three observed peaks 207.4, 236.8, and 267.8 ~$\mu$Hz with the theoretical models, the spacing between the three peaks is well matched with the \deltanu\ of $l=0$ modes of higher ages. Thus these three peaks represent \deltanu\ and it is measured to be 30~$\mu$Hz. This approach allowed us to identify \deltanu\ with confidence for four of our stars on the main sequence, N77-342: 30~$\mu$Hz, N77-210: 23~$\mu$Hz, N77-255: 26~$\mu$Hz and N77-325: 32~$\mu$Hz, as can be seen in Fig.~\ref{fig:model_two_stars} where we show the observed and calculated theoretical frequencies and also indicate the individual observed frequencies where a measured \deltanu\ is found. It is harder to identify the observed \deltanu\ using the approach of an echelle diagram because of the low-frequency coverage of TESS FFI and the lack of many excited frequencies.  We illustrate one example where the echelle diagram can be used, and it and the autocorrelation function, constructed by convolving the oscillation spectrum of the star N77-255, are given in Fig.~\ref{fig:echelle_255}. The peak of the autocorrelation function corresponds to \deltanu\ which is at 26~$\mu$Hz, and this confirms the \deltanu\ identified with the models. A Gaussian fit to the autocorrelation provided the error of 2~$\mu$Hz. This typical error is taken for all the stars. The observed \deltanu\ for the stars are tabulated in Table~\ref{tab:mass_lum_osc_star}}.
 
For the other 12 stars, either no frequency separation agreed with the theoretical one, or the existence of only one or two frequencies did not allow us to confirm the signature. In the next section we proceed with the use of the \deltanu\ signatures of these four stars.   
We note that the stars N77-342, N77-210, and N77-255 have been identified as $\delta$ Scuti~/~$\gamma$ Doradus~/~SX Phoenicis stars in the Gaia DR3 variable catalogue \cite{eyer_2022}.

\subsubsection{Using \deltanu\ and mass to identify the cluster parameters}

For each of the four stars, we have identified an observable signature of \deltanu\ with an uncertainty.  Using the \lum\ and \teff\ constraints we also have derived an upper and lower limit of the mass (for each isochrone).  
We plot these  \deltanu\ and mass values for the four stars in each of the panels in Fig.~\ref{fig:deltanu_two_stars}.
Each star is indicated by a different pattern, and the ellipses represent the full span of the observed \deltanu\ and mass value, while the colour code is for one of the three values of \ebvbkg.   We remind readers that for a higher \ebvbkg, we obtain a lower value of \ebv, which reduces the \lum\ and \teff\ of the star.  This would result in a lower fitted mass value to an isochrone of a given age and \mh.  This is precisely what we see in each of these panels, with the mass values shifting towards lower values as the \ebvbkg\ increases.  

As an example in the top left panel we show \deltanu\ and the range of masses corresponding to an age of 0.6~Ga and \mh\ = 0.06~dex for the four stars.  If we suppose \ebvbkg\ = 0.29~mag (green ellipses), the star N77-342 has a mass of between 1.89 and 1.93~\msol.   
We also overplot the theoretical values of \deltanu\ versus mass for the combination of age = 0.6~Ga and \mh\ = 0.06~dex, which is illustrated by the black dots in that same panel. 
As one can see in the top left panel, the isochrone does not match any of the four stars. So this combination of age and metallicity can be discarded. Likewise, by inspecting the other panels, we can discard 0.6, 0.8, and 0.9~Ga ages for all combinations of metallicity. In the case of 1.0~Ga and 1.1~Ga (the fourth and fifth rows), the four stars have masses that have theoretical \deltanu\ matches with the observed ones which for \mh = 0.10 and 0.16~dex, respectively.

{
Now for these combinations of age and metalliticy, we {visually} compare the observed and the theoretical frequencies directly in Fig.~\ref{fig:model_two_stars} and Fig.~\ref{fig:model_two_stars_1.1gyr}. Here we specifically compare the $l=0$ theoretical modes with the observed frequencies. 
For star N77-342, the observed peaks 207.4, 236.8, and 267.8~$\mu$Hz matches with $l=0$ and $n=3, 4, 5$ frequencies for model with age = 1.0~Ga and [M/H] = 0.1~dex. The observed frequencies are in agreement with the theoretical frequencies for all four stars for models with age = 1.0~Ga \& [M/H] = 0.10~dex and  age = 1.1~Ga \& [M/H] = 0.16~dex.}
For all other models of 1.0 and 1.1~Ga, the best-fitted masses and the observed \deltanu\ are in agreement with the unique isochrone for all four stars, but the individual observed frequencies {visually} differ from the theoretical ones and therefore we chose not to use these as our reference model.

In summary, we have identified two tailored MESA isochrones with ages of 1.0~Ga (\mh\ = 0.1~dex) and 1.1~Ga (\mh\ = 0.16~dex) in agreement with the observed \deltanu\ and individual frequencies. 
For these isochrones, the best corresponding \ebvbkg\ are 0.34 and 0.29~mag, respectively.
However, we were not able to further refine the \ebvbkg\ from the analysis. 
We, therefore, adopt \ebvbkg\ to be 0.29 $\pm$ 0.05~mag.

{Furthermore, when we compare these isochrones to the full set of stars in the cluster, the 1.0~Ga isochrone is fitted best to the individual star's $L$ and \teff\ (Fig.~\ref{fig_gaia_dr3_vsini}}).} { The best-fitted mass determined using asteroseismic modelling and the mass determined from isochrone fitting for all the oscillating stars along with its \lum\ and \teff\ is given in Table~\ref{tab:mass_lum_osc_star}.}
Our analysis points towards a cluster age of 1.0 $\pm$ 0.1~Ga with a \mh\ to be 0.10 $\pm$ 0.05~dex.

\begin{figure*}
    \centering
    \includegraphics[width=\textwidth]{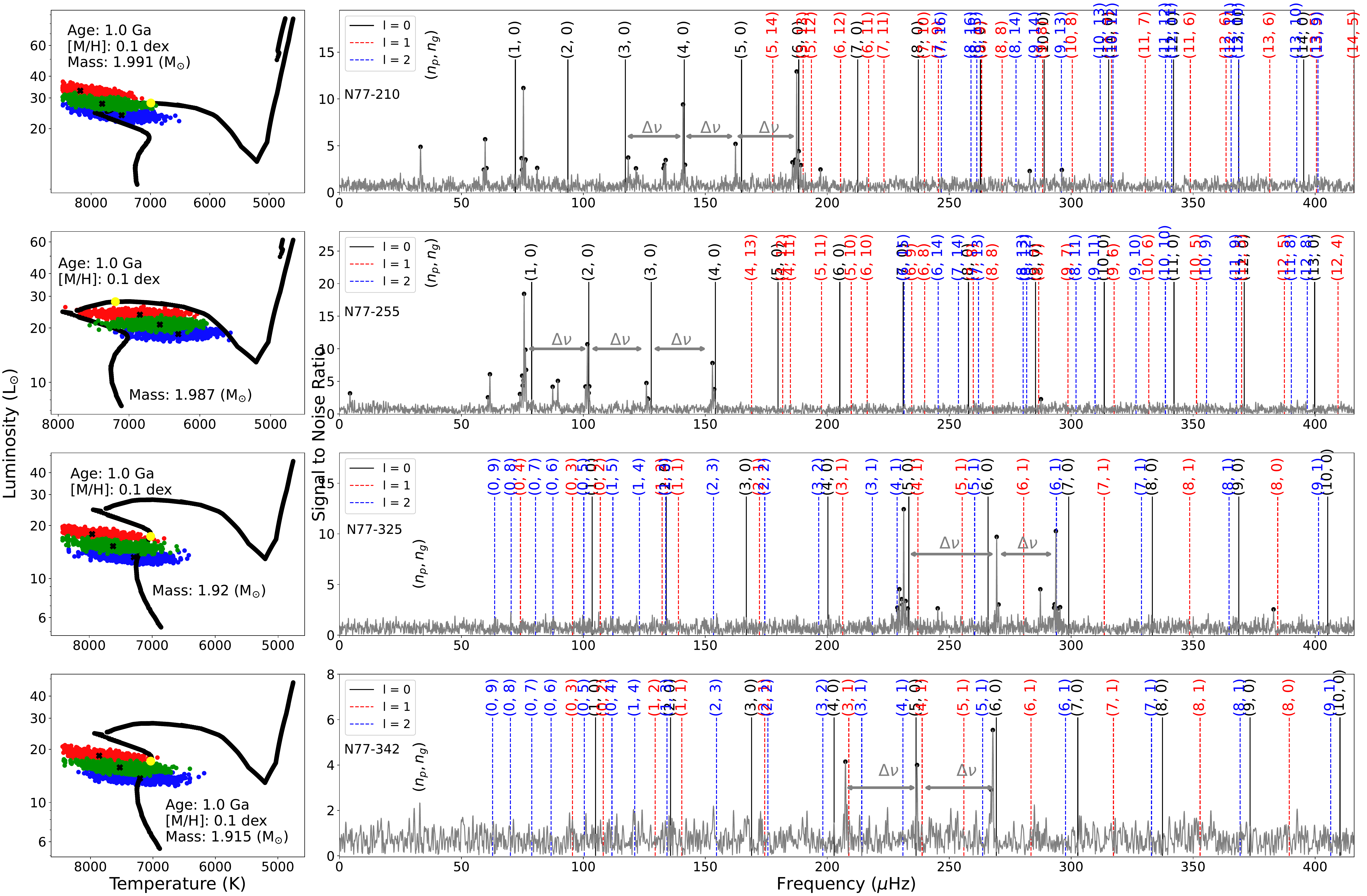} \\
    \caption{
    See caption of Fig.~\ref{fig:N77_342_lower_age} for details.
    {\sl Left:} A MESA isochrone (black) with an age of 1.0~Ga and \mh\ = 0.1~dex for the 
     stars {N77-210, N77-255, N77-325, and N77-342 (top to bottom)}. {The scattered red, green and blue points represent the LT error of stars for \ebvbkg\ = 0.24, 0.29, and 0.34~mag, respectively.}
    {\sl Right:} Observed frequencies (grey) and the corresponding pulsation models along with the proposed observed \deltanu. 
    }
    \label{fig:model_two_stars}
\end{figure*}

\begin{figure*}
    \centering
    \includegraphics[width=0.9\textwidth]{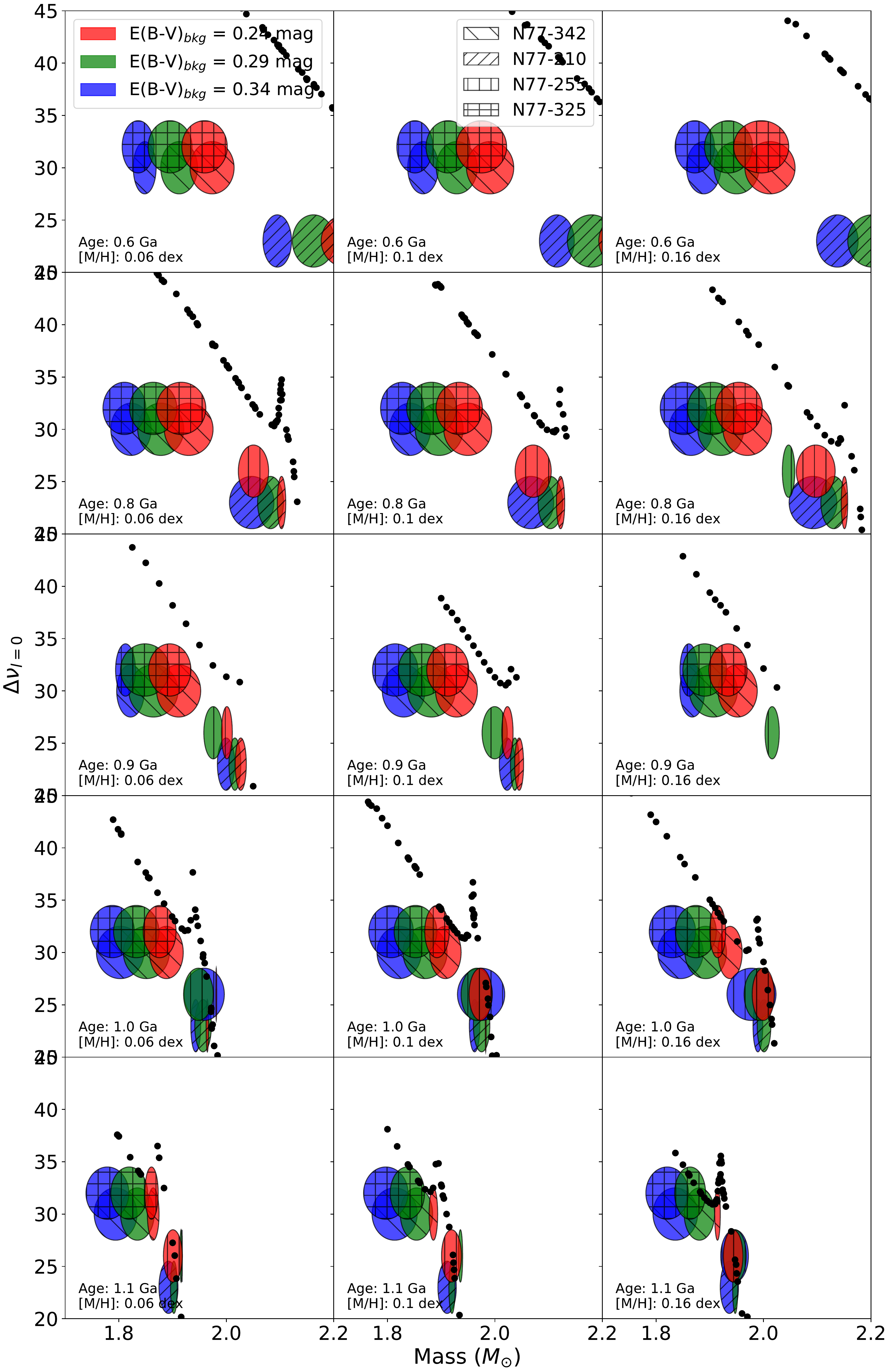}
    \caption{Mass and \deltanu\ for different isochrone models are shown in the figure along with the observed \deltanu\ of {four stars N77-210, N77-255, N77-325, and N77-342}, which are marked in as ellipses. The \deltanu\ of pulsation models of different mass are shown (black filled circles). The height and width of the ellipses represent the error in the \deltanu\ measurements and the mass ranges. }
    \label{fig:deltanu_two_stars}
\end{figure*}

\begin{figure*}
    \centering
    \includegraphics[width=\textwidth]{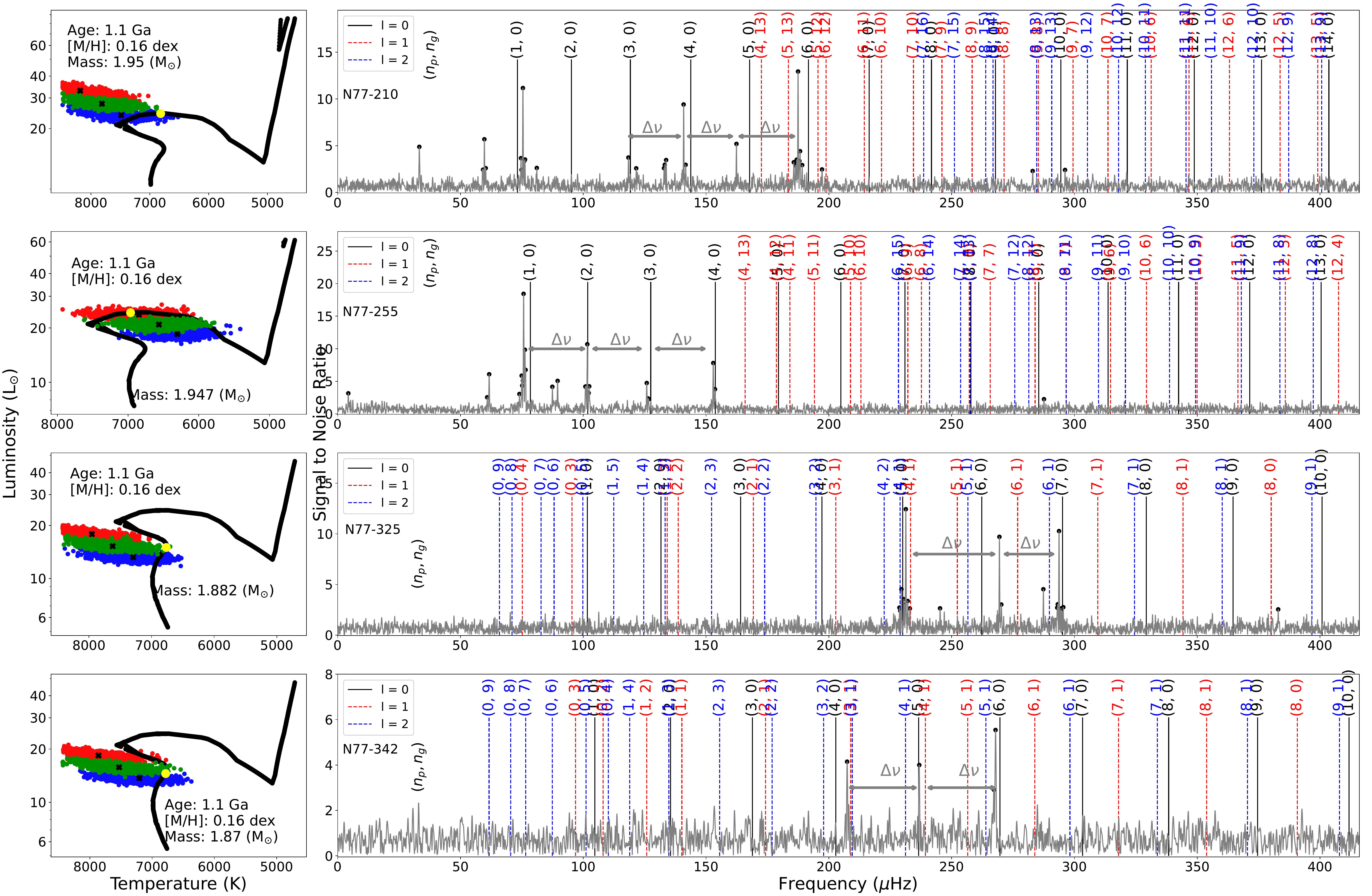} \\

\caption{
    See caption of Fig.~\ref{fig:N77_342_lower_age} for details.
    {\sl Left:} Our isochrone (black) with an age of 1.1~Ga and \mh\ = 0.16~dex for the 
     stars N77-210, N77-255, N77-325, and N77-342 (top to bottom). {The scattered red, green and blue points represent the LT error of stars for \ebvbkg\ = 0.24, 0.29, and 0.34~mag, respectively.} {\sl Right:} Observed frequencies (grey) and the corresponding pulsation models are shown for the two stars.}
    \label{fig:model_two_stars_1.1gyr}
\end{figure*}

\begin{table*}[]
\caption{Properties of oscillating stars. The error in \lum\ and \teff\ is the square root of diagonal elements of covariance matrix and the $c_{LT}$ represent the correlation coefficient between \lum\ and \teff. 
{\deltanu\ measured with a typical error of 2~$\mu$Hz.}
}
\begin{tabular}{lcccccc}
\toprule
Source & Mass$^{*}$ &     Mass   &        L  & \teff\ & $c_{LT}$ & \deltanu\ \\
      &(\msol) & (\msol)  & (\lsol) & (K) & &($\mu$Hz) \\  \midrule
N77-135 &2.003 & $1.99 \pm 0.01 $ & $33.58 \pm 1.10 $ & $6657 \pm 288 $ & $0.10$ & \\
N77-148 &1.993 & $2.02 \pm 0.01 $ & $40.24 \pm 1.48 $ & $7284 \pm 331 $ & $0.38$ & \\
N77-151 &1.991 & $1.99 \pm 0.01 $ & $31.00 \pm 1.07 $ & $7244 \pm 324 $ & $0.33$ & \\
N77-177 &1.991 & $2.02 \pm 0.01 $ & $38.99 \pm 1.30 $ & $6945 \pm 318 $ & $0.16$ & \\
N77-186 &1.984 & $1.99 \pm 0.01 $ & $32.69 \pm 1.18 $ & $7419 \pm 319 $ & $0.43$ & \\
N77-199 &1.961 & $1.97 \pm 0.02 $ & $26.74 \pm 1.12 $ & $7662 \pm 340 $ & $0.58$ & \\
N77-202 &1.988 & $1.99 \pm 0.01 $ & $31.50 \pm 1.35 $ & $7695 \pm 338 $ & $0.65$ & \\
N77-210 &1.991 & $1.98 \pm 0.02 $ & $27.87 \pm 1.18 $ & $7803 \pm 321 $ & $0.62$ & 23 \\
N77-240 &1.960 & $1.95 \pm 0.01 $ & $21.11 \pm 0.73 $ & $7299 \pm 318 $ & $0.38$ & \\
N77-250 &1.991 & $1.98 \pm 0.02 $ & $28.18 \pm 1.06 $ & $7471 \pm 328 $ & $0.48$ & \\
N77-255 &2.001 & $1.97 \pm 0.03 $ & $20.86 \pm 0.66 $ & $6588 \pm 271 $ & $-0.03$ & 26\\
N77-319 &1.851 & $1.88 \pm 0.04 $ & $15.54 \pm 0.52 $ & $6964 \pm 305 $ & $0.19$ & \\
N77-320 &1.961 & $1.96 \pm 0.01 $ & $22.54 \pm 0.80 $ & $7422 \pm 306 $ & $0.42$ & \\
N77-325 &1.860 & $1.85 \pm 0.04 $ & $15.33 \pm 0.62 $ & $7624 \pm 323 $ & $0.58$ & 32\\
N77-342 &1.915 & $1.87 \pm 0.04 $ & $15.83 \pm 0.58 $ & $7531 \pm 314 $ & $0.51$ & 30\\
N77-438 &1.79 & $1.82 \pm 0.05 $ & $13.44 \pm 0.45 $ & $6924 \pm 300 $ & $0.14$\\ \hline
\end{tabular}
\\$^{*}$ Mass determined using asteroseismic modelling.
\label{tab:mass_lum_osc_star}
\end{table*}

{
\section{Rotation}
\label{sec:rotation}

{ Rotation in stars can influence the stellar structure, which could affect the star's age and the observed pulsation frequencies \citep{eggenberger_2010}. \cite{brandt_2015} included rotation in  isochrones and found older ages for the Hyades cluster compared to non-rotating isochrones. Rotation has stronger effects on stars in the main sequence for early type stars. The temperature and flux can have alterations due to rotational effects that may cause an observer to perceive a star to be brighter or fainter in reality. These brighter or fainter main sequence stars will lead to uncertainties in the age \citep{maeder_2000,heger_2000a}.  It is therefore important to investigate the effect of rotating models on our results.

Gaia DR3 provides \vsini~ measurements for 179 member stars in the cluster, see 
Fig.~\ref{fig_gaia_dr3_vsini}.
Both N77-210 and N77-342 have \vsini~ of 195 $\pm$ 11 and 124 $\pm$ 16  km~s$^{-1}$, respectively, while the \vsini~ measurements in the cluster go up to 246 km~s$^{-1}$, clearly indicating substantial rotation for some stars.  
{We therefore repeated our previous analysis but this time we  considered rotation in the models.}

\begin{figure}
    \centering
   \includegraphics[width=8.5cm]{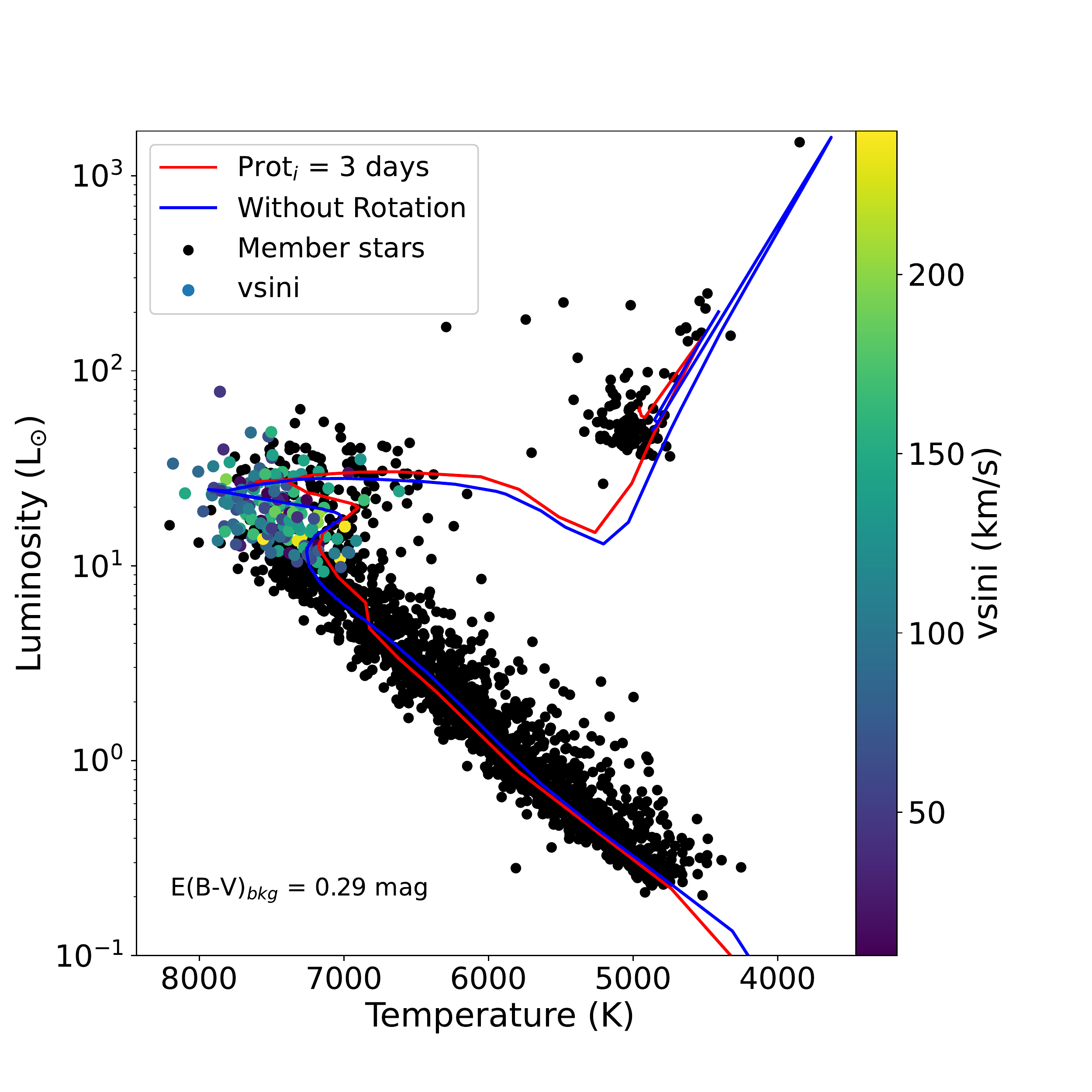}
\caption{\vsini~ measurements for member star from Gaia DR3 data. MESA isochrones with rotation (Prot$_{i}$ = 3 days) and without are shown as red and blue solid lines, respectively. The isochrones have an age of 1.0~Ga with a \mh\ of 0.1~dex, assuming \ebvbkg = 0.29~mag. The MESA isochrone with rotation that we constructed has missing masses in the RGB branch and therefore we do not see the full red giant branch in this figure.}
    \label{fig_gaia_dr3_vsini}
\end{figure}

To study the effects of rotation in our analysis, we created new sets of isochrones with different initial rotation values using MESA. 
We used the inlist\footnote{\url{https://cococubed.com/mesa_market/inlists.html}} file from \cite{gossage_2021}, where the authors implemented a description for magnetic breaking.  To check the model set-up, we first calibrated a solar-metallicity model on the Sun, and we adjusted the mixing-length parameter and free diffusion parameter to 2.1 and 1 $\times$ 10$^{4}$ cm$^{2}$s$^{-1}$, respectively \citep{gossage_2021} until we successfully recovered the \lum, \teff~ and rotation period of the Sun.
For the computation of rotation for the cluster NGC 2477, we adopted the same {mixing-length and diffusion parameter as the Sun} along with a metallicity of the cluster of 0.1~dex, with the helium and iron content identical to the ones used in the BASTI isochrone (Table~\ref{tab:iso_modelling}). 
We computed isochrones with different initial rotation values (Prot$_{i}$) of 3, 6, and 12~days. {Due to gravity-darkening effects, the inclination angle of the star along with rotation will change the observed \lum\ and \teff. The isochrones computed in this study are with intrinsic brightness \citep{paxton_2019}. The inclination angle can make a difference in \lum\ and \teff\ but this difference is within the \lum\ and \teff\ error of the stars. Since the inclination angle of the stars is unknown, we have not further computed isochrones with different projections.} Fig.~\ref{fig:rotation_models} illustrates the model surface average rotation as a function of mass for different Prot$_{i}$ at an age of 1.0~Ga.   We also show the measured \vsini\ where the mass is derived using the 1.0~Ga by comparing the \lum\ and \teff\ isochrone with Prot$_{i}$ = 3 days.
The effect of the position of these stars by using masses from the other models would be to shift them slightly to the left (increasing Prot$_{i}$ decreases mass). As the inclination angles for these stars are unknown, it is not possible to estimate the actual rotational velocity, but we know that the observed values set a lower bound.  We therefore propose a value of Prot$_{i}$ = 3 days to best model our stars, but we also calculated models with the other values for comparison.    

 \begin{figure}
\centering
   \includegraphics[width=8.5cm]{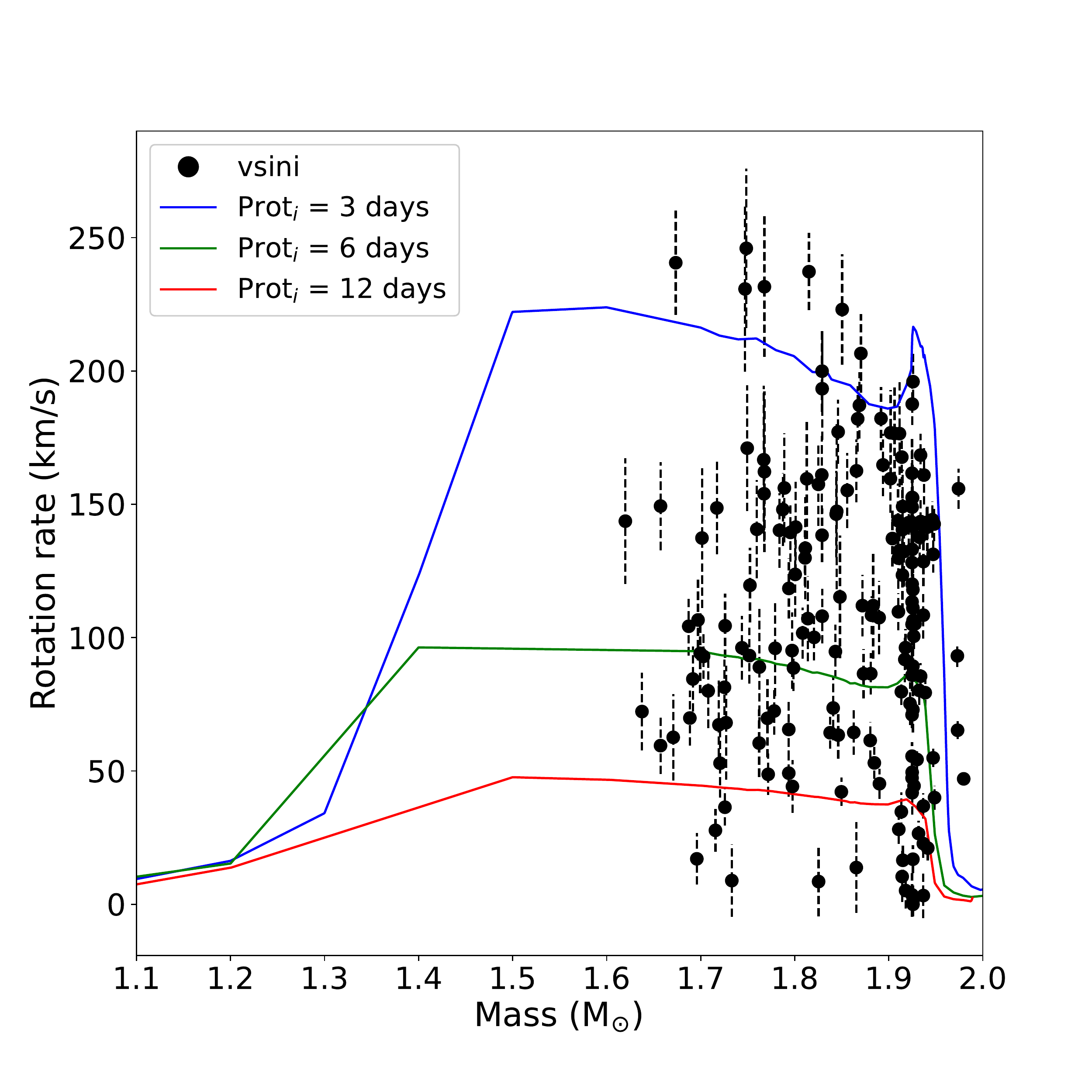}
     \caption{\vsini~ from Gaia DR3 (points) and theoretical average surface  rotation from mesa models for a 1.0~Ga isochrone with \mh = 0.1~dex as a function of mass (solid lines), for 3 initial of Prot$_{i}$ = 3, 6, and 12~days shown in blue, green and red respectively. }
     \label{fig:rotation_models}
\end{figure}

We follow the same methodology described in sections \ref{ssec:meth_of_search} and \ref{ssec:obs_osc_pul_model} and we calculate several isochrones for different combinations of age and metallicity and many pulsation models within the observed \teff, \lum\ constraint. 
In Fig.~\ref{fig:rotation_deltanu} we compare the observed and computed \deltanu-mass diagrams for the same four stars of interest while varying Prot$_{i}$ and age.   
It is very clear for this figure that the only matches between the observations and models is for an age $>0.9$~Ga and Prot$_i = 3$ days, with a better fit for the four stars at 1.0~Ga.      
Using this optimal cluster combination, we again show the four stars position in the HR diagram and a comparison of the observed and theoretical frequencies in 
Fig.~\ref{fig:rotation_frequencies}, where we see see good agreement and we  note that mass tends to reduce by 0.05 - 0.10~\msol\ compared to the non-rotating models.

In summary, by including rotation in the models we arrive at a similar conclusion to the one we found without including rotation, that is an optimal age of 1.0~Ga for \mh\ = 0.10~dex, and \ebvbkg = 0.29 $\pm$ 0.05~mag, but additionally we found that a Prot$_i = 3$ days (faster initial rotation) seems to be more favorable than slower initial rotation. The isochrone also well fits \lum\ and \teff\ of all the stars in the cluster (Fig.~\ref{fig_gaia_dr3_vsini}).
{ In this study, we have not computed the 2D effects, such as deformations due to the rotation. Along with rotational splitting, these can have effects on the computation of frequencies and mode identification \citep{reese_2013}. These effects in determining the age of the cluster should be studied with 2D stellar models in future.}

\begin{figure*}
\centering
   \includegraphics[width=\textwidth]{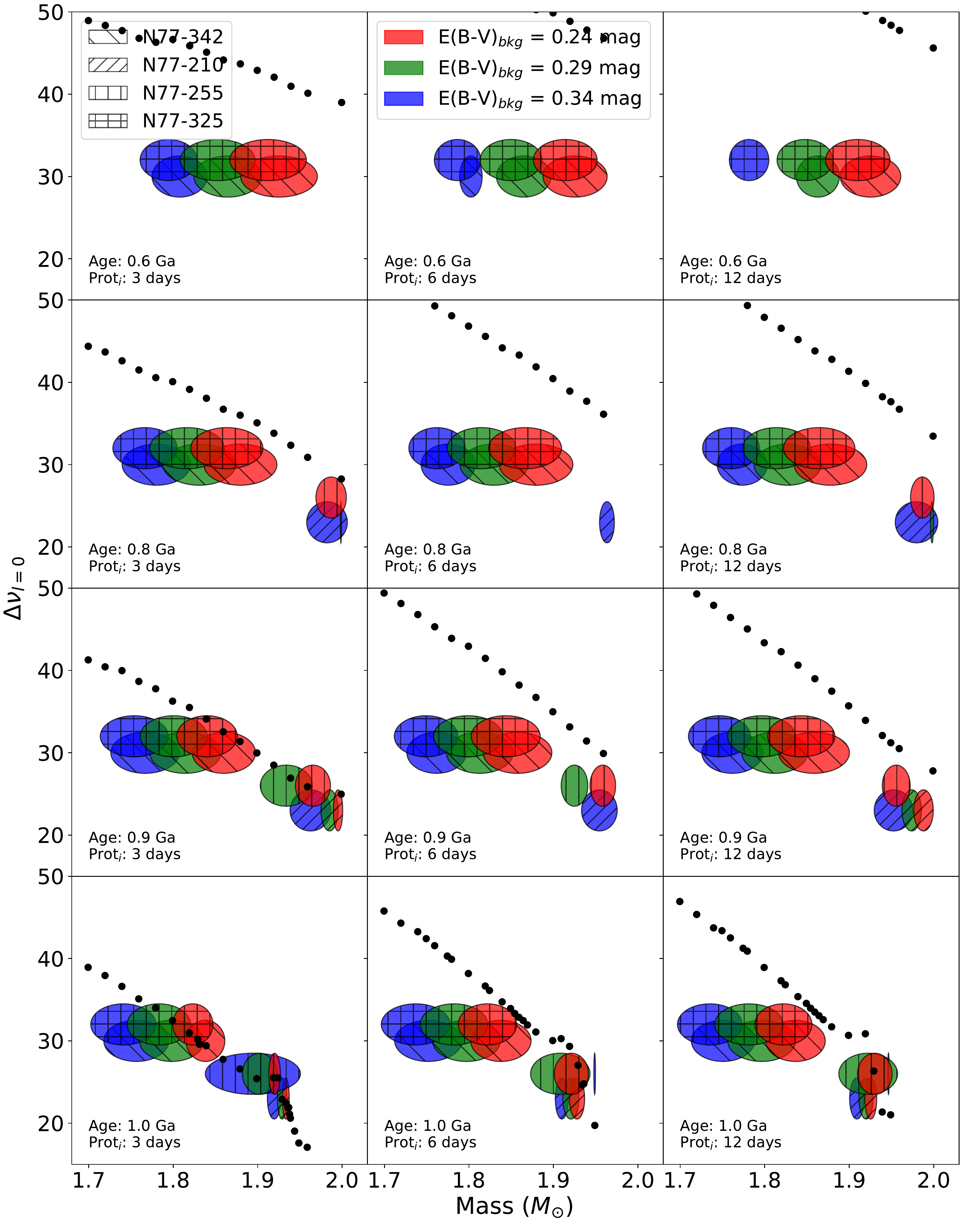}
     \caption{{Theoretical masses versus \deltanu\ for isochrones of different Prot$_{i}$ (3, 6, and 12~days) and age (0.6, 0.8, 0.9, and 1.0~Ga) with same \mh\ = 0.1~dex.  The observed \deltanu\ and inferred masses of  the four stars N77-210, N77-255, N77-325, and N77-342}  are shown as ellipses. See caption of Fig.~\ref{fig:deltanu_two_stars} for details. }
     \label{fig:rotation_deltanu}
\end{figure*}

\begin{figure*}
\centering
    \includegraphics[width=\textwidth]{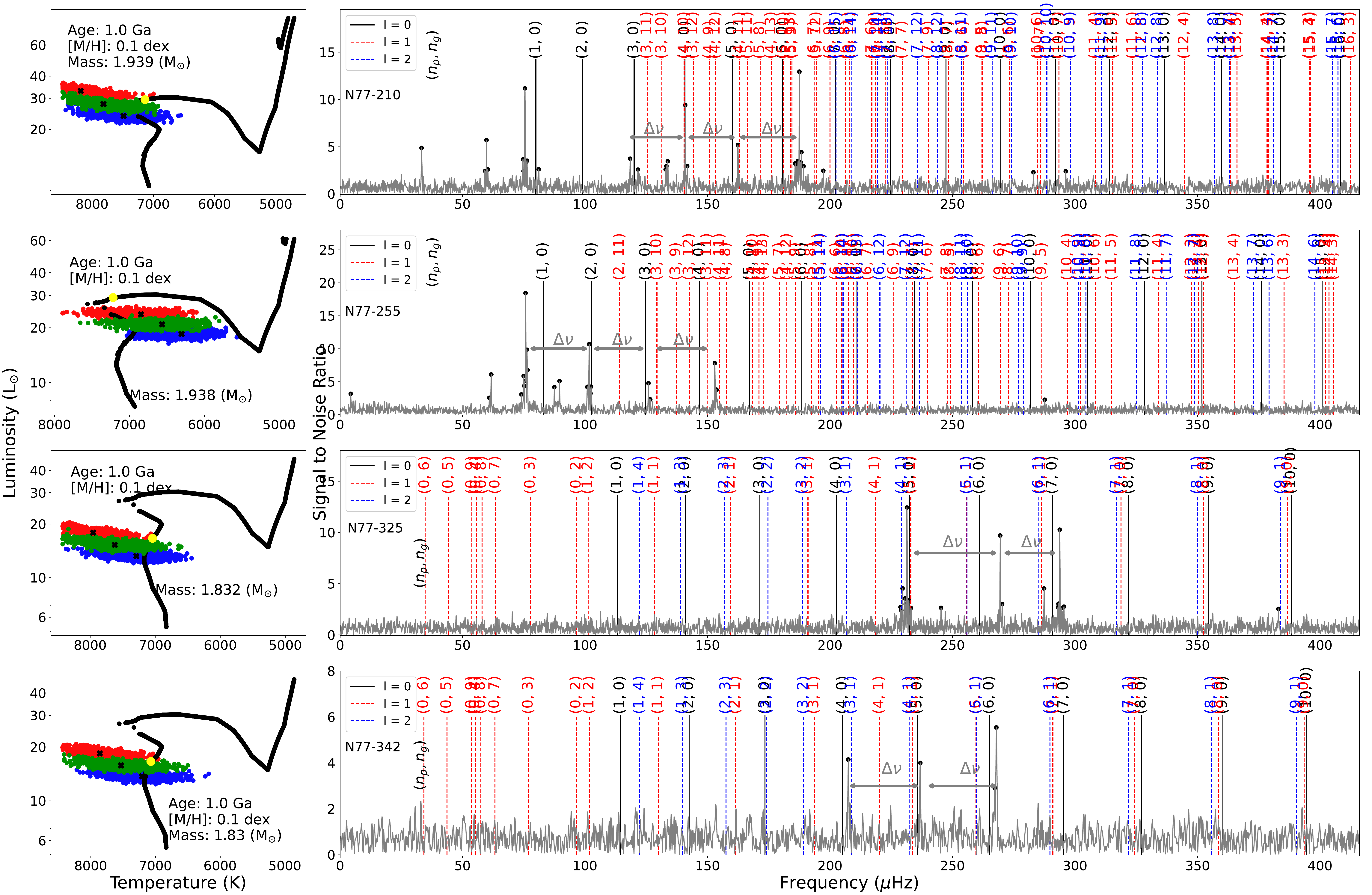} \\   
     \caption{
    See caption of Fig.~\ref{fig:N77_342_lower_age} for details.
    {\sl Left:} Our isochrone using rotation (black) with an age of 1.0~Ga, \mh\ = 0.1~dex and Prot$_{i}$ = 3 days for the 
     stars N77-210, N77-255, N77-325, and N77-342 (top to bottom). {The scattered red, green and blue points represent the LT error of stars for \ebvbkg\ = 0.24, 0.29, and 0.34~mag, respectively.} {\sl Right:} Observed frequencies (grey) and the corresponding pulsation models are shown for the four stars.}
     \label{fig:rotation_frequencies}
\end{figure*}
}
}

\section{Conclusions and remarks}\label{sec:conclusions}

This work focuses on the analysis of the cluster \ngc.  
We aimed to combine the most recent observations available in the literature with those from Gaia and then to use the TESS FFI to search for and analyse the oscillating stars in the cluster.   The objective was to investigate if we could use TESS observations to better constrain the ages of clusters.  
As TESS is an all-sky survey, this would open the possibility to better age-date clusters and to test asteroseismic models.

We reviewed and analysed the most recent spectroscopic, photometric, and extinction observations for the cluster.  
The available literature in the data gives a wide range of metallicities, but in this work we conclude that the \mh\ of the cluster is best described by 0.10 $\pm$ 0.05~dex.   
Using 2D dust maps and the data from \gaia\ DR2, we have analysed the presence of dust in the direction of the cluster. 
We confirmed that the differential reddening that is observed across the cluster in the 2D maps 
is not due to the background extinction, \ebvbkg.
To derive individual reddening values for each of the cluster members, 
we analysed several extinction maps and sources of individual measurements, and we finally
adopted the 2D extinction map from  \cite{schlegel_1998} after correcting for the background extinction.  

We then performed our own fitting of the cluster properties using isochrones that are available in the literature    
(PARSEC, BASTI, and MIST).
This analysis provided the minimum and maximum age of the cluster, 0.6 to 1.1~Ga, which we used later as a constraint in the 
asteroseismic analysis.  We  derived the intrinsic properties $L$, \teff, and mass for all of the member stars.

We analysed the TESS FFI and extracted the light curves of many variable sources in the cluster using the  {\tt tessipack} software \citep{palakkatharappil_2021}. 
Of the variable sources we found, see Table~\ref{tab:sourceinfo}, we confirm  16 non-contaminated oscillating stars and five binary systems.  By non-contaminated we imply that we can confirm that the oscillating signature is arising from the source that we analysed, and not a nearby star. The frequencies of the oscillating stars are then extracted using signal simulations.

We then used the MESA and GYRE stellar structure, evolution and oscillation codes in order to interpret the oscillations of the  uncontaminated stars.   We created a grid of models within the general constraints given by the cluster:  age between 0.6 and 1.1~Ga, $0.06~<$~\mh~$<~0.16$ and $0.24~<$~ \ebvbkg~$<~0.34$~mag. 
We ran pulsation models for the set of oscillating stars and compared the theoretical frequencies with the observed ones. {We identified the frequency separation \deltanu\ for four A-F type stars by comparing the observed with theoretical spectra. }
Then using the identified \deltanu\ for these stars and imposing that the best matched theoretical models have the same age, \mh, and \ebvbkg, we found that the best theoretical match implied an age of the cluster to be 1.0 $\pm$ 0.1~Ga.   We could not constrain the metallicity or extinction any further with the seismic information.   Including rotation in the models led to the same conclusion.

{In this work, we have used standard stellar structure and evolution, and pulsation models to interpret the frequency content of the oscillating stars.  %
We have adopted one value for the initial helium mass fraction, in line with the chemical enrichment law used in the BASTI isochrones (Table~\ref{tab:iso_modelling}). This is a necessary input ingredient in specifying the initial conditions of a stellar model.    
We cannot discard that the effect of changing this value could impact the age of the star.  Additionally, the effects of mixing length and overshooting parameters could also impact the frequencies \citep{miszuda_2021}, we have, however, run several test models and concluded that the impact of the mixing-length changes would be minimal for the oscillating stars in this work.  Further exploration of these is outside the scope of this work.}

In conclusion, in this work we have shown that using the {30 and 10 minutes} cadence TESS FFI data we can identify and exploit the oscillating stars in clusters and that further constraints on the age can be provided by asteroseismology.

\begin{acknowledgements}
The authors would like to thank Dr. Frederic Thevenin and Dr. Mathias Schultheis for their valuable discussions.
\end{acknowledgements}

%
  \bibliographystyle{aa} 
  \bibliography{bib.bib} 
 
\appendix
\section{Additional figures and tables}
  \begin{figure*}
 \centering
    \includegraphics[width=\textwidth]{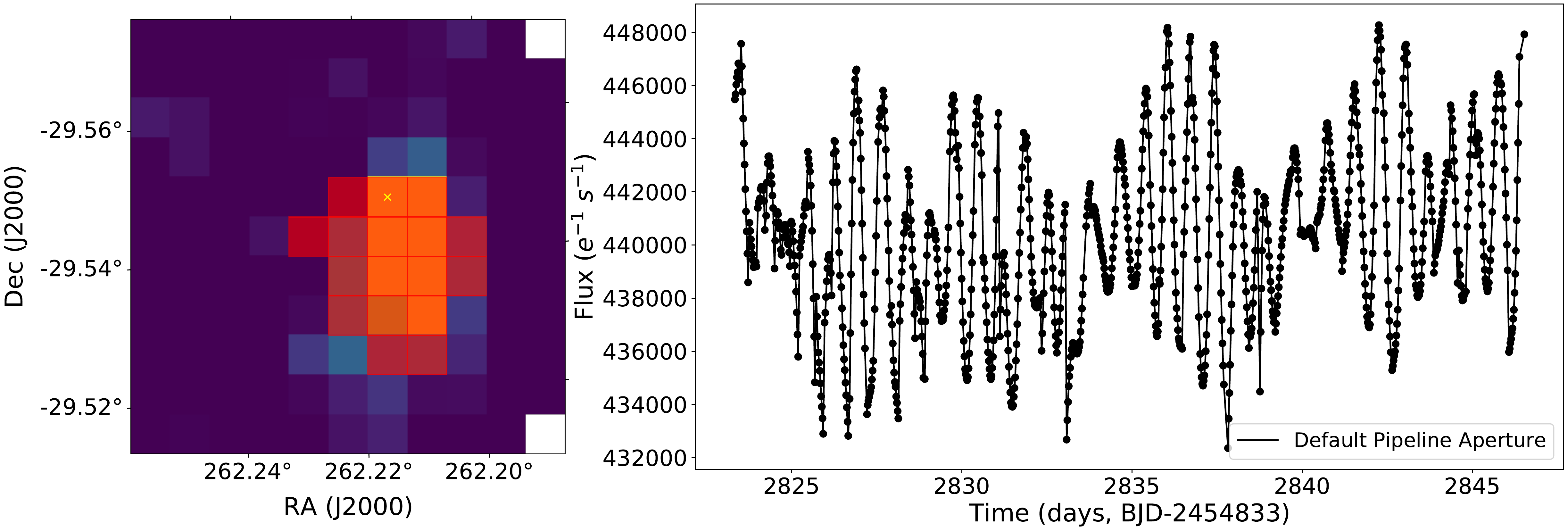}
    \includegraphics[width=\textwidth]{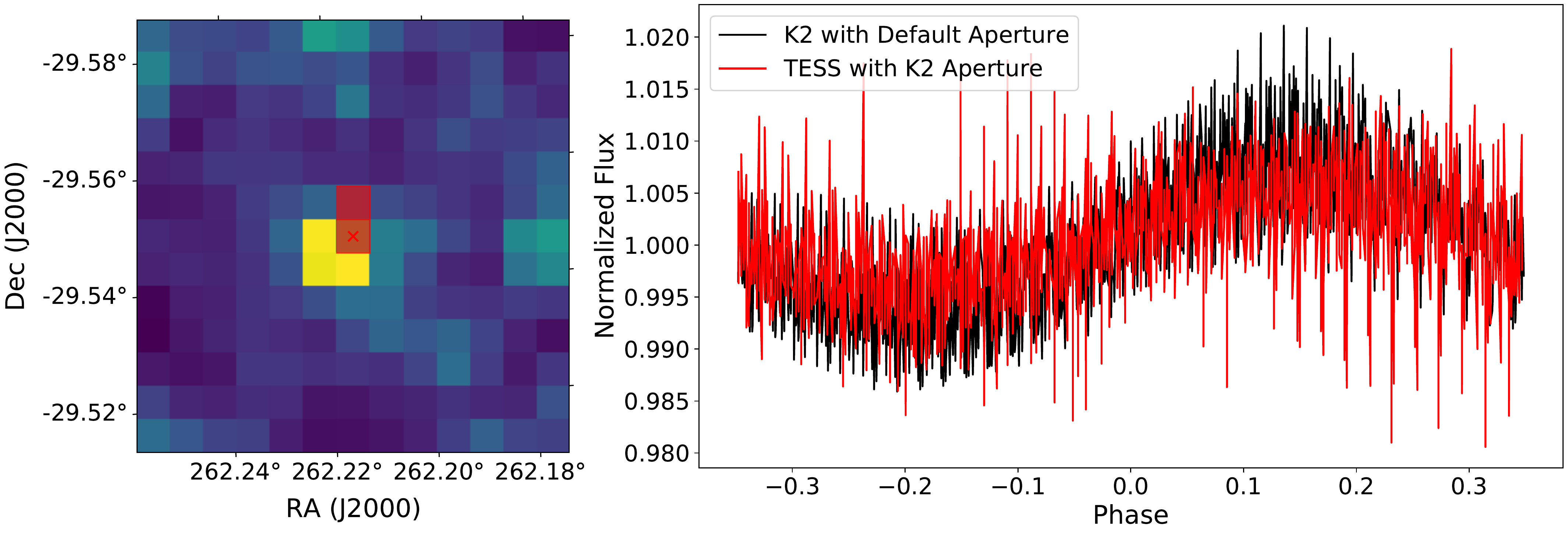}
      \caption{Comparison of light curves for TIC 187129574.
      {\sl Left panels:} TIC 187129574 target pixel files from K2 (top) and TESS FFI (bottom) overlayed with the K2 aperture. 
      {\sl Top Right}: Light curve from K2 data generated with the K2 pipeline.
      {\sl Bottom Right:} Folded light curve generated using the {\tt tessipack} package with the K2 aperture overlayed with the folded light curve from K2.}
      \label{fig:TIC_187129574_TESS}
 \end{figure*}

\begin{figure*}[h]
    \centering
    \includegraphics[width=\textwidth]{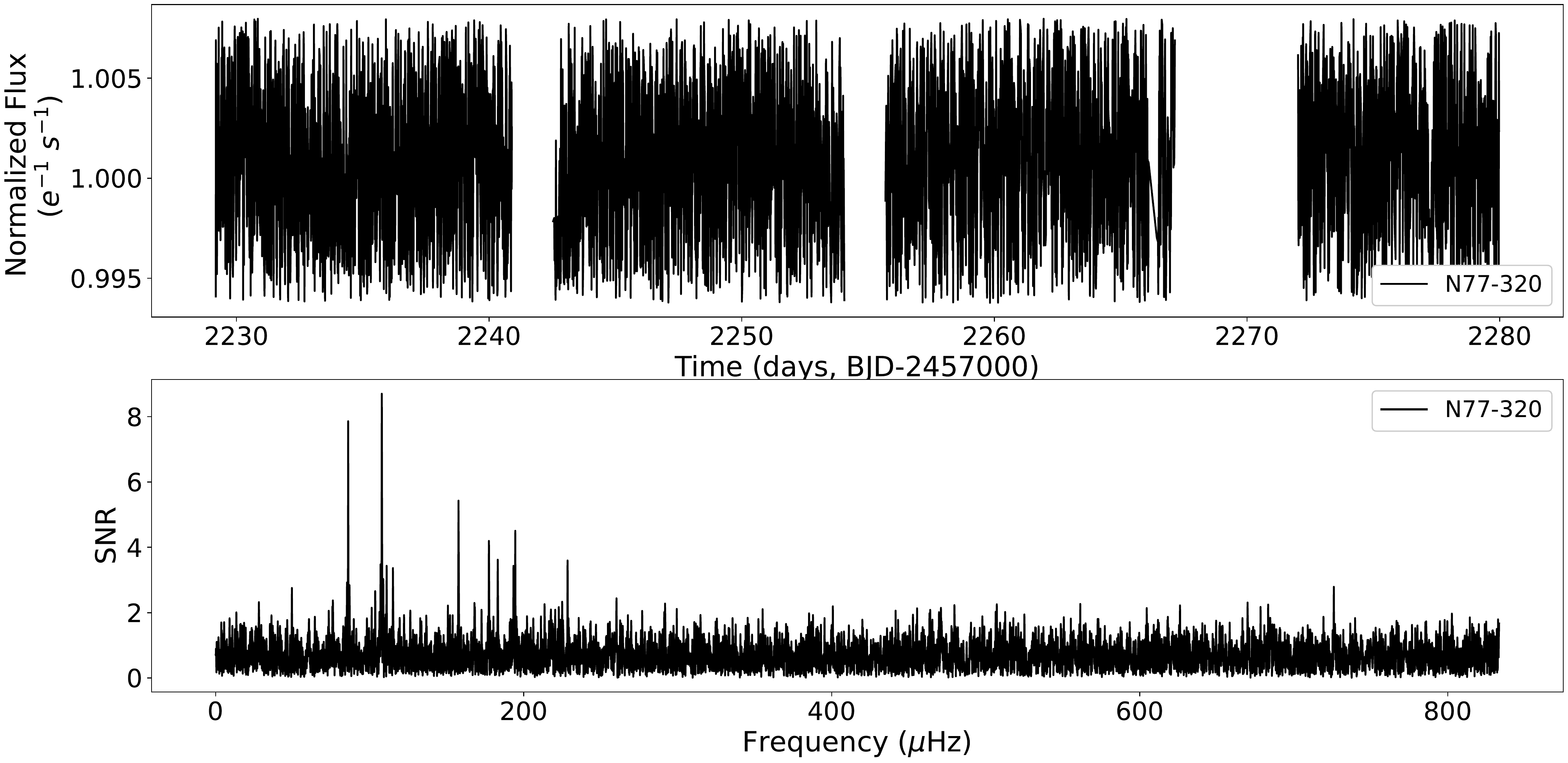}
    \caption{Light curve and periodogram for the oscillating star N77-320.}
    \label{fig:n77_320_osc}
\end{figure*}

\begin{figure*}[h]
    \centering
    \includegraphics[width=\textwidth]{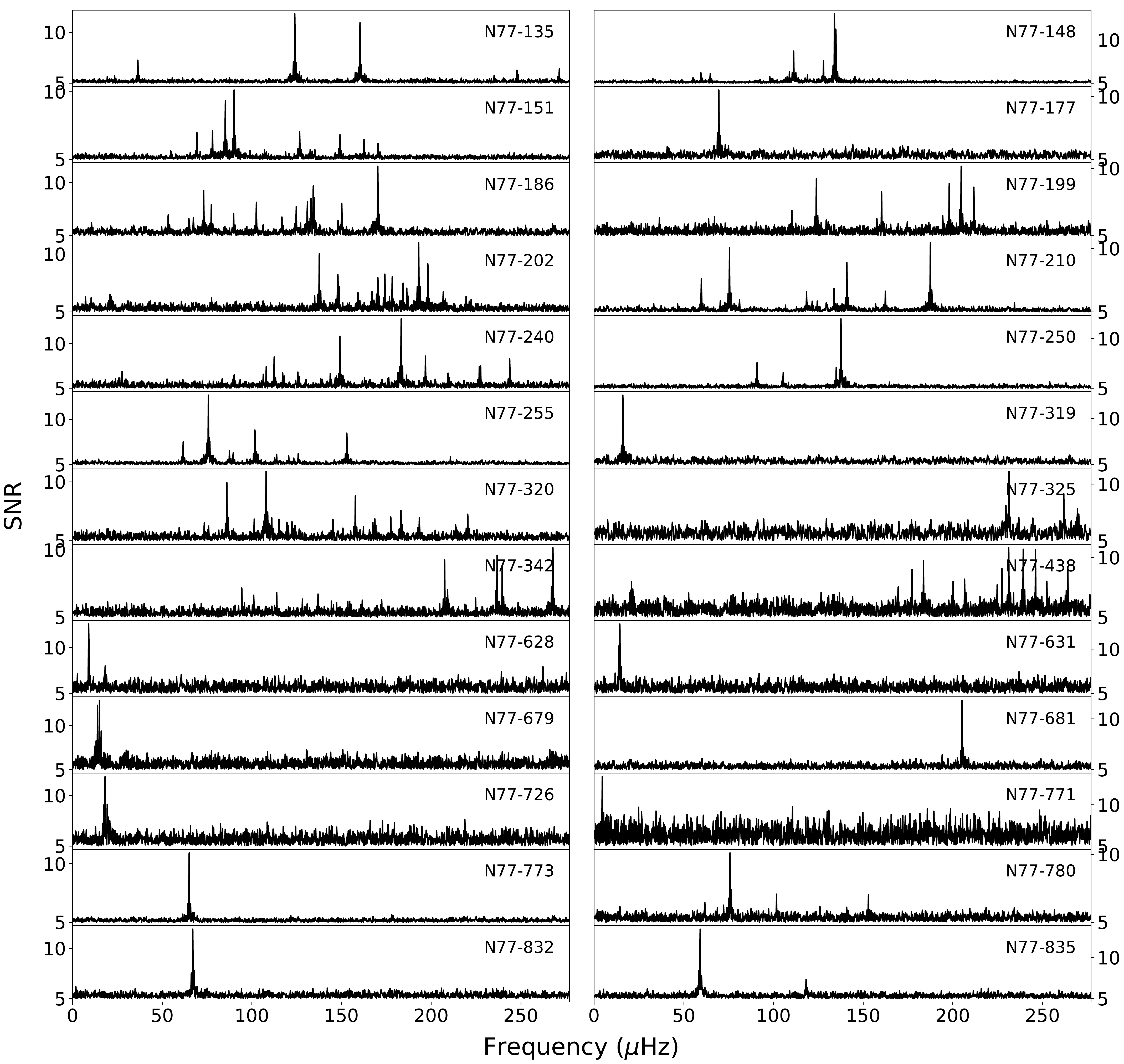}
    \caption{Periodogram for uncontaminated variable stars in the cluster.}
    \label{fig:all_osc_star}
\end{figure*}

\begin{table}[h]
\begin{center}
\caption{Identified individual frequencies in $\mu$Hz for oscillating stars. The fraction of times that the signal is above the S/N is given in $f_{det}$ column (see. Sect.~\ref{ssec:extraction_of_freq}). }
\begin{tabular}{lccc}
\toprule
  Source &  Frequency &    S/N &  $f_{det}$ \\
\midrule
  N77-135 &   19.417587 &   2.319588 &  1.000 \\
 N77-135 &   39.154242 &   2.284336 &  1.000 \\
 N77-135 &  112.038565 &   2.651557 &  1.000 \\
 N77-135 &  123.889674 &   6.148571 &  1.000 \\
 N77-135 &  159.488584 &   2.470550 &  1.000 \\
 N77-135 &  160.263464 &   6.701729 &  1.000 \\
 N77-135 &  197.959108 &   2.697782 &  1.000 \\
 N77-135 &  204.659543 &   3.926213 &  1.000 \\
 N77-135 &  211.724628 &   2.693564 &  1.000 \\
 N77-135 &  231.324539 &   2.206540 &  1.000 \\
 N77-135 &  233.330112 &   2.265976 &  1.000 \\
 N77-135 &  247.779349 &   2.904632 &  1.000 \\
 N77-148 &   32.718338 &   2.434745 &  1.000 \\
 N77-148 &   33.265163 &   2.493691 &  1.000 \\
 N77-148 &   54.226773 &   2.904365 &  1.000 \\
 N77-148 &   55.047009 &   3.891692 &  1.000 \\
 N77-148 &   59.467175 &   5.397474 &  1.000 \\
 N77-148 &   60.469687 &   3.917086 &  1.000 \\
 N77-148 &   63.932909 &   2.714217 &  1.000 \\
 N77-148 &   64.707577 &   4.764354 &  1.000 \\
 N77-148 &   97.881603 &   6.126372 &  1.000 \\
 N77-148 &   98.701840 &   3.468055 &  1.000 \\
 N77-148 &  106.630797 &   5.016427 &  1.000 \\
 N77-148 &  107.724446 &   3.823833 &  1.000 \\
 N77-148 &  108.043427 &   2.720897 &  1.000 \\
 N77-148 &  108.863664 &   7.774854 &  1.000 \\
 N77-148 &  109.456057 &   2.908640 &  1.000 \\
 N77-148 &  109.957313 &   2.492356 &  1.000 \\
 N77-148 &  110.413000 &   5.064274 &  1.000 \\
 N77-148 &  110.686412 &   5.082064 &  1.000 \\
 N77-148 &  110.914256 &   3.777943 &  1.000 \\
 N77-148 &  111.233237 &  20.651092 &  1.000 \\
 N77-148 &  111.552218 &   4.566577 &  1.000 \\
 N77-148 &  111.780061 &   3.893439 &  1.000 \\
 N77-148 &  112.053474 &   6.938028 &  1.000 \\
 N77-148 &  112.463592 &   8.430005 &  1.000 \\
 N77-148 &  112.873711 &   2.414219 &  1.000 \\
 N77-148 &  113.192692 &   2.581871 &  1.000 \\
 N77-148 &  113.466104 &   2.675118 &  1.000 \\
 N77-148 &  113.785085 &   3.760001 &  1.000 \\
 N77-148 &  117.248307 &   2.477503 &  1.000 \\
 N77-148 &  118.934350 &   4.525638 &  1.000 \\
...&...&...&  \\
 \bottomrule
\label{tab:oscillation_source}
\end{tabular}
\end{center}
{The entire table is also available in machine-readable form in the electronic edition of A $\&$ A.}
\end{table}

\begin{figure*}[h]
    \centering
    \includegraphics[width=\textwidth]{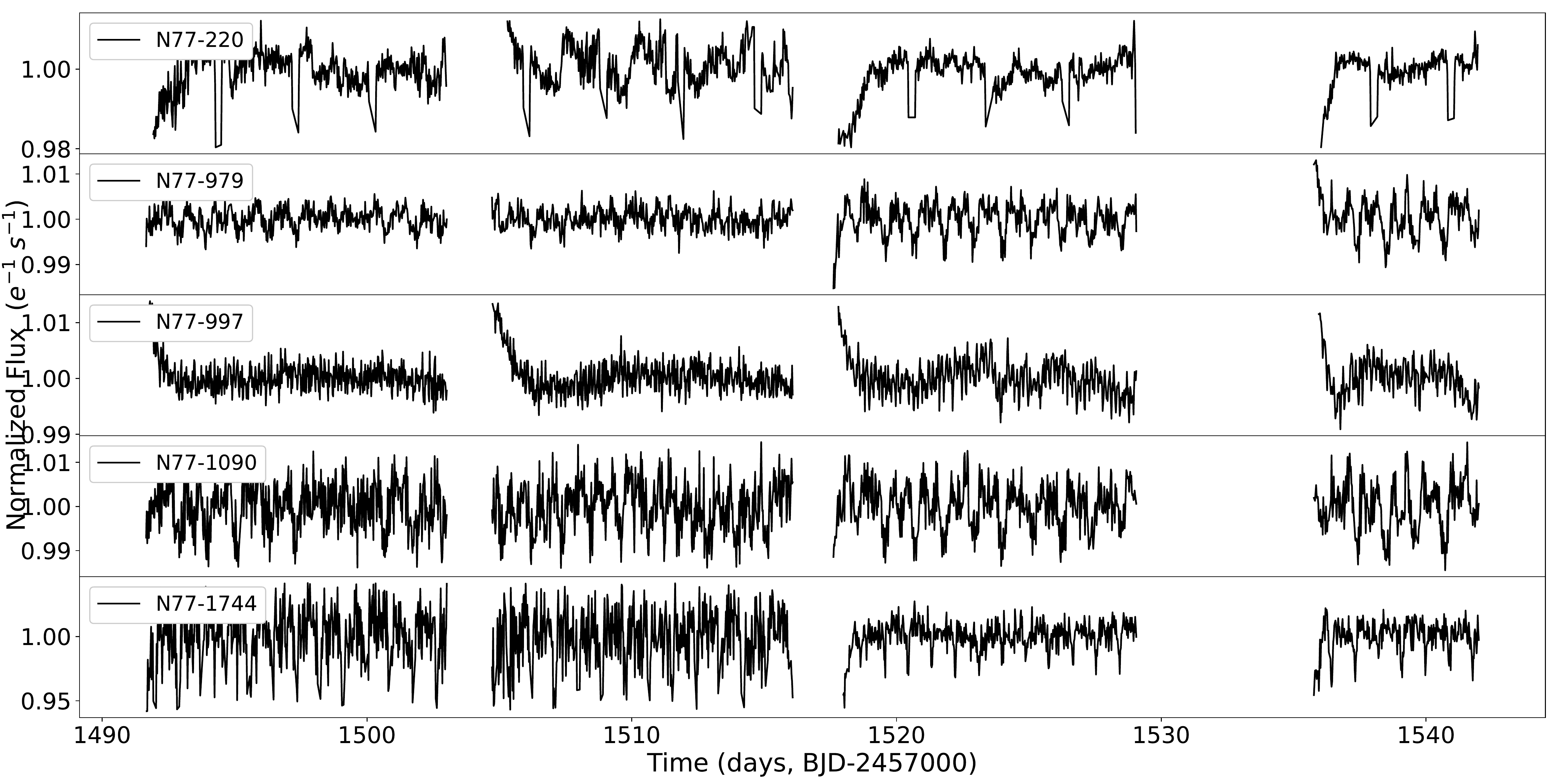}
    \caption{Light curves obtained from TESS FFI of identified candidate binary stars in NGC 2477. Data from two sectors (7 and 8) are  used to create light curve.}
    \label{fig:all_binary_star}
\end{figure*}

\begin{table*}[]
\caption{Summary of Adopted Physics}
\begin{tabular}{@{}llll@{}}
\toprule
                        & \multicolumn{2}{c}{Adopted Parameters}                                                                                                                                                                                                                  \\ 
Ingredient               & Mesa Models without Rotation                                                                                                      & Mesa Models with Rotation  \\ \midrule
Equation of state        & OPAL & OPAL\\
Opacity                  & GS98 \citep{grevesse_1998} & AGSS09 \citep{heger_2000b} \\
Reaction Rates           & JINA REACLIB & JINA REACLIB \\
Boundary conditions      & T\_tau (Eddington)& T\_tau (Eddington) \\
Diffusion                & False & True \\
Rotation                 & False & Implemented\\
MLT                      & $\alpha_{MLT}$ = 1.9179 & $\alpha_{MLT}$ = 2.1 \\
Overshoot                & Not implemented &\begin{tabular}[c]{@{}l@{}} time-dependent, diffusive,\\ $f_{ov,core}$ = 0.0160  , $f_{ov,sh}$ =  0.002, \\$f_{ov,env}$ =  0.002 \citep{herwig_2000}    \end{tabular} \\
Rotational mixing        & Not implemented  & \begin{tabular}[c]{@{}l@{}}Include SH, ES, GSF, SSI, and DSI \\ $f_{\nu}$ = 0.05 and $f_{c}$ = 1/30\end{tabular} \\
Magnetic effects         & Not implemented & \begin{tabular}[c]{@{}l@{}}Magnetic braking  \\ \citep{garraffo_2018, gossage_2021}\end{tabular} \\
Mass loss: low mass star & \begin{tabular}[c]{@{}l@{}}$\eta_{R}$ = 0 for the RGB\\ $\eta_{R}$ = 0 for the AGB\end{tabular} & \begin{tabular}[c]{@{}l@{}}$\eta_{R}$ = 0.1 for the RGB\\ $\eta_{R}$ = 0.1 for the AGB\end{tabular} \\ \bottomrule
\end{tabular}
\\
Note: The details of notations and parameters are identical to those in \citep{paxton_2011} and \citep{choi_2016a}.
\label{tab:meas_parameters}
\end{table*}

 \end{document}